\documentclass[12pt,a4paper]{article}
\usepackage[utf8]{inputenc}
\usepackage[english]{babel}
\usepackage{amsmath}
\allowdisplaybreaks[3]
\usepackage{amsfonts}
\usepackage{amssymb}
\usepackage{graphicx}
\usepackage[left=2cm,right=2cm,top=2cm,bottom=2cm]{geometry}
\usepackage{array}
\usepackage{slashed}
\usepackage{multirow}
\usepackage{xcolor}

\usepackage[backend=bibtex,style=numeric,citestyle=numeric-comp,sorting=none,doi=false]{biblatex}
\usepackage{hyperref}
\hypersetup{linktocpage=false,colorlinks=true,linkcolor=blue,citecolor=blue}

\newcommand{\csw}{c_\mathrm{SW}}

\newcommand{\Tr}{\mathrm{Tr}}
\newcommand{\intdk}{\int\limits_{-\pi}^\pi\frac{\mathrm{d}^4k}{(2\pi)^4}}
\newcommand{\intd}[1]{\int\limits_{-\pi}^\pi\frac{\mathrm{d}^4 #1 }{(2\pi)^4}}

\title{Stout smearing and Wilson flow in \\lattice perturbation theory}

\author{Maximilian Ammer$\,^{a}$, Stephan D\"urr$\,^{a,b}$}

\date{}

\begin{document}
\maketitle

\begin{center}
${}^a${\sl Department of Physics, University of Wuppertal, 42119 Wuppertal, Germany}\\
${}^b${\sl J\"ulich Supercomputing Centre, Forschungszentrum J\"ulich, 52425 J\"ulich, Germany}
\end{center}

\vspace{10pt}

\begin{abstract}
We present the expansion of stout smearing and the Wilson flow in lattice perturbation theory to order $g_0^3$, which is suitable for one-loop calculations.
As the Wilson flow is generated by infinitesimal stout smearing steps, the results are related to each other by taking the appropriate limits. We show how to apply perturbative stout smearing or Wilson flow to the Feynman rules of any lattice fermion action and illustrate them by
calculating the self-energy of the clover-improved Wilson fermion.
\end{abstract}
%
\section{Introduction\label{sec:intro}}
Link smearing is a standard recipe for reducing UV fluctuations in lattice gauge theory calculations, e.g.\ lattice QCD.
It is particularly useful when applied to the covariant derivative in a lattice fermion action.
Long ago it was noticed that the taste breaking of staggered fermions gets reduced by link
smearing \cite{Blum:1996uf,Orginos:1999cr,Hasenfratz:2001hp}, and similarly the additive mass shift
(taken as a measure for the severity of chiral symmetry breaking) of Wilson/clover fermions
gets reduced by link smearing \cite{DeGrand:1998jq,Bernard:1999kc,Stephenson:1999ns,Zanotti:2001yb,DeGrand:2002vu}.

In recent years the ``stout smearing'' procedure, proposed in Ref.~\cite{Morningstar:2003gk}, has become very popular,
since it yields a smeared link $U^{(n)}_\mu(x)$ which depends in a smooth (differentiable) manner on the unsmeared gauge
link $U_\mu(x)$ and its neighbors forming the ``staple'' around it.
This allows {one} to use smeared fermion actions in the HMC/RHMC algorithm \cite{Duane:1987de,Gottlieb:1987mq,Clark:2006fx},
which crucially relies on the differentiability with respect to the gauge field.

A related recipe is known as ``gradient flow'' or ``Wilson flow'' \cite{Luscher:2010iy,Luscher:2011bx,Luscher:2013cpa}.
The flow time $t$ (with the dimension of a length squared) {parametrizes} the smooth deformation from the unsmeared $U_\mu(x)$ to the
smeared/flowed $U_\mu(x,t)$, and the recipe was originally formulated as a differential equation (of the diffusion type) in $t$.
In fact, a sequence of stout smearings whose cumulative smearing parameter matches the flow time is the simplest possible integration
scheme (``forward Euler'') to implement the gradient flow -- see the discussion around Eq.~(\ref{eq:stout_gradflow_equivalence}) below for details.

The main difference, in contemporary use, between ``stout smearing'' and ``gradient/Wilson flow'' concerns what is kept fixed
in the continuum limit (i.e.\ if the lattice spacing is sent to zero, $a\to 0$, while the box size $L$ and the renormalized
quark masses $m_q$, all in physical units, remain constant).
With ``stout smearing'' it is common practice to keep both the smearing parameter ${\varrho}$ and the number of smearings $n_\mathrm{stout}$ fixed.
With ``gradient/Wilson flow'' it is common practice to keep the flow time $t$ (in physical units) constant as the continuum limit is taken,
whereupon the flow time in lattice units $t/a^2$ (dimensionless) grows in proportion to the {cutoff} squared.

The rationale behind the former choice is that it results in an {ultralocal} modification of the action.
Accordingly, a study with, say, seven stout smearings is guaranteed to yield the same continuum limit as with an unsmeared action
(see e.g.\ Ref.~\cite{BMW:2008jgk} for an example of this philosophy).
The rationale behind the latter choice is that $t^{-1/2}$ introduces a second regulator, independent of the {cutoff} $a^{-1}$,
which persists (and stays finite) in the continuum limit.
One ends up, with a flow-scale regulated theory (or a perfectly legitimate renormalization scheme and scale), which, contrary to the lattice
regulation, shows universal features \cite{Suzuki:2013gza,Luscher:2013vga,Makino:2014taa,Luscher:2014kea,Monahan:2015lha,Ramos:2015dla,Suzuki:2020zue,Nada:2020jay}.
Hence, in the future, one might calculate a light quark mass or $\alpha_\mathrm{s}$ with staggered fermions at the fixed flow scale $t^{-1/2}=3\,\mathrm{GeV}$.
Upon taking the continuum limit, all memory of the gluon/fermion action combination, which has been used in the computation, is lost, and one ends up with
a genuine physics result in the universal flow scheme.
Hence, no conversion to the $\overline{\mathrm{MS}}$ scheme is needed any more,
the advantage being that the renormalization scheme employed is well defined beyond perturbation theory.
There are encouraging signs that the continuum community is getting aware of the advantages of the flow scheme and provides the necessary
computational tools \cite{Harlander:2016vzb,Harlander:2018zpi,Artz:2019bpr,Harlander:2020duo,Harlander:2022tgk,Borgulat:2023xml}.

So far, we discussed the smoothing of gauge fields.
A second vital ingredient in the field-theoretic setup of many lattice QCD computations is the concept of Symanzik improvement \cite{Symanzik:1983dc}.
Specifically for Wilson fermions this program offers the opportunity of relegating the leading cutoff effects from $O(a)$ to $O(\alpha_\mathrm{s}^k a)$ or $O(a^2)$,
provided the coefficient $\csw$ in front of the Sheikholeslami-Wohlert term is properly chosen \cite{Sheikholeslami:1985ij,Heatlie:1990kg,Sommer:1997jg}.
This improvement program may be carried out either perturbatively or nonperturbatively, but even in the latter case the perturbative results are vital
for cross-checks and to stabilize the nonperturbative fits \cite{Sommer:1997jg}.

In order to combine the merits of link smoothing and Symanzik improvement, a valid framework for carrying out perturbative calculations with smoothed actions is mandatory.
Early steps in this direction have been taken in Refs.~\cite{Bernard:1999kc,Hasenfratz:2001tw,DeGrand:2002va,DeGrand:2002vu}.
More recently, the mutual benefits which come from combining link smearing with clover improvement for undoubled fermion actions
have been emphasized in Ref.~\cite{Capitani:2006ni}.
What is still missing is a universal framework that allows for an easy derivation of Feynman rules for fermion actions with arbitrary smearing recipes,
i.e.\ an arbitrary $(\varrho,n_\mathrm{stout})$ combination or an arbitrary flow time $t/a^2\in\mathbb{R}$.
In the present paper we try to fill this gap.
In particular we try to convince the reader that a ``brute force'' approach for deriving the Feynman rules (i.e.\ inserting the perturbative expansion
of the smeared/flowed variable into the underlying action) is not recommendable for $n_\mathrm{stout}>1$.
We advocate a more elegant method, based on an $SU(N_c)$ expansion, which keeps intermediate expressions much shorter.

The remainder of this article is organized as follows.
In Sec.~2 we work out the perturbative expansion of stout smearing, and in Sec.~3 we do the same thing for the gradient/Wilson flow.
As a by-product the perturbative matching between these recipes is discussed, see Eq.~(\ref{eq:stout_gradflow_equivalence}).
In Sec.~4 we discuss the procedure which yields, for any smoothing recipe, the Feynman rules in a manageable form.
As an application of this procedure we study in Sec.~5 the self-energy of a stout smeared clover fermion with variable $\varrho$ and $1 \leq n_\mathrm{stout} \leq 4$
or under the gradient flow.
Finally, Sec{tion}~6 gives a short summary and an outlook.
Some preliminary work on this topic is found in Ref.~\cite{Ammer:2021koh}.

%
\section{Perturbative expansion of stout smearing\label{sec:pert_stout}}
%
\subsection{Definition of stout smearing\label{subsec:stout}}
%
Stout smearing is a unitary transformation of the link variable $U_\mu(x)$, which depends on the plaquettes containing $U_\mu(x)$. Thereby the configuration of link variables is smoothed. Often multiple smearing steps are performed to increase the smoothing effect. We write  $U^{(n)}_\mu(x)$ for the link variable after $n$ smearing steps. The  transformation to obtain  $U^{(n+1)}_\mu(x)$ is given by
\begin{align}
U^{(n+1)}_\mu(x)&=\exp\big\{i\varrho Q^{(n)}_\mu(x)\big\}U^{(n)}_\mu(x)\label{eq:U(n+1)}\\
U^{(0)}_\mu(x)&=U_\mu(x),
\label{eq:U(0)=U}
\end{align}
with the smearing parameter $\varrho$ and the {H}ermitian operator $Q^{(n)}_\mu(x)$, which is constructed from link variables of the previous smearing step according to
\begin{align}
Q^{(n)}_\mu(x)&=\frac{1}{2i}\left(W^{(n)}_\mu(x)-\frac{1}{N_c}\Tr\left[W^{(n)}_\mu(x)\right]\right)\\
W^{(n)}_\mu(x)&=S^{(n)}_\mu(x)U^{(n)\dagger}_\mu(x)-U^{(n)}_\mu(x)S^{(n)\dagger}_\mu(x)\\
S^{(n)}_\mu(x)&=\sum\limits_{\nu\neq\mu}\left(U^{(n)}_\nu(x)U^{(n)}_\mu(x+\hat{\nu})U^{(n)\dagger}_\nu(x+\hat{\mu})+U^{(n)\dagger}_\nu(x-\hat{\nu})U^{(n)}_\mu(x-\hat{\nu})U^{(n)}_\nu(x+\hat{\mu}-\hat{\nu})\right).
\end{align}
The superscript $(n)$ denotes quantities that are constructed from link variables after $n$ smearing steps. The quantity  $S^{(n)}_\mu(x)$ is referred to as the sum of staples around the link $U^{(n)}_\mu(x)$ and  $iQ^{(n)}_\mu(x)$ is constructed as the projection of the product $S^{(n)}_\mu(x)U^{(n)\dagger}_\mu(x)$ onto the algebra $\mathfrak{su}(N_c)$.

%
\subsection{Stout smearing in perturbation theory\label{subsec:pert_exp_stout}}
%
In lattice perturbation theory the group element $U_\mu(x)\in SU(N_c)$ is written as the exponential of a linear combination of generators $T^a$ and gluon fields $A_\mu^a(x)$
\begin{align}
U_\mu(x)&=e^{ig_0T^aA^a_\mu(x)},
\end{align}
where the summation over repeated adjoint color indices $a,b,c,\hdots\in\{1,\hdots,N_c^2-1\}$ is implied.
Expanding the action to the desired order in the bare coupling $g_0$ then gives the Feynman rules. For one-loop calculations  we need to expand up to third order in $g_0$
\begin{align}
U_\mu(x)&=1+ig_0T^aA^a_\mu(x)-\frac{g_0^2}{2}T^aT^b A_\mu^a(x) A_\mu^b(x)
-i\frac{g_0^3}{6}T^aT^bT^cA^a_\mu(x)A^b_\mu(x)A^c_\mu(x)+\mathcal{O}(g_0^4).
\label{eq:unsmeared expansion}
\end{align}
The Fourier transform of the gluon fields is chosen to be
\begin{align}
A^a_\mu(x)=\intdk e^{i(x+\hat{\mu}/2)k} A^a_\mu(k)
\;.
\end{align}
The shift $x+\hat{\mu}/2$ is needed to eliminate overall phase factors in our results. 
In order for $U_\mu(x)$ to be unitary, the exponent $ig_0T^aA^a_\mu(x)$ has to be anti-{H}ermitian.
This means the gluon fields $A^a_\mu(x)$ are real (because the generators are {H}ermitian). 
This implies that the momentum space field $A_\mu^a(k)$ fulfils
\begin{align}
A^a_\mu(k)=A^a_\mu(-k)
\,.
\end{align} 
As we are going to be working in momentum space, it is helpful to keep in mind that taking the {H}ermitian conjugate in position space involves the transformation $k\to -k $ in momentum space. \\

For a link variable after stout smearing things are more complicated.
As the smeared link variable $U_\mu^{(n)}(x)$ is again a group element of $SU(N_c)$ \cite{Morningstar:2003gk} we expect it to have an exponential representation similar to the unsmeared one
\begin{align}
U^{(n)}_\mu(x)&=e^{ig_0T^a\tilde{A}^{(n)a}_\mu(g_0,x)}\\
&=1+ig_0T^a\tilde{A}^{(n)a}_\mu(g_0,x)-\frac{g_0^2}{2}T^aT^b\tilde{A}_\mu^{(n)a}(g_0,x) \tilde{A}_\mu^{(n)b}(g_0,x)
\nonumber\\
&-i\frac{g_0^3}{6}T^aT^bT^c\tilde{A}^{(n)a}_\mu(g_0,x)\tilde{A}^{(n)b}_\mu(g_0,x)\tilde{A}^{(n)c}_\mu(g_0,x)+\mathcal{O}(T^4).
\label{eq:smeared_expansion_in_T}
\end{align}
This expansion in terms of generators $T$ does, however, no longer coincide with the expansion in the bare coupling $g_0$. The new fields $\tilde{A}^{(n)a}_\mu(g_0,x)$ are themselves functions of $g_0$.
By expanding 
$\tilde{A}^{(n)a}_\mu(g_0,x)$
in powers of $g_0$ we  will be able to determine its coefficients by comparison with the perturbative expansion of the right{-}hand side of Eq.~(\ref{eq:U(n+1)}). 
This also means that, as we will see below, we will be able to infer the general structure of the coefficients of $\tilde{A}^{(n)a}_\mu(g_0,x)$ from the perturbative expansion of only the first smearing step
\begin{align}
U^{(1)}_\mu(x)&=\exp\big\{i\varrho Q^{(0)}_\mu(x)\big\}U^{(0)}_\mu(x),
\label{eq:U(1)}
\end{align}
which in turn allows us to use the following strategy: Expand Eq.~(\ref{eq:U(1)}) to the desired order in $g_0$ and replace $A^{(0)}_\mu(x)$ by the expansion of 
$\tilde{A}^{(n)a}_\mu(g_0,x)$. This will generate new contributions to higher orders in $g_0$ and give the results for a general smearing step $n+1>1$.\\

We start by defining expansions of $S_\mu(x)$ and  $Q_\mu(x)$ in $g_0$ (we drop the superscript $(0)$ for now):
\begin{align}
S_\mu(x)&=6+ig_0T^aS^a_{1\mu}(x)-g_0^2T^aT^bS^{ab}_{2\mu}(x)
-ig_0^3T^aT^bT^c
S^{abc}_{3\mu}(x)+\mathcal{O}(g_0^4)
\\
Q_\mu(x)&=g_0T^aQ^a_{1\mu}(x)+ig_0^2\Big(T^aT^b-T^bT^a\Big)Q^{ab}_{2\mu}(x)
\nonumber\\
&+g_0^3\Big(T^aT^bT^c+T^cT^bT^a-\frac{1}{N_c}\Tr\big[T^aT^bT^c+T^cT^bT^a\big]\Big)
Q^{abc}_{3\mu}(x)+\mathcal{O}(g_0^4)
\;.
\end{align}
Using the Baker-Campbell-Hausdorff formula
\begin{align}
e^Xe^Y=e^{X+Y+\frac 12 [X,Y]+\frac 1{12}([X,[X,Y]]+[Y,[Y,X]])+\hdots}
\;,
\end{align}
we therefore get
\begin{align}
\exp&\big\{i\varrho Q_\mu(x)\big\}\exp\big\{ig_0T^aA^a_\mu(x)\big\}
=\exp\bigg\{i g_0 T^aA^a_\mu(x) +i\varrho g_0T^aQ^a_{1\mu}(x)-\varrho g_0^2[T^a,T^b]Q^{[ab]}_{2\mu}(x)
\nonumber\\&
+i\varrho g_0^3\Big(T^aT^bT^c+T^cT^bT^a-\frac{1}{N_c}\Tr\big[T^aT^bT^c+T^cT^bT^a\big]\Big)
Q^{abc}_{3\mu}(x)
\nonumber\\&
+\frac{1}{2}\Big(
-\varrho g_0^2[T^a,T^b]Q^a_{1\mu}(x)A^b_\mu(x)
-i\varrho g_0^3[[T^a,T^b],T^c]Q^{[ab]}_{2\mu}(x)A^c_\mu(x)
\Big)
\nonumber\\&
+\frac{1}{12}\Big(
-i\varrho^2g_0^3 [T^a,[T^b,T^c]]Q^{a}_{1\mu}(x)Q^{b}_{1\mu}(x)A^c_\mu(x)
-i\varrho g_0^3 [T^a,[T^b,T^c]]A^{a}_{\mu}(x)A^{b}_{\mu}(x)Q^c_{1\mu}(x)
\Big)
\nonumber\\&
+\mathcal{O}(g_0^4)\bigg\}
\label{eq:U1_after_BCH}
\\
&=1+ig_0T^a\big(A^a_\mu(x)+\varrho Q^a_{1\mu}(x)\big)
-\frac {1}{2} g_0^2T^aT^b\big(A^a_\mu(x)+\varrho Q^a_{1\mu}(x)\big)\big(A^b_\mu(x)+\varrho Q^b_{1\mu}(x)\big)
\nonumber\\
&-\varrho g_0^2[T^a,T^b]\Big(\frac{1}{2} Q^a_{1\mu}(x)A_\mu^b(x)+Q^{ab}_{2\mu}(x)\Big)
\nonumber\\
& -i\frac{g_0}{6}^3T^aT^bT^c 
\Big(A^a_\mu(x)+\varrho Q^a_{1\mu}(x)\Big)
\Big(A^b_\mu(x)+\varrho Q^b_{1\mu}(x)\Big) 
\Big(A^c_\mu(x)+\varrho Q^c_{1\mu}(x)\Big)
\nonumber\\
& -i\frac{g_0^3}{2} \{T^a,[T^b,T^c]\} \Big(A^a_\mu(x)+\varrho Q^a_{1\mu}(x)\Big)\bigg(\frac{1}{2}\varrho Q^b_{1\mu}(x)A^c_\mu(x)+\varrho Q^{bc}_{2\mu}(x)\bigg)\nonumber\\
&+i\frac{g_0^3}{2}[T^a,[T^b,T^c]]A^a_\mu(x)\varrho Q^{[bc]}_{2\mu}(x)
\nonumber\\&
+\frac{1}{12}\Big(
-i\varrho^2g_0^3 [T^a,[T^b,T^c]]Q^{a}_{1\mu}(x)Q^{b}_{1\mu}(x)A^c_\mu(x)
-i\varrho g_0^3 [T^a,[T^b,T^c]]A^{a}_{\mu}(x)A^{b}_{\mu}(x)Q^c_{1\mu}(x)
\Big)
\nonumber\\&
+i\varrho g_0^3\Big(T^aT^bT^c+T^cT^bT^a-\frac{1}{N_c}\Tr\big[T^aT^bT^c+T^cT^bT^a\big]\Big)
Q^{abc}_{3\mu}(x)
+\mathcal{O}(g_0^4)
\label{eq:U1_after_BCH_and_exp}
\end{align}
where we have used
\begin{align}
e^{ax+bx^2+cx^3}=1+ax+\big(\tfrac{1}{2}a^2+b\big)x^2+\big(\tfrac{1}{6}a^3+\tfrac 12 (ab+ba) +c\big)x^3+\mathcal{O}(x^4)
\end{align}
in the second step.
Based on this, it is convenient to define fields $A^{(n)a}$, $A^{(n)ab}$, $\bar{A}^{(n)abc}$, and $\hat{A}^{(n)abc}$ at each order in $g_0$ for a general  $n$ such that
\begin{align}
&U^{(n)}_\mu(x)=1+ig_0T^aA^{(n)a}_\mu(x)-\frac{g_0^2}{2}\Big(T^aT^bA^{(n)a}_\mu(x)A^{(n)b}_\mu(x)+[T^a,T^b] A_\mu^{(n)ab}(x)\Big)
\nonumber\\
&-i\frac{g_0^3}{6}\bigg(T^aT^bT^c A^{(n)a}_\mu(x)A^{(n)b}_\mu(x)A^{(n)c}_\mu(x)
+\frac{3}{2}\{T^a,[T^b,T^c]\}A^{(n)a}_\mu(x) A_\mu^{(n)bc}(x)
\nonumber\\&
+[T^a,[T^b,T^c]]\bar{A}^{(n)abc}_{\mu}(x)
+\Big(T^aT^bT^c+T^cT^bT^a-\frac{1}{N_c}\Tr\big[T^aT^bT^c+T^cT^bT^a\big]\Big)   \hat{A}^{(n)abc}_\mu(x)\bigg)
\nonumber\\&
+\mathcal{O}(g_0^4)
\;.
\label{eq:smearing_expansion_in_g0}
\end{align}
We can now also incorporate $\bar{A}$ into $\hat{A}$ because\footnote{We use the following shorthand notation for {symmetrized} and {antisymmetrized} expressions: \\$A^{\{ab\}}=\frac{1}{2}(A^{ab}+A^{ba})$ and $A^{[ab]}=\frac{1}{2}(A^{ab}-A^{ba})$.}
\begin{align}
[T^a,[T^b,T^c]]\bar{A}^{(n)abc}_{\mu}(x)&=
2[T^a,T^bT^c]\bar{A}^{(n)a[bc]}_{\mu}(x)
=2(T^aT^bT^c+T^cT^bT^a)\bar{A}^{(n)a[bc]}_{\mu}(x)
\end{align}
and
\begin{align}
\Tr[T^aT^bT^c+T^cT^bT^a]\bar{A}^{(n)a[bc]}_{\mu}(x)&
=\Tr[T^aT^bT^c-T^bT^cT^a]\bar{A}^{(n)a[bc]}_{\mu}(x)
=0
\;.
\end{align}
Thus we can write
\begin{align}
U^{(n)}_\mu(x)&=1+ig_0T^aA^{(n)a}_\mu(x)-\frac{g_0^2}{2}\Big(T^aT^bA^{(n)a}_\mu(x)A^{(n)b}_\mu(x)+[T^a,T^b] A_\mu^{(n)ab}(x)\Big)
\nonumber\\
-i\frac{g_0^3}{6}&\bigg(T^aT^bT^c A^{(n)a}_\mu(x)A^{(n)b}_\mu(x)A^{(n)c}_\mu(x)
+\frac{3}{2}\{T^a,[T^b,T^c]\}A^{(n)a}_\mu(x) A_\mu^{(n)bc}(x)
\nonumber\\&
+\Big(T^aT^bT^c+T^cT^bT^a-\frac{1}{N_c}\Tr\big[T^aT^bT^c+T^cT^bT^a\big]\Big)   A^{(n)abc}_\mu(x)\bigg)
+\mathcal{O}(g_0^4)
\end{align}
with 
\begin{align}
A^{(n)abc}_\mu(x)=\hat{A}^{(n)abc}_\mu(x)+2\bar{A}^{(n)a[bc]}_\mu(x)
\end{align}

We define their Fourier transforms as
\begin{align}
A^{(n)a}_\mu(x)&=\intd{k}e^{i(x+\hat{\mu}/2)k}A^{(n)a}_\mu(k)\\
A^{(n)ab}_\mu(x)&=\intd{k_1}\intd{k_2}e^{i(x+\hat{\mu}/2)(k_1+k_2)}A^{(n)ab}_\mu(k_1,k_2)\\
A^{(n)abc}_\mu(x)&=\intd{k_1}\intd{k_2}\intd{k_3}e^{i(x+\hat{\mu}/2)(k_1+k_2+k_3)}A^{(n)abc}_\mu(k_1,k_2,k_3)
\;.
\end{align} 
These fields are ultimately functions of the original gluon fields
\begin{align}
A^{(n)a}_\mu(x)&=\sum\limits_\nu\sum\limits_y \tilde{g}^{(n)}_{\mu\nu}(\varrho,y)A^{(0)a}_\nu(x+y)\\
A^{(n)ab}_\mu(x)&=\sum\limits_{\nu\rho}\sum\limits_{yz} \tilde{g}^{(n)}_{\mu\nu\rho}(\varrho,y,z)A^{(0)a}_\nu(x+y)A^{(0)b}_\rho(x+z) \\
 A^{(n)abc}_\mu(x)&=\sum\limits_{\nu\rho\sigma}\sum\limits_{yzr} \tilde{g}^{(n)}_{\mu\nu\rho\sigma}(\varrho,y,z,r)A^{(0)a}_\nu(x+y)A^{(0)b}_\rho(x+z)A^{(0)c}_\sigma(x+r) \\
 A^{(n)a}_\mu(k)&=\sum\limits_\nu\tilde{g}^{(n)}_{\mu\nu}(\varrho,k)A^{(0)a}_\nu(k)\\
 A^{(n)ab}_\mu(k_1,k_2)&=\sum\limits_{\nu\rho}\tilde{g}^{(n)}_{\mu\nu\rho}(\varrho,k_1,k_2)A^{(0)a}_\nu(k_1)A^{(0)b}_\rho(k_2)\\
 A^{(n)abc}_\mu(k_1,k_2,k_3)&=\sum\limits_{\nu\rho\sigma}\tilde{g}^{(n)}_{\mu\nu\rho\sigma}(\varrho,k_1,k_2,k_3)A^{(0)a}_\nu(k_1)A^{(0)b}_\rho(k_2)A^{(0)c}_\sigma(k_3)
 \label{eq:def_g_tilde}
\end{align}
with some complicated functions simply denoted by $\tilde{g}$ here. Sections \ref{subsec:LO}, \ref{subsec:NLO}, and \ref{subsec:NNLO}
will be largely concerned with determining the momentum space $\tilde{g}$ explicitly.\\

{
We have used \emph{Mathematica} \cite{Wolfram2022} to supplement and
double check our analytical derivations as well as to derive Feynman
rules and perform the calculations of the fermion self energy presented
in Section \ref{sec:self_energy} \cite{anc}.
}
%
\subsubsection{Connecting the perturbative and $SU(N_c)$ expansions\label{sec:connecting_expansions}}
%
In order to relate the smeared gluon field $\tilde{A}^{(n)a}_\mu(g_0,x)$ to the perturbative functions $A_\mu^{(n)a}(x)$, $A_\mu^{(n)ab}(x)$, and $A_\mu^{(n)abc}(x)$, we need to expand the products of color matrices such that only terms with a single $T^a$ remain.
We use
\begin{align}
[T^a,T^b]&=if^{abc}T^c\;, & f^{abc}&=-2i\Tr\big(T^a[T^b,T^c]\big)
\\
\{T^a,T^b\}&=\frac{1}{N_c}\delta^{ab}+d^{abc}T^c\;, & d^{abc}&=2\Tr\big(T^a\{T^b,T^c\}\big)
\end{align}
and
\begin{align}
&T^aT^b=\frac{1}{2N_c}\delta^{ab}+\frac{1}{2}\big(d^{abc}+if^{abc}\big)T^c\\
&\Tr\big[T^aT^bT^c\big]=\frac{1}{2}\Tr\big[T^a\{T^b,T^c\}+T^a[T^b,T^c]\big]
=\frac{1}{4}\big(d^{abc}+if^{abc}\big)
\,.
\end{align}
With the above color identities we can now rewrite the following expressions in a way that only one generator $T$ remains:
\begin{align}
& [T^a,T^b]A^{(n)[ab]}_\mu(x)=if^{abc}T^aA^{(n)[bc]}_\mu(x)
\\
&\bigg(T^aT^bT^c+T^cT^bT^a-\frac{1}{N_c}\Tr\big[T^aT^bT^c+T^cT^bT^a\big]\bigg)A^{(n)abc}_\mu(x)
\nonumber\\&
=\bigg(\frac{1}{2}\Big(\{T^a,\{T^b,T^c\}\}+[T^a,[T^b,T^c]]\Big)-\frac{1}{N_c}\Tr\big[T^aT^bT^c+T^cT^bT^a\big]\bigg)A^{(n)abc}_\mu(x)
\\&
=T^a\bigg[\frac{1}{N_c}\delta^{ae}\delta^{bc}
+\frac{1}{2}\big(d^{bcd}d^{eda}-f^{bcd}f^{eda}\big)\bigg] A^{(n)ebc}_\mu(x)
\;.
\end{align}
Thus we can rearrange our perturbative expansion (\ref{eq:smearing_expansion_in_g0}) to look like 
\begin{align}
&U^{(n)}_\mu(x)=1
+ig_0T^a
\nonumber\\&\times
\Bigg\{
A^{(n)a}_\mu(x)
-\frac{g_0}{2}f^{abc}A^{(n)[bc]}_\mu(x)
-\frac{g_0^2}{6}\bigg[\frac{1}{N_c}\delta^{ae}\delta^{bc}
+\frac{1}{2}\big(d^{bcd}d^{eda}
-f^{bcd}f^{eda}\big)\bigg] A^{(n)ebc}_\mu(x)
+\mathcal{O}(g_0^3)
\Bigg\}
\nonumber\\
&-\frac{g_0^2}{2} T^aT^b \Bigg\{
A_\mu^{(n)a}(x)A_\mu^{(n)a}(x)-\frac{g_0}{2}
\Big(
A^{(n)a}_\mu(x)f^{bcd}+f^{acd}A^{(n)b}_\mu(x)\Big)A^{(n)[cd]}_\mu(x)
+\mathcal{O}(g_0^2) \Bigg\}\nonumber \\
&-i\frac{g_0^3}{6}T^aT^bT^c\Bigg\{
A_\mu^{(n)a}(x)A_\mu^{(n)b}(x)A_\mu^{(n)c}(x)
+\mathcal{O}(g_0)\Bigg\}+\mathcal{O}(T^4).
\label{eq:smeared_exp_in_T_with_As}
\end{align} 
We can see that the expressions in the large braces correspond to the first, second, and third order of an expansion of the exponential function, each cut off to give an overall highest order of $g_0^3$.
Hence we can give the first few orders of the smeared gluon field as
\begin{align}
\tilde{A}_\mu^{(n)a}(g_0,x)&=A^{(n)a}_\mu(x)
-\frac{g_0}{2}f^{abc}A^{(n)[bc]}_\mu(x)
-\frac{g_0^2}{6}
\bigg[\frac{1}{N_c}\delta^{ae}\delta^{bc}
+\frac{1}{2}\big(d^{bcd}d^{eda}-f^{bcd}f^{eda}\big)\bigg] A^{(n)ebc}_\mu(x)
\nonumber\\&
+\mathcal{O}(g_0^3)
\,.
\end{align}
Thus, we have expressed the new gluon field in terms of the perturbative fields.
%
\subsection{Leading order\label{subsec:LO}}
As we have seen, at leading order the perturbative and $SU(N_c)$ expansions are identical: $\tilde{A}^{(n)a}_\mu(g_0,x)=A^{(n)a}_\mu(x)+\mathcal{O}(g_0)$.
The sum of staples at leading order is:
\begin{align}
S^{(n)}_\mu(x)&= 6+ig_0T^a\bigg(\sum\limits_{\nu=1}^4\Big[A_\nu^{(n)a}(x)+A_\mu^{(n)a}(x+\hat{\nu})-A_\nu^{(n)a}(x+\hat{\mu})-A_\nu^{(n)a}(x-\hat{\nu})
\nonumber \\
&+A_\mu^{(n)a}(x-\hat{\nu})+A_\nu^{(n)a}(x+\hat{\mu}-\hat{\nu})\Big]-2A_\mu^{(n)a}(x)\bigg)+\mathcal{O}(g_0^2)
\end{align}
where the sum now includes the $\nu=\mu$ term, which is compensated by the subtraction of $2A_\mu^{(n)a}(x)$. Therefore
\begin{align}
Q^{(n)a}_{1\mu}(x)=&\sum\limits_{\nu=1}^4\Big[A_\nu^{(n)a}(x)+A_\mu^{(n)a}(x+\hat{\nu})-A_\nu^{(n)a}(x+\hat{\mu})-A_\nu^{(n)a}(x-\hat{\nu})
\nonumber \\
&+A_\mu^{(n)a}(x-\hat{\nu})+A_\nu^{(n)a}(x+\hat{\mu}-\hat{\nu})-2A_\mu^{(n)a}(x)\Big]
\\
&=-6A^{(n)a}_\mu(x)+S^{(n)a}_{1\mu}(x)
\,.
\end{align}
The smeared gluon field in position space can then be written as a convolution \cite{Bernard:1999kc}:
\begin{align}
A_\mu^{(n+1)a}(x)&=A^{(n)a}_\mu(x)+\varrho\, Q^{(n)a}_{1\mu}(x)
\\&
=A^{(n)a}_\mu(x)+\varrho\sum\limits_{\nu=1}^4\sum\limits_y g_{\mu\nu}(y) A_\nu^{(n)a}(x+y)
\\&
=:\sum\limits_\nu\sum\limits_y \tilde{g}_{\mu\nu}(\varrho,y) A_\nu^{(n)a}(x+y)
\end{align}
with
$
\tilde{g}_{\mu\nu}(\varrho,y) =\delta_{\mu\nu}+\varrho g_{\mu\nu}(y)
$
 and
\begin{align}
g_{\mu\nu}(y) &=\delta(y,0)-\delta(y,\hat{\mu})-\delta(y,-\hat{\nu})+\delta(y,\hat{\mu}-\hat{\nu})
+\delta_{\mu\nu}\bigg[\sum\limits_{\tau=1}^4\big(\delta(y,\hat{\tau})+\delta(y,-\hat{\tau})\big)-8\delta(y,0)\bigg].
\end{align}
Performing a Fourier transformation using $\delta(x,y)=\intd{p}  e^{i (x-y)p}$, we get:
\begin{align}
& e^{iyp}\Bigg(1-e^{-ip_\mu}-e^{ip_\nu}+e^{-i(p_\mu-p_\nu)}+\delta_{\mu\nu}\bigg[-8+ \sum\limits_{\tau=1}^4\big(e^{-ip_\tau}+e^{ip_\tau}\big)\bigg]\Bigg)
\nonumber\\
&=e^{iyp}e^{-\frac i2(p_\mu-p_\nu)}\Big(-\delta_{\mu\nu}\hat{p}^2+\hat{p}_\mu\hat{p}_\nu\Big)
\end{align}
with $\hat{p}_\mu=2\sin(\tfrac{1}{2}p_\mu)$ and $\hat{p}^2=\sum_{\mu=1}^4 \hat{p}_\mu^2$.
Now we can express the Fourier transform of the first order smeared gluon field as
\begin{align}
A_\mu^{(n+1)a}(x)&=\sum_{\nu,y}\intdk\intd{p}
e^{iyp}e^{-\frac i2(p_\mu-p_\nu)}\Big(\delta_{\mu\nu}+\varrho\Big(-\delta_{\mu\nu}\hat{p}^2+\hat{p}_\mu\hat{p}_\nu\Big)\Big)
e^{i(x+y+\hat{\nu}/2)k}A^{(n)a}_\nu(k)
\\
&=\sum_{\nu}\intdk
e^{i(x+\hat{\mu}/2)k} \Big(\delta_{\mu\nu}+\varrho\Big(-\delta_{\mu\nu}\hat{k}^2+\hat{k}_\mu\hat{k}_\nu\Big)\Big)A^{(n)a}_\nu(k)
\;,
\end{align}
where we have used $\sum\limits_x e^{-i x(p-q)}=(2\pi)^4\delta(p-q)$ and performed the integration over $p$.
Thus at leading order the result in momentum space is
\begin{align}
A_\mu^{(n+1)a}(k)=A_\mu^{(n)a}(k)+\varrho\sum\limits_\nu g_{\mu\nu}(k)A_\nu^{(n)a}(k)
\end{align}
with 
\begin{align}
g_{\mu\nu}(k)&=-\delta_{\mu\nu}\hat{k}^2+\hat{k}_\mu\hat{k}_\nu
\end{align}
Expressing it in terms of the function $f(k)=1-\varrho\hat{k}^2$, it becomes easier to iterate
\begin{align}
A_\mu^{(n+1)a}(k)=\sum\limits_\nu \left(f(k)\delta_{\mu\nu}-(f(k)-1)\frac{\hat{k}_\mu\hat{k}_\nu}{\hat{k}^2}\right)A^{(n)a}_\nu(k).
\end{align}
In this form, the form factor retains the same structure after multiple iterations, only the powers of $f$ increase \cite{Bernard:1999kc}\cite{Capitani:2006ni}:
\begin{align}
A_\mu^{(n)a}(k)&=\sum\limits_\nu \bigg(f^n(k)\delta_{\mu\nu}-(f^n(k)-1)\frac{\hat{k}_\mu\hat{k}_\nu}{\hat{k}^2}\bigg)A^{(0)a}_\nu(k)
\\&
=:\sum\limits_\nu \tilde{g}^n_{\mu\nu}(\varrho,k)A^{(0)a}_\nu(k)
\label{eq:g_tilde_LO}
\end{align}                                                                                                                                                                                                                                                                                                                                                                                                                                                                                                                                                                                                                                                                                                                                                                                                                                                                                                                                                                                                                                                                                                                                                                                                                                          
We note that $\tilde{g}^n_{\mu\nu}(\varrho,k)$ is not exactly the Fourier transform of $\tilde{g}^n_{\mu\nu}(\varrho,y)$ but rather fulfils the same purpose in momentum space of relating the newly smeared field to the one of the previous smearing step.
%
\subsection{Next-to-leading order\label{subsec:NLO}}
%
At higher orders the expansions deviate, and Eqs. (\ref{eq:U1_after_BCH}) and (\ref{eq:U1_after_BCH_and_exp}) would need to be modified for a general smearing step $n\to n+1$  to include the higher order perturbative fields $A^{(n)[ab]}_\mu(x)$, $\bar{A}^{(n)abc}_\mu(x)$, and $\hat{A}^{(n)abc}_\mu(x)$. This is why we start again by considering the first smearing step.
At order $g_0^2$, the sum of staples $S^{(0)}_\mu(x)$ is given by:
\begin{align}
S^{(0)ab}_{2\mu}(x)&=\sum\limits_{\nu\neq\mu}\bigg[ A_\nu^{a}(x)A_\mu^{b}(x+\hat{\nu})- A_\nu^{a}(x)A_\nu^{b}(x+\hat{\mu})- A_\mu^{a}(x+\hat{\nu})A_\nu^{b}(x+\hat{\mu})
\nonumber\\
&- A_\nu^{a}(x-\hat{\nu})A_\mu^{b}(x-\hat{\nu})- A_\nu^{a}(x-\hat{\nu})A_\nu^{b}(x+\hat{\mu}-\hat{\nu})
\nonumber\\
&+ A_\mu^{a}(x-\hat{\nu})A_\nu^{b}(x+\hat{\mu}-\hat{\nu}) 
+\frac {1}{2} \Big\{A_\nu^{a}(x)A_\nu^{b}(x)
+A_\mu^{a}(x+\hat{\nu})A_\mu^{b}(x+\hat{\nu})
\nonumber \\
&+A_\nu^{a}(x+\hat{\mu})A_\nu^{b}(x+\hat{\mu})
+A_\nu^{a}(x-\hat{\nu})A_\nu^{b}(x-\hat{\nu})
+A_\mu^{a}(x-\hat{\nu})A_\mu^{b}(x-\hat{\nu})
\nonumber \\
&+A_\nu^{a}(x+\hat{\mu}-\hat{\nu})A_\nu^{b}(x+\hat{\mu}-\hat{\nu})\Big\}
\bigg].
\end{align}
The part in braces $\{\hdots\}$ is symmetric in $a\leftrightarrow b$ and will therefore disappear in $A^{(1)[ab]}_\mu(x)$.
We get
\begin{align}
Q^{(0)ab}_{2\mu}(x)&=\frac {1}{2}\Big(-S^{(0)a}_{1\mu}(x)A^{(0)b}_\mu(x)+ S^{(0)ab}_{2\mu}(x)\Big)
\end{align}
which leaves 
\begin{align}
A_\mu^{(1)[ab]}(x)&=2\varrho \Big(\frac{1}{2}Q^{(0)[a}_{1\mu}(x)A^{(0)b]}_\mu(x)
+Q^{(0)[ab]}_{2\mu}(x)\Big)
=\varrho S^{(0)[ab]}_{2\mu}(x)
\;,
\end{align}
which we can write as a double convolution:
\begin{align}
 S^{(0)[ab]}_{2\mu}(x)&=\sum\limits_{y,z}\sum\limits_{\nu,\rho=1}^4g_{\mu\nu\rho}(y,z) \cdot A^{[a}_\nu(x+y)A^{b]}_\rho(x+z)
 \\
 &=\sum\limits_{y,z}\sum\limits_{\nu,\rho=1}^4\frac{1}{2}\big(
 g_{\mu\nu\rho}(y,z)- g_{\mu\rho\nu}(z,y)\big) \cdot A^{a}_\nu(x+y)A^{b}_\rho(x+z)
\end{align}
 with
 \begin{align}
g_{\mu\nu\rho}(y,z)
&=\delta_{\mu\nu}\left[\delta(y,-\hat{\rho})\delta(z,\hat{\mu}-\hat{\rho})-\delta(y,\hat{\rho})\delta(z,\hat{\mu})\right]\nonumber\\
&+\delta_{\mu\rho}\left[\delta(y,0)\delta(z,\hat{\nu})-\delta(y,-\hat{\nu})\delta(z,-\hat{\nu})\right]\nonumber\\
&-\delta_{\nu\rho}\left[\delta(y,0)\delta(z,\hat{\mu})+\delta(y,-\hat{\nu})\delta(z,\hat{\mu}-\hat{\nu})\right]
\;.
 \end{align}
Fourier transforming gives us
\begin{align}
A_\mu^{(1)[ab]}(k_1,k_2)=\varrho \sum\limits_{\nu,\rho} g_{\mu\nu\rho}(k_1,k_2) A^a_\nu(k_1)A^b_\rho(k_2)\;,
\end{align}
where
\begin{align}
g_{\mu\nu\rho}(k_1,k_2)
&=2i\Big(
\delta_{\mu\rho}c(k_{1\mu})s( 2 k_{2\nu}+k_{1\nu})
-\delta_{\mu\nu}c(k_{2\mu})s(2 k_{1\rho}+k_{2\rho})
\nonumber\\&
+\delta_{\nu\rho}s( k_{1\mu}-k_{2\mu})c( k_{1\nu}+k_{2\nu})\Big)
\;.
\end{align}
Here we use the common shorthand for the trigonometric functions
\begin{align}
s(k_\mu)&=\sin\big(\tfrac{1}{2}k_\mu\big) &
c(k_\mu)&=\cos\big(\tfrac{1}{2}k_\mu\big) \\
\bar{s}(k_\mu)&=\sin(k_\mu) & \bar{c}(k_\mu)&=\cos(k_\mu)\\
s^2(k)&=\sum\limits_\mu s^2(k_\mu) &&
\end{align}
Although $g_{\mu\nu\rho}(k_1,k_2)$ is imaginary (and therefore so is $ A_\mu^{(1)[ab]}(k_1,k_2)$), it also fulfils \linebreak
$g_{\mu\nu\rho}(-k_1,-k_2)=-g_{\mu\nu\rho}(k_1,k_2)$, which means that $A_\mu^{(1)[ab]}(x)$ is real.\\

If we were to repeat the previous calculation for a smearing step $n\to n+1$, we would not only have to replace the $A^{(0)a}_\mu(x)$ by $A^{(n)a}_\mu(x)$ but the expansion of $S^{(n)ab}_{2\mu}(x)$ would also contain terms linear in $A^{(n)[ab]}_{\mu}(x)$.
Instead, it is easier to realize that we can get the same contribution by replacing 
\begin{align}
T^aA^{(0)a}_\mu(x)\to T^a\tilde{A}^{(n)a}_\mu(g_0,x) = T^a A^{(n)a}_\mu(x)-\frac{g_0}{2i}[T^a,T^b]A^{(n)[ab]}_\mu(x)+\mathcal{O}(g_0^2)
\end{align}
in the \emph{leading order} calculation.
Then we get for a general smearing step 
\begin{align}
A_\mu^{(n+1)[ab]}(k_1,k_2)=&
\sum\limits_\nu
\tilde{g}_{\mu\nu}(\varrho,k_1+k_2)
A_\nu^{(n)[ab]}(k_1,k_2)
+\varrho \sum\limits_{\nu,\rho} g_{\mu\nu\rho}(k_1,k_2) A^{(n)a}_\nu(k_1)A^{(n)b}_\rho(k_2)
\end{align}
By iterating once
\begin{align}
A_\mu^{(2)[ab]}(k_1,k_2)=&
\sum\limits_\nu
\tilde{g}_{\mu\nu}(\varrho,k_1+k_2)
\varrho \sum\limits_{\rho\sigma} g_{\nu\rho\sigma}(k_1,k_2) A^{(0)a}_\rho(k_1)A^{(0)b}_\sigma(k_2)
\nonumber\\&
+\varrho \sum\limits_{\nu,\rho} g_{\mu\nu\rho}(k_1,k_2) A^{(1)a}_\nu(k_1)A^{(1)b}_\rho(k_2)
\end{align}
and twice
\begin{align}
A_\mu^{(3)[ab]}(k_1,k_2)=&
\sum\limits_\nu
\tilde{g}^2_{\mu\nu}(\varrho,k_1+k_2)
\varrho \sum\limits_{\rho\sigma} g_{\nu\rho\sigma}(k_1,k_2) A^{(0)a}_\rho(k_1)A^{(0)b}_\sigma(k_2)
\nonumber\\&
+\sum\limits_\nu
\tilde{g}_{\mu\nu}(\varrho,k_1+k_2)
\varrho \sum\limits_{\rho\sigma} g_{\nu\rho\sigma}(k_1,k_2) A^{(1)a}_\rho(k_1)A^{(1)b}_\sigma(k_2)
\nonumber\\&
+\varrho \sum\limits_{\nu,\rho} g_{\mu\nu\rho}(k_1,k_2) A^{(2)a}_\nu(k_1)A^{(2)b}_\rho(k_2)\;,
\end{align}
we can infer the general form
\begin{align}
A_\mu^{(n)[ab]}(k_1,k_2)=&
\sum\limits_{\nu\rho\sigma}
\varrho\,g_{\nu\rho\sigma}(k_1,k_2)
\sum_{m=0}^{n-1}
\tilde{g}^{n-m-1}_{\mu\nu}(\varrho,k_1+k_2)
 A^{(m)a}_\rho(k_1)A^{(m)b}_\sigma(k_2)
 \;.
\end{align}
Finally, we can also express the first order fields in the sum in terms of the original unsmeared ones
\begin{align}
A_\mu^{(n)[ab]}(k_1,k_2)&=
\sum\limits_{\alpha\beta\gamma\nu\rho}
\varrho\,g_{\alpha\beta\gamma}
\Bigg[\sum_{m=0}^{n-1}
\tilde{g}^{n-m-1}_{\mu\alpha}(\varrho,k_1+k_2)
\tilde{g}^m_{\beta\nu}(\varrho,k_1)\tilde{g}^m_{\gamma\rho}(\varrho,k_2)\Bigg]
 A^{(0)a}_\nu(k_1)A^{(0)b}_\rho(k_2)
\nonumber\\&
=:\sum\limits_{\nu\rho}\tilde{g}^{(n)}_{\mu\nu\rho}(\varrho,k_1,k_2)
 A^{(0)a}_\nu(k_1)A^{(0)b}_\rho(k_2)
 \;.
 \label{eq:g_tilde_NLO}
\end{align}
%
\subsection{Next-to-next-to-leading order\label{subsec:NNLO}}
%
As before, we start again by considering the first smearing step and {generalize} later to an arbitrary number of smearing steps. At third order we have defined two perturbative fields $\bar{A}$ and $\hat{A}$ in Eq.~(\ref{eq:smearing_expansion_in_g0}), where the former can be read off of Eq.~(\ref{eq:U1_after_BCH_and_exp}) to be 
\begin{align}
\bar{A}^{(1)abc}_{\mu}(x)&=-3\varrho A^a_\mu(x)Q^{[bc]}_{2\mu}(x)
+\frac{1}{2}\varrho^2 Q^a_{1\mu}(x)Q^b_{1\mu}(x)A^c_\mu(x)
+\frac{1}{2}\varrho A^a_{\mu}(x)A^b_\mu(x)Q^c_{1\mu}(x)
\\
&=-\frac{3}{2} A^a_\mu(x)A^{(1)[bc]}_\mu(x)+\frac{3}{2}\varrho A^a_\mu(x)Q^b_{1\mu}(x)A^c_\mu(x)
+\frac{1}{2}\varrho^2 Q^a_{1\mu}(x)Q^b_{1\mu}(x)A^c_\mu(x)
\nonumber\\&
+\frac{1}{2}\varrho A^a_{\mu}(x)A^b_\mu(x)Q^c_{1\mu}(x)
\\
&=-\frac{3}{2} A^a_\mu(x)A^{(1)[bc]}_\mu(x)
+\frac{1}{2} A^a_\mu(x)A^{(1)b}_{\mu}(x)A^c_\mu(x)
+\frac{1}{2} A^{(1)a}_{\mu}(x)A^{(1)b}_{\mu}(x)A^c_\mu(x)
\,.
\end{align}
We have dropped all terms that were symmetric in $b\leftrightarrow c$ in the last step.
Thus $\bar{A}$ is completely given by already known quantities, which is not surprising as it stems only from the commutator terms of the {Baker-Campbell-Hausdorff} formula. 

From the third order in the expansion of $Q_\mu(x)$ we get 
\begin{align}
\hat{A}^{(1)abc}_\mu(x)
=-6\varrho Q^{abc}_{3\mu}(x)
\end{align}
with
\begin{align}
Q^{abc}_{3\mu}(x)&=A^a_\mu(x)A^b_{\mu}(x)A^c_\mu(x)
-\frac{1}{2}S^a_{1\mu}(x)A^b_{\mu}(x)A^c_\mu(x)
+S^{ab}_{2\mu}(x)A^c_\mu(x)-\frac{1}{2}S^{abc}_{3\mu}(x)
\\&
=-2A^a_\mu(x)A^b_{\mu}(x)A^c_\mu(x)
-\frac{1}{2}Q^a_{1\mu}(x)A^b_{\mu}(x)A^c_\mu(x)
+S^{ab}_{2\mu}(x)A^c_\mu(x)-S^{abc}_{3\mu}(x)
\;.
\end{align}
We revisit the sum of staples at second order and define
\begin{align}
S^{ab}_{2\mu}(x)&= s_{1\mu}^{ab}(x)+\frac 12 s_{2\mu}^{ab}(x)
\end{align}
with
\begin{align}
s_{1\mu\nu}^{ab}(x)&=\sum\limits_\nu\Big[ A_\nu^a(x)A_\mu^b(x+\hat{\nu})- A_\nu^a(x)A_\nu^b(x+\hat{\mu})- A_\mu^a(x+\hat{\nu})A_\nu^b(x+\hat{\mu})\nonumber\\
&- A_\nu^a(x-\hat{\nu})A_\mu^b(x-\hat{\nu})- A_\nu^a(x-\hat{\nu})A_\nu^b(x+\hat{\mu}-\hat{\nu})+ A_\mu^a(x-\hat{\nu})A_\nu^b(x+\hat{\mu}-\hat{\nu})\Big]\\
s_{2\mu}^{ab}(x)&=\sum\limits_\nu \Big[A_\nu^a(x)A_\nu^b(x)+A_\mu^a(x+\hat{\nu})A_\mu^b(x+\hat{\nu})+A_\nu^a(x+\hat{\mu})A_\nu^b(x+\hat{\mu})\nonumber \\
&+A_\nu^a(x-\hat{\nu})A_\nu^b(x-\hat{\nu})+A_\mu^a(x-\hat{\nu})A_\mu^b(x-\hat{\nu}) +A_\nu^a(x+\hat{\mu}-\hat{\nu})A_\nu^b(x+\hat{\mu}-\hat{\nu})
\nonumber\\&
-2\delta_{\mu\nu}A^a_\mu(x)A^b_\mu(x)
\Big]
\;.
\end{align}
Their Fourier transforms are
\begin{align}
s_{1\mu}^{ab}(x)&=\sum\limits_{yz}\sum\limits_{\nu\rho}a_{1\mu\nu\rho}(y,z)A^a_\nu(x+y)A^b_\rho(x+z)\nonumber\\
&=\sum\limits_{\nu\rho}\int \frac{\mathrm{d}^4k_1\,\mathrm{d}^4k_2}{(2\pi)^8}   e^{-i(x+\frac 12 \hat{\mu})(k_1+k_2) }\cdot a_{1\mu\nu\rho}(k_1,k_2) A^a_\nu(k_1)A^b_\rho(k_2)
\\
s_{2\mu}^{ab}(x)&=\sum\limits_{y}\sum\limits_{\nu}a_{2\mu\nu}(y)A^a_\nu(x+y)A^b_\nu(x+y)
\nonumber\\&
=\sum\limits_{\nu}\int \frac{\mathrm{d}^4k_1\,\mathrm{d}^4k_2}{(2\pi)^8}  e^{-i(x+\frac {1}{2} \hat{\mu})(k_1+k_2) }\cdot a_{2\mu\nu}(k_1,k_2) A^a_\nu(k_1)A^b_\nu(k_2)
\end{align}
with
\begin{align}
a_{1\mu\nu\rho}(y,z)
&=\delta_{\mu\nu}\left[\delta(y,-\hat{\rho})\delta(z,\hat{\mu}-\hat{\rho})-\delta(y,\hat{\rho})\delta(z,\hat{\mu})\right]
\nonumber\\
&+\delta_{\mu\rho}\left[\delta(y,0)\delta(z,\hat{\nu})-\delta(y,-\hat{\nu})\delta(z,-\hat{\nu})\right]
\nonumber\\
&-\delta_{\nu\rho}\left[\delta(y,0)\delta(z,\hat{\mu})+\delta(y,-\hat{\nu})\delta(z,\hat{\mu}-\hat{\nu})\right]
\\
a_{2\mu\nu}(y)
&=\delta_{\mu\nu}\sum\limits_{\alpha=1}^4 \Big[\delta(y,\hat{\alpha})+ \delta(y,-\hat{\alpha})-2\delta(y,0)\Big]
+6\delta_{\mu\nu}\delta(y,0)
\nonumber\\&
+\delta(y,0)+ \delta(y,\hat{\mu}) + \delta(y,-\hat{\nu})
+\delta(y,\hat{\mu}-\hat{\nu})
 \end{align}
and
\begin{align}
a_{1\mu\nu\rho}(k_1,k_2)
&=-\delta_{\mu\nu}\left[2i\sin\left(k_{1\rho}+\tfrac 12 k_{2\rho}\right)e^{\frac{i}{2} k_{2\mu}}\right]
+\delta_{\mu\rho}\left[2i \sin\left(k_{2\nu}+\tfrac 12 k_{1\nu}\right)e^{-\frac{i}{2} k_{1\mu}}\right]
\nonumber\\&
-\delta_{\nu\rho}\left[2\cos\left(\tfrac 12 k_{1\nu}+\tfrac 12 k_{2\nu}\right)e^{-\frac{i}{2}(k_{1\mu}-k_{2\mu})} \right]
\\&
=-h_{\mu\nu\rho}(k_1,k_2)+g_{\mu\nu\rho}(k_1,k_2)
\;.
\end{align}
Here $h_{\mu\nu\rho}$ and $g_{\mu\nu\rho}$ are essentially the real and imaginary parts of $a_{1\mu\nu\rho}$, i.e.\ 
\begin{align}
g_{\mu\nu\rho}(k_1,k_2)
&=2i\Big(-
\delta_{\mu\nu}c(k_{2\mu})s(2 k_{1\rho}+k_{2\rho})
+\delta_{\mu\rho}c(k_{1\mu})s( 2 k_{2\nu}+k_{1\nu})
\nonumber\\&
+\delta_{\nu\rho}s( k_{1\mu}-k_{2\mu})c( k_{1\nu}+k_{2\nu})\Big)
\\
h_{\mu\nu\rho}(k_1,k_2)
&=2\Big(
-\delta_{\mu\nu}s(k_{2\mu})s(2 k_{1\rho}+k_{2\rho})
-\delta_{\mu\rho}s(k_{1\mu})s( 2 k_{2\nu}+k_{1\nu})
\nonumber\\&
+\delta_{\nu\rho}c( k_{1\mu}-k_{2\mu})c( k_{1\nu}+k_{2\nu})\Big)
\;.
\end{align}
We know the {antisymmetric} $g_{\mu\nu\rho}$ already from the second order.\\

Furthermore,
\begin{align}
a_{2\mu\nu}(k_1,k_2)
&=6\delta_{\mu\nu}+\delta_{\mu\nu}\sum\limits_{\alpha=1}^4 \bigg[2\cos\big(k_{1\alpha}+k_{2\alpha}\big)-2\bigg]
+4\cos\big(\tfrac{1}{2} (k_{1\mu}+k_{2\mu})\big)
\cos\big(\tfrac{1}{2}(k_{1\nu}+k_{2\nu})\big)
\nonumber\\&
=6\delta_{\mu\nu}-\delta_{\mu\nu}4s^2(k_1+k_2)
+4c (k_{1\mu}+k_{2\mu})
c(k_{1\nu}+k_{2\nu})
\\&
=:6\delta_{\mu\nu}-g^{(i)}_{\mu\nu}(k_1+k_2)
\end{align}
where we have defined $g^{(i)}_{\mu\nu}(k)$ which is a variation of $g_{\mu\nu}(k)$ with cosines in the second term instead of sines.
\\

For the sum of staples at third order we define
\begin{align}
S^{abc}_{3\mu}(x)=&s_{1\mu}^{abc}(x)+\frac 12 s_{2\mu}^{abc}(x) +\frac 16 s_{3\mu}^{abc}(x)
\end{align}
with
\begin{align}
s_{1\mu}^{abc}&(x)=\sum\limits_\nu\Big[-A_\nu^a(x)A_\mu^b(x+\hat{\nu})A_\nu^c(x+\hat{\mu})
-A_\nu^a(x-\hat{\nu})A_\mu^b(x-\hat{\nu})A_\nu^c(x+\hat{\mu}-\hat{\nu})\Big]
\\
 s_{2\mu}^{abc}&(x) = \sum\limits_\nu\Big[\big\{
A^a_\nu(x)A^b_\mu(x+\hat{\nu})A^c_\mu(x+\hat{\nu})+A^a_\nu(x)A^b_\nu(x+\hat{\mu})A^c_\nu(x+\hat{\mu})
\nonumber\\&
+A^a_\mu(x+\hat{\nu})A^b_\nu(x+\hat{\mu})A^c_\nu(x+\hat{\mu})
- A^a_\nu(x-\hat{\nu})A^b_\mu(x-\hat{\nu})A^c_\mu(x-\hat{\nu})
\nonumber\\&
-A^a_\nu(x-\hat{\nu})A^b_\nu(x+\hat{\mu}-\hat{\nu})A^c_\nu(x+\hat{\mu}-\hat{\nu})
 + A^a_\mu(x-\hat{\nu})A^b_\nu(x+\hat{\mu}-\hat{\nu})A^c_\nu(x+\hat{\mu}-\hat{\nu})\big\}
\nonumber\\&
+\big\{ -A^a_\mu(x+\hat{\nu})A^b_\mu(x+\hat{\nu})A^c_\nu(x+\hat{\mu})
+A^a_\nu(x)A^b_\nu(x)A^c_\mu(x+\hat{\nu})
 -A^a_\nu(x)A^b_\nu(x)A^c_\nu(x+\hat{\mu})
\nonumber\\&
+ A^a_\mu(x-\hat{\nu})A^b_\mu(x-\hat{\nu})A^c_\nu(x+\hat{\mu}-\hat{\nu})
+ A^a_\nu(x-\hat{\nu})A^b_\nu(x-\hat{\nu})A^c_\mu(x-\hat{\nu})
\nonumber\\&
+  A^a_\nu(x-\hat{\nu})A^b_\nu(x-\hat{\nu})A^c_\nu(x+\hat{\mu}-\hat{\nu})
\big\}
\Big]
\\
s_{3\mu}^{abc}&(x)=\sum\limits_\nu\Big[
-2\delta_{\mu\nu} A^a_\mu(x) A^b_\mu(x) A^c_\mu(x)+
A^a_\nu(x)A^b_\nu(x)A^c_\nu(x)+A^a_\mu(x+\hat{\nu})A^b_\mu(x+\hat{\nu})A^c_\mu(x+\hat{\nu})
\nonumber\\&
- A^a_\nu(x+\hat{\mu}) A^b_\nu(x+\hat{\mu}) A^c_\nu(x+\hat{\mu})
- A^a_\nu(x-\hat{\nu}) A^b_\nu(x-\hat{\nu}) A^c_\nu(x-\hat{\nu}) 
\nonumber\\&
+ A^a_\mu(x-\hat{\nu})A^b_\mu(x-\hat{\nu})A^c_\mu(x-\hat{\nu})
+ A^a_\nu(x+\hat{\mu}-\hat{\nu}) A^b_\nu(x+\hat{\mu}-\hat{\nu}) A^c_\nu(x+\hat{\mu}-\hat{\nu})\Big]
\;.
\end{align}
They can be written as 
\begin{align}
s_{1\mu}^{abc}(x)&=\sum\limits_{y,z,r}\sum\limits_{\nu\rho\sigma}b_{1\mu\nu\rho\sigma}(y,z,r)A^a_\nu(x+y)A^b_\rho(x+z)A^c_\sigma(x+r)
\\
s_{2\mu}^{abc}(x)&=\sum\limits_{y,z}\sum\limits_{\nu\rho}b_{21\mu\nu\rho}(y,z)A^a_\nu(x+y)A^b_\rho(x+z)A^c_\rho(x+z)
\nonumber\\&
+\sum\limits_{y,z}\sum\limits_{\nu\rho}b_{22\mu\nu\rho}(y,z)A^a_\nu(x+y)A^b_\nu(x+y)A^c_\rho(x+z)
\\
s_{3\mu}^{abc}(x)&=\sum\limits_{y}\sum\limits_{\nu}b_{3\mu\nu}(y)A^a_\nu(x+y)A^b_\nu(x+y)A^c_\nu(x+y)
\end{align}
with
\begin{align}
b_{1\mu\nu\rho\sigma}(y,z,r)&=
-\delta_{\mu\rho}\delta_{\nu\sigma}
\Big(\delta(y,0)\delta(z,\hat{\nu})\delta(r,\hat{\mu})
+\delta(y,-\hat{\nu})\delta(z,-\hat{\nu})\delta(r,\hat{\mu}-\hat{\nu})\Big)
\\
b_{21\mu\nu\rho}(y,z)&=
\delta_{\mu\nu}\Big(
	 \delta(y,\hat{\rho})\delta(z,\hat{\mu})
	+\delta(y,-\hat{\rho})\delta(z,\hat{\mu}-\hat{\rho})
\Big)
\nonumber\\&
+\delta_{\mu\rho}\Big(
	\delta(y,0)\delta(z,\hat{\nu})
	-\delta(y,-\hat{\nu})\delta(z,-\hat{\nu})
\Big)
\nonumber\\&
+\delta_{\nu\rho}\Big(
	\delta(y,0)\delta(z,\hat{\mu})
	-\delta(y,-\hat{\nu})\delta(z,\hat{\mu}-\hat{\nu})
\Big)
\\
b_{22\mu\nu\rho}(y,z)&=
\delta_{\mu\nu}\Big(
	 -\delta(y,\hat{\rho})\delta(z,\hat{\mu})
	+\delta(y,-\hat{\rho})\delta(z,\hat{\mu}-\hat{\rho})
\Big)
\nonumber\\&
+\delta_{\mu\rho}\Big(
	\delta(y,0)\delta(z,\hat{\nu})
	+\delta(y,-\hat{\nu})\delta(z,-\hat{\nu})
\Big)
\nonumber\\&
+\delta_{\nu\rho}\Big(
	-\delta(y,0)\delta(z,\hat{\mu})
	+\delta(y,-\hat{\nu})\delta(z,\hat{\mu}-\hat{\nu})
\Big)
\\
b_{3\mu\nu}(y)&=6\delta_{\mu\nu}\delta(y,0)+g_{\mu\nu}(y)
\;.
\end{align}
 Fourier transforming the $s_{i\mu}^{abc}(x)$ leads to
 \begin{align}
 s_{1\mu}^{abc}(x)&=\sum\limits_{\nu\rho\sigma}
\int\frac{\mathrm{d}^4k_1\,\mathrm{d}^4k_2\,\mathrm{d}^4k_3}{(2\pi)^{12}}
e^{-i\left(x+\frac 12 \hat{\mu}\right)(k_1+k_2+k_3)}
 b_{1\mu\nu\rho\sigma}(k_1,k_2,k_3) A^a_\nu(k_1) A^b_\rho(k_2)A^c_\sigma(k_3)
\\
s_{2\mu}^{abc}(x)&=\sum\limits_{\nu\rho}
\int\frac{\mathrm{d}^4k_1\,\mathrm{d}^4k_2\,\mathrm{d}^4k_3}{(2\pi)^{12}}
 e^{-i\left(x+\frac 12 \hat{\mu}\right)(k_1+k_2+k_3)}
 b_{21\mu\nu\rho}(k_1,k_2+k_3) A^a_\nu(k_1) A^b_\rho(k_2)A^c_\rho(k_3)
\nonumber\\
&+\sum\limits_{\nu\rho}
\int\frac{\mathrm{d}^4k_1\,\mathrm{d}^4k_2\,\mathrm{d}^4k_3}{(2\pi)^{12}}
 e^{-i\left(x+\frac 12 \hat{\mu}\right)(k_1+k_2+k_3)}
 b_{22\mu\nu\rho}(k_1+k_2,k_3) A^a_\nu(k_1) A^b_\nu(k_2)A^c_\rho(k_3)
 \\
s_{3\mu}^{abc}(x)&=\sum\limits_{\nu}
\int\frac{\mathrm{d}^4k_1\,\mathrm{d}^4k_2\,\mathrm{d}^4k_3}{(2\pi)^{12}}
 e^{-i\left(x+\frac 12 \hat{\mu}\right)(k_1+k_2+k_3)}
 g_{\mu\nu}(k_1+k_2+k_3) A^a_\nu(k_1) A^b_\nu(k_2)A^c_\nu(k_3)
 \end{align}
 with
 \begin{align}
 b_{1\mu\nu\rho\sigma}(k_1,k_2,k_3)
 &=-2\delta_{\mu\rho}\delta_{\nu\sigma}\cos\big(\tfrac 12 k_{1\nu} +k_{2\nu} +\tfrac 12 k_{3\nu}\big)e^{\frac i2 (k_{1\mu}-k_{3\mu})}
\\
 b_{21\mu\nu\rho}(k_1,k_2+k_3)
 &=\delta_{\mu\nu}2\cos\big( k_{1\rho}+\tfrac{1}{2}( k_{2\rho}+k_{3\rho})\big)e^{-\frac i2(k_{2\mu}+k_{3\mu})}
 \nonumber\\ &
+\delta_{\mu\rho}(-2i)\sin\big(\tfrac 12 k_{1\nu}+k_{2\nu}+k_{3\nu}\big)e^{\frac i2k_{1\mu}}\nonumber\\
 &+\delta_{\nu\rho}(-2i)\sin\big(\tfrac{1}{2} (k_{1\nu}+ k_{2\nu}+ k_{3\nu})\big)e^{\frac i2(k_{1\mu}-k_{2\mu}-k_{3\mu})}
\\
 b_{22\mu\nu\rho}(k_1+k_2,k_3)
 &=\delta_{\mu\nu}2i\sin\big( k_{1\rho}+ k_{2\rho}+\tfrac{1}{2} k_{3\rho}\big)e^{-\frac{i}{2}k_{3\mu}}
 \nonumber\\ &
+\delta_{\mu\rho}2\cos\big(\tfrac{1}{2} (k_{1\nu}+k_{2\nu})+k_{3\nu}\big)e^{\frac{i}{2}(k_{1\mu}+k_{2\mu})}\nonumber\\
 &+\delta_{\nu\rho}2i\sin\big(\tfrac{1}{2} (k_{1\nu}+ k_{2\nu}+ k_{3\nu})\big)e^{\frac i2(k_{1\mu}+k_{2\mu}-k_{3\mu})}
\\
 b_{3\mu\nu}(k_1+k_2+k_3)&=
6\delta_{\mu\nu}+g_{\mu\nu}(k_1+k_2+k_3)
\;.
\end{align}
We can {symmetrize} these expressions according to
\begin{align}
\big(T^a T^bT^c+T^cT^bT^a\big)v_{i\mu}^{abc}=\big(T^a T^bT^c+T^cT^bT^a\big)\frac{1}{2}\big(v_{i\mu}^{abc}+v_{i\mu}^{cba}\big)
\end{align}
and observe that under the integrals
\begin{align}
b_{1\mu\nu\rho\sigma}(k_1,k_2,k_3) A^c_\nu(k_1) A^b_\rho(k_2)A^a_\sigma(k_3) 
&=b_{1\mu\sigma\rho\nu}(k_3,k_2,k_1) A^a_\nu(k_1) A^b_\rho(k_2)A^c_\sigma(k_3) 
\\
b_{21\mu\nu\rho}(k_1,k_2+k_3) A^c_\nu(k_1) A^b_\rho(k_2)A^a_\rho(k_3)
&=
b_{21\mu\rho\nu}(k_3,k_1+k_2) A^a_\nu(k_1) A^b_\nu(k_2)A^c_\rho(k_3)
\\
b_{22\mu\nu\rho}(k_1+k_2,k_3) A^c_\nu(k_1) A^b_\nu(k_2)A^a_\rho(k_3)
&=
b_{22\mu\rho\nu}(k_2+k_3,k_1) A^a_\nu(k_1) A^b_\rho(k_2)A^c_\rho(k_3)
\;.
\end{align}
Thus we need 
 \begin{align}
 b_{1\mu\nu\rho\sigma}(k_1,k_2,k_3)+ b_{1\mu\sigma\rho\nu}(k_3,k_2,k_1)
 &=-4\delta_{\mu\rho}\delta_{\nu\sigma}c( k_{1\nu} +2k_{2\nu} + k_{3\nu})c(k_{1\mu}-k_{3\mu})
 \\
 &=: g^{(iii)}_{\mu\nu\rho\sigma}(k_1,k_2,k_3)
\\
 b_{21\mu\nu\rho}(k_1,k_2+k_3)+ b_{22\mu\rho\nu}(k_2+k_3,k_1)
& =
4\delta_{\mu\nu}c(2k_{1\rho}+k_{2\rho}+k_{3\rho})c(k_{2\mu}+k_{3\mu})
\nonumber\\&
+4\delta_{\mu\rho}s(k_{1\mu})s(k_{1\nu}+2(k_{2\nu}+k_{3\nu}))
\nonumber\\&
+4\delta_{\nu\rho}s(k_{1\mu}-k_{2\mu}-k_{3\mu})s(k_{1\nu}+k_{2\nu}+k_{3\nu})
\\
&=:g^{(ii)}_{\mu\nu\rho}(k_1,k_2+k_3)
\\
 b_{22\mu\nu\rho}(k_1+k_2,k_3)+ b_{21\mu\rho\nu}(k_3,k_1+k_2)
& =g^{(ii)}_{\mu\rho\nu}(k_3,k_1+k_2)
\;.
\end{align}
Now we can write
\begin{align}
\hat{A}^{(1)abc}_\mu(k_1,k_2,k_3)
&={\varrho}\sum\limits_{\nu\rho\sigma}\bigg(
3\big(h_{\mu\nu\rho}(k_1,k_2)-g_{\mu\nu\rho}(k_1,k_2)\big)\delta_{\mu\sigma}
+\frac{3}{2} g_{\mu\nu}(k_1)\delta_{\mu\rho}\delta_{\mu\sigma}
\nonumber\\&
+\frac{3}{2}\Big(g^{(i)}_{\mu\nu}(k_1,k_2)\delta_{\nu\rho}\delta_{\mu\sigma}
+g^{(ii)}_{\mu\nu\rho}(k_1,k_2+k_3)\delta_{\rho\sigma}
+g^{(iii)}_{\mu\nu\rho\sigma}(k_1,k_2,k_3)\Big)
\nonumber\\&
+\frac{1}{2}g_{\mu\nu}(k_1+k_2+k_3)\delta_{\nu\rho}\delta_{\nu\sigma}
\bigg)A^a_\nu(k_1)A^b_\rho(k_2)A^c_\sigma(k_3)
\;.
\end{align}
For a general smearing step $n>1$ we have to take into account that in the previous step $A^{(n-1)[ab]}_\mu\neq 0$, $\bar{A}^{(n-1)abc}_{\mu}\neq 0$, and $\hat{A}^{(n-1)abc}_{\mu}\neq 0$. 
Again we replace the unsmeared gluon field $A^{(0)}$ in the previous orders by the smeared one $\tilde{A}^{(n)}$ and collect all terms of order $g_0^3$:
\begin{align}
&U^{(n+1)}_\mu(k)=1+ig_0\sum\limits_\nu \tilde{g}_{\mu\nu}(\varrho,k)\bigg(
T^a A^{(n)a}_\nu(k)
-\frac{g_0}{2i}[T^a,T^b]A^{(n)[ab]}_\nu(k_1,k_2)
\nonumber\\&
-\frac{g_0^2}{6}\Big(T^aT^bT^c+T^cT^bT^a-\frac{1}{N_c}\Tr[T^aT^bT^c+T^cT^bT^a]\Big)A^{(n)abc}_\nu(k_3,k_4,k_5)
\bigg)
\nonumber\\&
-\frac{g_0^2}{2}\sum\limits_{\nu\rho}
 \Big(\tilde{g}_{\mu\nu}(\varrho,k_1) \tilde{g}_{\mu\rho}(\varrho,k_2)
+2g_{\mu\nu\rho}(k_1,k_2)\Big)
\bigg(T^a A^{(n)a}_\nu(k_1)
-\frac{g_0}{2i}[T^a,T^b]A^{(n)[ab]}_\nu(k_{11},k_{12})\bigg)
\nonumber\\&
\times
\bigg(T^a A^{(n)a}_\rho(k_2)
-\frac{g_0}{2i}[T^a,T^b]A^{(n)[ab]}_\rho(k_{21},k_{22})\bigg)
\nonumber\\&
-i\frac{g_0^3}{6}\sum\limits_{\nu\rho\sigma}\bigg(T^a T^bT^c 
\tilde{g}_{\mu\nu}(\varrho,k_1) \tilde{g}_{\mu\rho}(\varrho,k_2)\tilde{g}_{\mu\sigma}(\varrho,k_3)
+\frac{3}{2}\{T^a,[T^b,T^c]\}\tilde{g}_{\mu\nu}(\varrho,k_1){\varrho} g_{\mu\rho\sigma}(k_2,k_3)
\nonumber\\&
+[T^a,[T^b,T^c]]\Big(-\frac{3}{2}\delta_{\mu\nu}\varrho g_{\mu\rho\sigma}(k_2,k_3)
+\frac{1}{2}\delta_{\mu\nu}\tilde{g}_{\mu\rho}(\varrho,k_2)\delta_{\mu\sigma}
+\frac{1}{2}\tilde{g}_{\mu\nu}(\varrho,k_1) \tilde{g}_{\mu\rho}(\varrho,k_2)\delta_{\mu\sigma}\Big)
\nonumber\\&
+\Big(T^aT^bT^c+T^cT^bT^a-\frac{1}{N_c}\Tr[T^aT^bT^c+T^cT^bT^a]\Big)\hat{G}_{\mu\nu\rho\sigma}(k_1,k_2,k_3)\bigg)
A^{(n)a}_\nu(k_1) A^{(n)b}_\rho(k_2) A^{(n)c}_\sigma(k_3)
\end{align}
with $k=k_1+k_2=k_3+k_4+k_5$, $k_1=k_{11}+k_{12}$, and $k_2=k_{21}+k_{22}$.
From the third term involving the product $\tilde{A}_\nu \tilde{A}_\rho$ the following terms appear at order $g_0^3$:
\begin{align}
&-i\frac{g_0^3}{4}\sum\limits_{\nu\rho}\Big(
\big( \tilde{g}_{\mu\nu}(\varrho,k_1) \tilde{g}_{\mu\rho}(\varrho,k_2+k_3)
+2g_{\mu\nu\rho}(k_1,k_2+k_3)\big)
T^a[T^b,T^c]A^{(n)a}_\nu(k_1)A^{(n)[bc]}_\rho(k_2,k_3)
\nonumber\\&
+\big( \tilde{g}_{\mu\nu}(\varrho,k_1+k_2) \tilde{g}_{\mu\rho}(\varrho,k_3)
+2g_{\mu\nu\rho}(k_1+k_2,k_3)\big)
[T^a,T^b]T^cA^{(n)[ab]}_\nu(k_1,k_2)A^{(n)c}_\rho(k_3)\Big)
\nonumber\\&
=-i\frac{g_0^3}{6}\bigg(\frac{3}{2}\{T^a,[T^b,T^c]\}\sum\limits_{\nu\rho}
 \tilde{g}_{\mu\nu}(\varrho,k_1) \tilde{g}_{\mu\rho}(\varrho,k_2+k_3)A^{(n)a}_\nu(k_1)A^{(n)[bc]}_\rho(k_2,k_3)
 \nonumber\\&
+3[T^a,[T^b,T^c]]\sum\limits_{\nu\rho}g_{\mu\nu\rho}(k_1,k_2+k_3)A^{(n)a}_\nu(k_1)A^{(n)[bc]}_\rho(k_2,k_3)
\bigg)
\;.
\end{align}
Then we have for our two third-order fields $\bar{A}$ and $\hat{A}$:
\begin{align}
\bar{A}^{(n+1)abc}_{\mu}(k_1,k_2,k_3)&=
\sum\limits_{\nu} \tilde{g}_{\mu\nu}(\varrho,k_1+k_2+k_3)\bar{A}^{(n)abc}_{\nu}(k_1,k_2,k_3)
\nonumber\\&
+3\varrho \sum\limits_{\nu\rho}g_{\mu\nu\rho}(k_1,k_2+k_3)A^{(n)a}_\nu(k_1)A^{(n)[bc]}_\rho(k_2,k_3)
\nonumber\\&
-\frac{3}{2}\varrho \sum\limits_{\nu\rho\sigma} \delta_{\mu\nu}g_{\mu\rho\sigma}(k_2,k_3)A^{(n)a}_\nu(k_1)A^{(n)b}_\rho(k_2)A^{(n)c}_\sigma(k_3)
\nonumber\\&
+\frac{1}{2} A^{(n)a}_\mu(k_1)A^{(n+1)b}_{\mu}(k_2)A^{(n)c}_\mu(k_3)
+\frac{1}{2} A^{(n+1)a}_{\mu}(k_1)A^{(n+1)b}_{\mu}(k_2)A^{(n)c}_\mu(k_3)
\end{align}
\begin{align}
\hat{A}^{(n+1)abc}_\mu(k_1,k_2,k_3)&=
\sum\limits_{\nu} \tilde{g}_{\mu\nu}(\varrho,k_1+k_2+k_3)\hat{A}^{(n)abc}_{\nu}(k_1,k_2,k_3)
\nonumber\\&
\varrho\sum\limits_{\nu\rho\sigma}\bigg(
3\big(h_{\mu\nu\rho}(k_1,k_2)+g_{\mu\nu\rho}(k_1,k_2)\big)\delta_{\mu\sigma}
+\frac{3}{2} g_{\mu\nu}(k_1)\delta_{\mu\rho}\delta_{\mu\sigma}
\nonumber\\&
+\frac{3}{2}\Big(g^{(i)}_{\mu\nu}(k_1,k_2)\delta_{\nu\rho}\delta_{\mu\sigma}
+g^{(ii)}_{\mu\nu\rho}(k_1,k_2+k_3)\delta_{\rho\sigma}
+g^{(iii)}_{\mu\nu\rho\sigma}(k_1,k_2,k_3)\Big)
\nonumber\\&
+\frac{1}{2}g_{\mu\nu}(k_1+k_2+k_3)\delta_{\nu\rho}\delta_{\nu\sigma}
\bigg)A^{(n)a}_\nu(k_1)A^{(n)b}_\rho(k_2)A^{(n)c}_\sigma(k_3)
\;,
\end{align}
and finally
\begin{align}
A^{(n+1)abc}_\mu(k_1,k_2,k_3)&=\hat{A}^{(n+1)abc}_\mu(k_1,k_2,k_3)+2\bar{A}^{(n+1)a[bc]}_{\mu}(k_1,k_2,k_3)
\\&
=\sum\limits_{\nu} \tilde{g}_{\mu\nu}(\varrho,k_1+k_2+k_3)A^{(n)abc}_{\nu}(k_1,k_2,k_3)
\nonumber\\&
+6\varrho \sum\limits_{\nu\rho}g_{\mu\nu\rho}(k_1,k_2+k_3)A^{(n)a}_\nu(k_1)A^{(n)[bc]}_\rho(k_2,k_3)
\nonumber\\&
+\sum\limits_{\nu\rho\sigma}\Bigg(\frac{1}{2}\varrho^2\Big(g_{\mu\nu}(k_1)g_{\mu\rho}(k_2)\delta_{\mu\sigma}-g_{\mu\nu}(k_1)\delta_{\mu\rho}g_{\mu\sigma}(k_3)\Big)
+\varrho G_{\mu\nu\rho\sigma}(k_1,k_2,k_3)
\Bigg)
\nonumber\\&
\times
A^{(n)a}_\nu(k_1)A^{(n)b}_\rho(k_2)A^{(n)c}_\sigma(k_3)
\end{align}
with
\begin{align}
G_{\mu\nu\rho\sigma}(k_1,k_2,k_3)&=
\frac{1}{2} g_{\mu\nu}(k_1)\delta_{\mu\rho}\delta_{\mu\sigma}
+\delta_{\mu\nu}g_{\mu\rho}(k_2)\delta_{\mu\sigma}
+3h_{\mu\nu\rho}(k_1,k_2)\delta_{\mu\sigma}
\nonumber\\&
+\frac{3}{2}\Big(g^{(i)}_{\mu\nu}(k_1,k_2)\delta_{\nu\rho}\delta_{\mu\sigma}
+g^{(ii)}_{\mu\nu\rho}(k_1,k_2+k_3)\delta_{\rho\sigma}
+g^{(iii)}_{\mu\nu\rho\sigma}(k_1,k_2,k_3)\Big)
\nonumber\\&
+\frac{1}{2}g_{\mu\nu}(k_1+k_2+k_3)\delta_{\nu\rho}\delta_{\nu\sigma}
\;.
\end{align}
After two smearing steps:
\begin{align}
A^{(2)abc}_\mu(k_1,k_2,k_3)&=\sum\limits_{\alpha} \tilde{g}_{\mu\alpha}(\varrho,k_1+k_2+k_3)\sum\limits_{\nu\rho\sigma}
G_{\alpha\nu\rho\sigma}(\varrho,k_1,k_2,k_3)A^{(0)a}_\nu(k_1)A^{(0)b}_\rho(k_2)A^{(0)c}_\sigma(k_3)
\nonumber\\&
+6\varrho \sum\limits_{\nu\rho}g_{\mu\nu\rho}(k_1,k_2+k_3)A^{(1)a}_\nu(k_1)A^{(1)[bc]}_\rho(k_2,k_3)
\nonumber\\&
+\sum\limits_{\nu\rho\sigma}
\tilde{G}_{\mu\nu\rho}(\varrho,k_1,k_2,k_3)A^{(1)a}_\nu(k_1)A^{(1)b}_\rho(k_2)A^{(1)c}_\sigma(k_3)
\;.
\end{align}
After three smearing steps:
\begin{align}
A^{(3)abc}_\mu(k_1,k_2,k_3)&=\sum\limits_{\alpha} \tilde{g}^2_{\mu\alpha}(\varrho,k_1+k_2+k_3)\sum\limits_{\nu\rho\sigma}
\tilde{G}_{\alpha\nu\rho\sigma}(\varrho,k_1,k_2,k_3)A^{(0)a}_\nu(k_1)A^{(0)b}_\rho(k_2)A^{(0)c}_\sigma(k_3)
\nonumber\\&
+\sum\limits_{\alpha} \tilde{g}_{\mu\alpha}(\varrho,k_1+k_2+k_3)\sum\limits_{\nu\rho\sigma}
\tilde{G}_{\alpha\nu\rho\sigma}(\varrho,k_1,k_2,k_3)A^{(1)a}_\nu(k_1)A^{(1)b}_\rho(k_2)A^{(1)c}_\sigma(k_3)
\nonumber\\&
+\tilde{G}_{\mu\nu\rho}(\varrho,k_1,k_2,k_3)A^{(2)a}_\nu(k_1)A^{(2)b}_\rho(k_2)A^{(2)c}_\sigma(k_3)
\nonumber\\&
+6\varrho\sum\limits_\alpha \tilde{g}_{\mu\alpha}(\varrho,k_1+k_2+k_3) \sum\limits_{\nu\rho}g_{\alpha\nu\rho}(k_1,k_2+k_3)A^{(1)a}_\nu(k_1)A^{(1)[bc]}_\rho(k_2,k_3)
\nonumber\\&
+6\varrho \sum\limits_{\nu\rho}g_{\mu\nu\rho}(k_1,k_2+k_3)A^{(2)a}_\nu(k_1)A^{(2)[bc]}_\rho(k_2,k_3)
\;.
\end{align}
After $n$ smearing steps:
\begin{align}
A^{(n)abc}_\mu(k_1,k_2,k_3)&=
6\varrho\sum\limits_{\alpha\nu\rho}g_{\alpha\nu\rho}(k_1,k_2+k_3)\sum\limits_{m=1}^{n-1}\tilde{g}^{n-m-1}_{\mu\alpha}({\varrho},k_1+k_2+k_3) A^{(m)a}_\nu(k_1)A^{(m)[bc]}_\rho(k_2,k_3)
\nonumber\\&
+\sum\limits_{\alpha\nu\rho\sigma}\Bigg(
\frac{1}{2}\varrho^2
\Big(g_{\alpha\nu}(k_1)g_{\mu\rho}(k_2)\delta_{\mu\sigma}-g_{\mu\nu}(k_1)\delta_{\mu\rho}g_{\mu\sigma}(k_3)\Big)
+\varrho G_{\alpha\nu\rho\sigma}(k_1,k_2,k_3)
\Bigg)
\nonumber\\&
\times\sum\limits_{m=0}^{n-1} \tilde{g}^{n-m-1} _{\mu\alpha}(\varrho,k_1+k_2+k_3)A^{(m)a}_\nu(k_1)A^{(m)b}_\rho(k_2)A^{(m)c}_\sigma(k_3)
\;.
\end{align}
Plugging in the relations from leading order and next-to-leading order, we express the end result as a function of the unsmeared fields:
\begin{align}
A^{(n)abc}_\mu&(k_1,k_2,k_3)=
\nonumber\\&
\sum\limits_{\nu\rho\sigma}\Bigg(
\sum\limits_{\alpha\beta\gamma}
6\varrho g_{\alpha\beta\gamma}(k_1,k_2+k_3)\sum\limits_{m=1}^{n-1}\tilde{g}^{n-m-1}_{\mu\alpha}(\varrho,k_1+k_2+k_3) \tilde{g}^m_{\beta\nu}(\varrho,k_1)\tilde{g}^{(m)}_{\gamma\rho\sigma}(\varrho,k_2,k_3)
\nonumber\\&
+\sum\limits_{\alpha\beta\gamma\delta}\bigg(
\frac{1}{2}\varrho^2
\Big(g_{\alpha\beta}(k_1)g_{\alpha\gamma}(k_2)\delta_{\alpha\delta}-g_{\alpha\beta}(k_1)\delta_{\alpha\gamma}g_{\alpha\delta}(k_3)\Big)
+\varrho G_{\alpha\beta\gamma\delta}(k_1,k_2,k_3)
\bigg)
\nonumber\\&
\times\sum\limits_{m=0}^{n-1} \tilde{g}^{n-m-1} _{\mu\alpha}(\varrho,k_1+k_2+k_3)
\tilde{g}^m_{\beta\nu}(\varrho,k_1)
\tilde{g}^m_{\gamma\rho}(\varrho,k_2)
\tilde{g}^m_{\delta\sigma}(\varrho,k_3)
\Bigg)
A^{(0)a}_\nu(k_1)A^{(0)b}_\rho(k_2)A^{(0)c}_\sigma(k_3)
\nonumber\\&
=:\sum\limits_{\nu\rho\sigma}\tilde{g}^{(n)}_{\mu\nu\rho\sigma}(\varrho,k_1,k_2,k_3)A^{(0)a}_\nu(k_1)A^{(0)b}_\rho(k_2)A^{(0)c}_\sigma(k_3)\;,
\label{eq:g_tilde_NNLO}
\end{align}
and we have found the last of the three $\tilde{g}$ functions from Eq.~(\ref{eq:def_g_tilde}).
\section{Perturbative expansion of the Wilson flow}
The Wilson flow is defined through
\begin{align}
\partial_t U_\mu(x,t)&=iQ_\mu(x,t) U_\mu(x,t)\\
U_\mu(x,0)&=U_\mu(x)
\end{align}
with the same $Q_\mu$ as in the definition of stout smearing but made up from the ``flowed'' fields $U_\mu(x,t)$ instead of the smeared fields. We already know the perturbative expansion of $Q_\mu$ 
\begin{align}
Q_\mu(x,t)&=g_0T^aQ^a_{1\mu}(x,t)+ig_0^2[T^a,T^b]Q^{ab}_{2\mu}(x,t)
\nonumber\\
&+g_0^3\Big(T^aT^bT^c+T^cT^bT^a-\frac{1}{N_c}\Tr\big[T^aT^bT^c+T^cT^bT^a\big]\Big)
Q^{abc}_{3\mu}(x,t)
\end{align}
We define the expansion of the flowed link variable in terms of fields $A^a_\mu(x,t)$, $A^{[ab]}_\mu(x,t)$, and $A^{abc}_\mu(x,t)$ such that similarly to the smeared case
\begin{align}
U_\mu(x,t)&=
1+ig_0T^aA^a_\mu(x,t)
-\frac{g_0^2}{2}\Big(T^aT^b A^a_\mu(x,t)A^b_\mu(x,t)+[T^a,T^b]A^{[ab]}_\mu(x,t)\Big)
\nonumber\\&
-i\frac{g_0^3}{6}\Bigg(T^aT^bT^c  A^a_\mu(x,t)A^b_\mu(x,t)A^c_\mu(x,t)
+\frac{3}{2}\{T^a,[T^b,T^c]\}A^a_\mu(x,t)A^{[ab]}_\mu(x,t)
\nonumber\\&
+\bigg(T^aT^bT^c+T^cT^bT^a-\frac{1}{N_c}\Tr\big[T^aT^bT^c+T^cT^bT^a\big]\bigg)A^{abc}_\mu(x,t)\Bigg)
+\mathcal{O}(g_0^4)
\;.
\end{align}
They fulfil the initial conditions
\begin{align}
A^a_\mu(x,0)&=A^a_\mu(x)\;,
&
A^{[ab]}_\mu(x,0)&=0\;,
&
A^{abc}_\mu(x,0)&=0\;,
\end{align}
and we define their Fourier transforms consistently:
\begin{align}
A^{a}_\mu(x,t)&=\intd{k}e^{i(x+\hat{\mu}/2)k}A^{a}_\mu(k,t)
\\
A^{ab}_\mu(x,t)&=\intd{k_1}\intd{k_2}e^{i(x+\hat{\mu}/2)(k_1+k_2)}A^{ab}_\mu(k_1,k_2,t)
\\
A^{abc}_\mu(x,t)&=\intd{k_1}\intd{k_2}\intd{k_3}e^{i(x+\hat{\mu}/2)(k_1+k_2+k_3)}A^{abc}_\mu(k_1,k_2,k_3,t)
\;.
\end{align} 

As in the stout smearing case, we can connect the perturbative fields to the flowed gluon field $\tilde{A}^a_\mu(g_0,x,t)$ in the $SU(N_c)$ expansion such that
\begin{align}
U_\mu(x,t)=e^{ig_0T^a\tilde{A}^a_\mu(g_0,x,t)}
\end{align}
and
\begin{align}
\tilde{A}_\mu^{a}(g_0,x,t)&=A^{a}_\mu(x,t)
-\frac{g_0}{2}f^{abc}A^{[bc]}_\mu(x,t)
-\frac{g_0^2}{6}
\bigg[\frac{1}{N_c}\delta^{ae}\delta^{bc}
+\frac{1}{2}\big(d^{bcd}d^{eda}-f^{bcd}f^{eda}\big)\bigg] A^{ebc}_\mu(x,t)
\nonumber\\&
+\mathcal{O}(g_0^3)
\,.
\label{eq:A_tilde_flow}
\end{align}

\subsection{Leading order}
At leading order the differential equation is
\begin{align}
\partial_t A^{a}_\mu(x,t)= Q_{1\mu}^a(x,t)=\sum\limits_{\nu,y} g_{\mu\nu}(y)A^a_\nu(x+y,t)
\end{align}
which in momentum space becomes
\begin{align}
\partial_t A^{a}_\mu(k,t)= \sum\limits_\nu g_{\mu\nu}(k)A^{a}_\nu(k,t)
\end{align}
with the same $g_{\mu\nu}$ as defined at leading order of the stout smearing.
With 
\begin{align}
g^2_{\mu\nu}(k)=\sum\limits_\alpha g_{\mu\alpha}(k)g_{\alpha\nu}(k)=-\hat{k}^2 g_{\mu\nu}(k)
\;,
\end{align}
the matrix exponential of $g$ is 
\begin{align}
\exp(g(k)t)_{\mu\nu}&=\delta_{\mu\nu}+g_{\mu\nu} t+\frac 12 g^2_{\mu\nu} t^2+\hdots\nonumber\\
&=\delta_{\mu\nu}-\frac{1}{\hat{k}^2}\left(e^{-\hat{k}^2 t}-1\right)g_{\mu\nu}(k)\;.
\end{align}
With the initial condition for $t=0$, we arrive at:
\begin{align}
A^{a}_{\mu}(k,t)&=\sum\limits_\nu\left(
e^{-\hat{k}^2t}\delta_{\mu\nu}-\left(e^{-\hat{k}^2t}-1\right)\frac{\hat{k}_\mu\hat{k}_\nu}{\hat{k}^2}\right)A^{a}_\nu(k,0)
\\
&=:\sum\limits_\nu B_{\mu\nu}(k,t)A^{a}_\nu(k,0)
\;.
\label{eq:B_LO}
\end{align}
It is easy to see that for an infinitesimal flow-time $t=\epsilon$, the exponentials become $e^{-\hat{k}^2\epsilon}=1-\epsilon\hat{k}^2+\mathcal{O}(\epsilon^2)$ and the flow transformation becomes the same as a stout smearing with smearing parameter $\varrho=\epsilon$, reiterating the fact that the Wilson flow is generated by infinitesimal stout smearing.\\
The same result is obtained by starting from $n$ stout smearings and taking the limit $n\to \infty$ while keeping the product $n\varrho=:t$ constant. Then $f(k)^n=(1-\tfrac{t}{n}\hat{k}^2)^n\to e^{-\hat{k}^2 t}$ \cite{Ammer:2021koh}. Moreover, this means for a small flow time $t<1$:
\begin{align}
B_{\mu\nu}(k,t)&=\tilde{g}_{\mu\nu}(t,k)+\frac{t^2}{2}(\hat{k}^2)^2g_{\mu\nu}(k)+\mathcal{O}(t^3)
\\&
=\tilde{g}^2_{\mu\nu}(t/2,k)+\frac{t^2}{4}(\hat{k}^2)^2g_{\mu\nu}(k)+\mathcal{O}(t^3)
\\&
=\tilde{g}^n_{\mu\nu}(t/n,k)+\frac{t^2}{2n}(\hat{k}^2)^2g_{\mu\nu}(k)+\mathcal{O}(t^3)
\label{eq:flow_time_correction_1}
\end{align}
Thus the error we get by approximating the flowed field by the field which has been smeared $n$ times with smearing parameter $\varrho=t/n$ is always of the order $t^2$ but decreases with $1/n$.
%
\subsection{Next-to-leading order}
At order $g_0^2$, the differential equation is
\begin{align}
T^aT^b \partial_t \big(A^a_\mu(x,t)A^b_\mu(x,t)\big)+[T^a,T^b]\partial_t A^{[ab]}_\mu(x) &=
2T^aT^bQ^a_{1\mu}(x,t)A^b_\mu(x,t)+2[T^a,T^b]Q^{ab}_{2\mu}(x,t)
\;.
\end{align}
With
\begin{align}
Q^{ab}_{2\mu}(x,t)&=\frac {1}{2}\Big(-S^{a}_{1\mu}(x,t)A^{b}_\mu(x,t)+ S^{ab}_{2\mu}(x,t)-6A^{[ab]}_\mu(x,t)\Big)
\end{align}
and the result from the leading order, we arrive at
\begin{align}
\partial_t A^{[ab]}_\mu(x) &=S^{[ab]}_{2\mu}(x,t)-6A^{[ab]}_\mu(x,t)
\\&
=\sum\limits_{\nu,y}g_{\mu\nu}(y)A^{[ab]}_\nu(x+y,t)+\sum\limits_{\nu\rho,yz}g_{\mu\nu\rho}(y,z)A^a_\nu(x+y,t)A^b_\rho(x+z,t)
\;.
\end{align}
In momentum space, we have
\begin{align}
\partial_t A^{ab}_\mu(k_1,k_2,t)&=
\sum\limits_\nu g_{\mu\nu}(k_1+k_2)A^{[ab]}_\nu(k_1,k_2,t)
+\sum\limits_{\nu\rho}g_{\mu\nu\rho}(k_1,k_2)A^{a}_\nu(k_1,t)A^{b}_\rho(k_2,t)
\end{align}
with the same $g_{\mu\nu\rho}$ we know from stout smearing. Now we have an inhomogenous differential equation and make the ansatz
\begin{align}
A^{[ab]}_\mu(k_1,k_2,t)=\sum\limits_\alpha B_{\mu\alpha}(k_1+k_2,t)C^{[ab]}_\alpha(k_1,k_2,t)
\end{align}
as we know that $B_{\mu\nu}$ solves the homogeneous part. For the inhomogeneous part we need to solve
\begin{align}
\sum\limits_\alpha B_{\mu\alpha}(k_1+k_2,t)\partial_t C^{[ab]}_\alpha(k_1,k_2,t)=\sum\limits_{\nu\rho}g_{\mu\nu\rho}(k_1,k_2)A^{a}_\nu(k_1,t)A^{b}_\rho(k_2,t)
\;,
\end{align}
or equivalently
\begin{align}
\partial_t C^{[ab]}_\mu(k_1,k_2,t)=\sum\limits_\alpha B_{\mu\alpha}(k_1+k_2,-t)\sum\limits_{\nu\rho}g_{\alpha\nu\rho}(k_1,k_2)A^{a}_\nu(k_1,t)A^{b}_\rho(k_2,t)
\;,
\end{align}
because $\big(B_{\mu\nu}(k,t)\big)^{-1}=B_{\mu\nu}(k,-t)$. Thus
\begin{align}
C^{[ab]}_\mu(k_1,k_2,t)=\sum\limits_{\alpha\nu\rho}g_{\alpha\nu\rho}(k_1,k_2)\int\limits_0^t B_{\mu\alpha}(k_1+k_2,-t')A^{a}_\nu(k_1,t')A^{b}_\rho(k_2,t')dt'
\end{align}
and 
\begin{align}
A^{[ab]}_\mu(k_1,k_2,t)&=
\sum\limits_{\alpha\beta\nu\rho}B_{\mu\alpha}(k_1+k_2,t)\int\limits_0^t B_{\alpha\beta}(k_1+k_2,-t')g_{\beta\nu\rho}(k_1,k_2)A^{a}_\nu(k_1,t')A^{b}_\rho(k_2,t')dt'
\nonumber\\&
=\sum\limits_{\alpha\nu\rho}g_{\alpha\nu\rho}(k_1,k_2)\int\limits_0^t B_{\mu\alpha}(k_1+k_2,t-t')A^{a}_\nu(k_1,t')A^{b}_\rho(k_2,t')dt'
\;,
\end{align}
where we have used $\sum_\alpha B_{\mu\alpha}(k,t_1)B_{\alpha\nu}(k,t_2)=B_{\mu\nu}(k,t_1+t_2)$.
We can also go further and express $A^{[ab]}_\mu$ in terms of the original gluon fields
\begin{align}
A^{[ab]}_\mu(k_1,k_2,t)&=
\sum\limits_{\alpha\beta\gamma\nu\rho}g_{\alpha\beta\gamma}(k_1,k_2)\bigg[\int\limits_0^t B_{\mu\alpha}(k_1+k_2,t-t')B_{\beta\nu}(k_1,t^\prime)B_{\gamma\rho}(k_2,t^\prime)dt^\prime\bigg]
\nonumber\\&\times 
A^{a}_\nu(k_1,0)A^{b}_\rho(k_2,0)
\nonumber\\&
=:\sum\limits_{\nu\rho}B_{\mu\nu\rho}(k_1,k_2,t)
A^{a}_\nu(k_1,0)A^{b}_\rho(k_2,0)
\;.
\label{eq:B_NLO}
\end{align}
Again, we can check that we recover $A^{(1)[ab]}_\mu(k_1,k_2)$ for an infinitesimal flow time $t=\epsilon$
\begin{align}
A^{[ab]}_\mu(k_1,k_2,\epsilon)&=
\sum\limits_{\alpha\beta\gamma\nu\rho}g_{\alpha\beta\gamma\delta}(k_1,k_2)\bigg[\epsilon B_{\mu\alpha}(k_1+k_2,0)B_{\beta\nu}(k_1,0)B_{\gamma\rho}(k_2,0)\bigg]
 A^{a}_\nu(k_1,0)A^{b}_\rho(k_2,0)
\\&=
\sum\limits_{\nu\rho}\epsilon g_{\mu\nu\rho}(k_1,k_2) A^{a}_\nu(k_1,0)A^{b}_\rho(k_2,0)
=A^{(1)[ab]}_\mu(k_1,k_2)\Big|_{{\varrho}=\epsilon}
\end{align}
If we expand further (for small $t<1$), we get
\begin{align}
B_{\mu\nu\rho}(k_1,k_2,t)&=\sum\limits_{\alpha\beta\gamma}g_{\alpha\beta\gamma}(k_1,k_2)\bigg(
\delta_{\mu\alpha}\delta_{\beta\nu}\delta_{\gamma\rho}t+\frac{1}{2}\big(g_{\mu\alpha}(k_1+k_2)\delta_{\beta\nu}\delta_{\gamma\rho}
+\delta_{\mu\alpha}g_{\beta\nu}(k_1)\delta_{\gamma\rho}
\nonumber\\&
+\delta_{\mu\alpha}\delta_{\beta\nu}g_{\gamma\rho}(k_2)\big)t^2+\mathcal{O}(t^3)
\bigg)
\\&
=\tilde{g}^{(1)}_{\mu\nu\rho}(t,k_1,k_2)+\frac{t^2}{2}(\hdots)+\mathcal{O}(t^3)
\\&
=\tilde{g}^{(n)}_{\mu\nu\rho}(t/n,k_1,k_2)+\frac{t^2}{2n}(\hdots)+\mathcal{O}(t^3)
\label{eq:flow_time_correction_2}
\end{align}
which shows the same behavior as the first order fields.
\subsection{Next-to-next-to leading order}
The differential equation at third order is
\begin{align}
&-i\frac{g_0^3}{6}\Bigg[T^aT^bT^c \partial_t\big(A^a_\mu(x,t)A^b_\mu(x,t)A^c_\mu(x,t)\big)+\frac{3}{2}\{T^a,[T^b,T^c]\}\partial_t\big(A^a_\mu(x,t)A^{[bc]}_\mu(x,t)\big)
\nonumber\\&
+\bigg(T^aT^bT^c+T^cT^bT^a-\frac{1}{N_c}\Tr\big(T^aT^bT^c+T^cT^bT^a\big)\bigg)\partial_t A^{abc}_\mu(x,t)\Bigg]
\nonumber\\&
=-ig_0^3\Bigg[\frac{1}{2}T^aT^bT^cQ^a_{1\mu}(x,t)A^b_\mu(x,t) A^c_\mu(x,t)+\frac{1}{2}T^a[T^b,T^c]Q^a_{1\mu}(x,t)A^{[bc]}_\mu(x,t)\nonumber\\&
+[T^a,T^b]T^cQ^{ab}_{2\mu}(x,t)A^c_\mu(x,t)
-\bigg(T^aT^bT^c+T^cT^bT^a-\frac{1}{N_c}\Tr\big(T^aT^bT^c+T^cT^bT^a\big)\bigg)Q^{abc}_{3\mu}(x,t)
\Bigg]
\;.
\end{align}
The right-hand side can be rewritten to look like
\begin{align}
&-i\frac{g_0^3}{6}\bigg(T^aT^bT^c\partial_t(A^a_\mu(x,t) A^b_\mu(x,t) A^c_\mu(x,t))+\frac{3}{2}\{T^a[T^b,T^c]\}\partial_t ( A^a_\mu(x,t)A^{[bc]}_\mu(x,t))
\nonumber\\&
- \bigg(T^aT^bT^c+T^cT^bT^a-\frac{1}{N_c}\Tr\big(T^aT^bT^c+T^cT^bT^a\big)\bigg)
\Big(-\frac{1}{2}Q_{1\mu}^a(x,t)A^b_\mu(x,t)A^c_\mu(x,t)
\nonumber\\&
-A^aQ_{1\mu}^bA^c_\mu(x,t)+3V^{\{ab\}}_{2\mu}(x,t)A^c_\mu(x,t)-3V^{abc}_{3\mu}(x,t)-6A^a_\mu(x,t)A^b_\mu(x,t)A^c_\mu(x,t)
\nonumber\\&
+6A^{abc}_\mu(x,t)\Big)\bigg)
\;.
\end{align}
Thus
\begin{align}
\partial_t A^{abc}_\mu(x,t)&=
\frac{1}{2}Q_{1\mu}^a(x,t)A^b_\mu(x,t)A^c_\mu(x,t)
+A^aQ_{1\mu}^bA^c_\mu(x,t)
-3V^{\{ab\}}_{2\mu}(x,t)A^c_\mu(x,t)
\nonumber\\&
+3V^{abc}_{3\mu}(x,t)+6A^a_\mu(x,t)A^b_\mu(x,t)A^c_\mu(x,t)-6A^{abc}_\mu(x,t)
\;.
\end{align}
In momentum space, we get
\begin{align}
\partial_t A^{abc}_\mu(k_1,k_2,k_3,t)&=\sum\limits_\nu	g_{\mu\nu}(k_1+k_2+k_3)A^{abc}_\nu(k_1,k_2,k_3,t)
\nonumber\\&
+6\sum\limits_{\nu\rho}g_{\mu\nu\rho}(k_1,k_2+k_3)A^a_\nu(k_1,t)A^{[bc]}_\rho(k_2,k_3,t)
\nonumber\\&
+\sum\limits_{\nu\rho\sigma}G_{\mu\nu\rho\sigma}(k_1,k_2,k_3)A^a_\nu(k_1,t)A^b_\rho(k_2,t)A^c_\sigma(k_3,t)
\;.
\end{align}
We make again the following ansatz 
\begin{align}
A^{abc}_\mu(k_1,k_2,k_3,t)=\sum\limits_\alpha B_{\mu\alpha}(k_1+k_2+k_3,t)C_\alpha^{abc}(k_1,k_2,k_3,t)
\end{align}
as we know from the leading order that $B_{\mu\nu}(k_1+k_2+k_3,t)$ satisfies the homogeneous part of the differential equation. Then we are left with the inhomogeneous part
\begin{align}
\sum\limits_\alpha  B_{\mu\alpha}(k_1+k_2+k_3,t)\partial_t  C^{abc}_\alpha(k_1,k_2,k_3,t)&=
6\sum\limits_{\nu\rho}g_{\mu\nu\rho}(k_1,k_2+k_3)A^a_\nu(k_1,t)A^{[bc]}_\rho(k_2,k_3,t)
\nonumber\\&
+\sum\limits_{\nu\rho\sigma}G_{\mu\nu\rho\sigma}(k_1,k_2,k_3)A^a_\nu(k_1,t)A^b_\rho(k_2,t)A^c_\sigma(k_3,t)
\;,
\end{align}
or equivalently
\begin{align}
\partial_t  C^{abc}_\alpha(k_1,k_2,k_3,t)&=
6\sum\limits_{\alpha\nu\rho}B_{\mu\alpha}(k_1+k_2+k_3,-t)g_{\alpha\nu\rho}(k_1,k_2+k_3)A^a_\nu(k_1,t)A^{[bc]}_\rho(k_2,k_3,t)
\nonumber\\&
+\sum\limits_{\alpha\nu\rho\sigma}B_{\mu\alpha}(k_1+k_2+k_3,-t)G_{\alpha\nu\rho\sigma}(k_1,k_2,k_3)A^a_\nu(k_1,t)A^b_\rho(k_2,t)A^c_\sigma(k_3,t)
\;.
\end{align}
Integrating this and putting everything together, we get
\begin{align}
A_\mu^{abc}&(k_1,k_2,k_3,t)=
\nonumber\\&
 \sum\limits_{\alpha\nu\rho}g_{\alpha\nu\rho}(k_1,k_2+k_3) 
\int\limits_0^t B_{\mu\alpha}(k_1+k_2+k_3,t-t^\prime)
A_\nu^{a}(k_1,t^\prime)A_\rho^{[bc]}(k_2,k_3,t^\prime)
\,dt^\prime
\nonumber\\&
+\sum\limits_{\alpha\nu\rho\sigma}G_{\alpha\nu\rho\sigma}(k_1,k_2,k_3) 
\int\limits_0^t B_{\mu\alpha}(k_1+k_2+k_3,t-t^\prime)
A_\nu^{a}(k_1,t^\prime)A_\rho^{b}(k_2,t^\prime)A_\sigma^{c}(k_3,t^\prime)
\,dt^\prime
\end{align}
and we can define a function $B_{\mu\nu\rho\sigma}(k_1,k_2,k_3,t)$ for the third order of the Wilson flow:
\begin{align}
A_\mu^{abc}&(k_1,k_2,k_3,t)=
\nonumber\\&
 \sum\limits_{\nu\rho\sigma}\Bigg(\sum\limits_{\alpha\beta\gamma}g_{\alpha\beta\gamma}(k_1,k_2+k_3) 
\int\limits_0^t B_{\mu\alpha}(k_1+k_2+k_3,t-t^\prime)B_{\beta\nu}(k_1,t^\prime)
B_{\gamma\rho\sigma}(k_2,k_3,t^\prime)
\,dt^\prime
\nonumber\\&
+\sum\limits_{\alpha\beta\gamma\delta}G_{\alpha\beta\gamma\delta}(k_1,k_2,k_3) 
\int\limits_0^t B_{\mu\alpha}(k_1+k_2+k_3,t-t^\prime)
B_{\beta\nu}(k_1,t^\prime)
B_{\gamma\rho}(k_2,t^\prime)
B_{\delta_\sigma}(k_3,t^\prime)
\,dt^\prime
\Bigg)
\nonumber\\&
\times
A_\nu^{a}(k_1,0)A_\rho^{b}(k_2,0)A_\sigma^{c}(k_3,0)
\nonumber\\&
=:\sum\limits_{\nu\rho\sigma}B_{\mu\nu\rho\sigma}(k_1,k_2,k_3,t)A_\nu^{a}(k_1,0)A_\rho^{b}(k_2,0)A_\sigma^{c}(k_3,0)
\label{eq:B_NNLO}
\;.
\end{align}
Again, we can check that the error one makes by approximating the Wilson flow by stout smearing is of the order $t^2/(2n)$:
\begin{align}
B_{\mu\nu\rho\sigma}(k_1,k_2,k_3,t)=\tilde{g}^{(n)}_{\mu\nu\rho\sigma}(t/n,k_1,k_2,k_3)+\frac{t^2}{2n}(\hdots)
\;.
\label{eq:flow_time_correction_3}
\end{align} 
Thus we have found three $B$ functions that are the Wilson flow equivalent to the $\tilde{g}$ functions from the expansion of stout smearing.
%
\subsection{Continuum limit}
%
In the continuum limit $a\to 0$, we replace $\cos(k_\mu)\to 1$ and $\sin(k_\mu)\to  k_\mu$, and expand to first order in the momenta.
We get
\begin{align}
B_{\mu\nu}(k,t)&\to 
\sum\limits_\nu\bigg(
e^{-k^2t}\delta_{\mu\nu}-\left(e^{-k^2t}-1\bigg)\frac{k_\mu k_\nu}{k^2}\right)
\\
g_{\mu\nu\rho}(k_1,k_2)&\to 
i \Big(-\delta_{\beta\nu}\big(2k_{1\rho}+k_{2\rho}\big)+\delta_{\beta\rho}\big(2k_{2\nu}+k_{1\nu}\big)
+\delta_{\nu\rho}\big(k_{1\mu}-k_{2\mu}\big)\Big)
\\
G_{\mu\nu\rho\sigma}(k_1,k_2,k_3)&\to 
6\big(\delta_{\mu\nu}\delta_{\rho\sigma}-\delta_{\mu\rho}\delta_{\nu\sigma}\big)
\;,
\end{align}
and the flowed field $\tilde{A}_\mu$ from Eq.~(\ref{eq:A_tilde_flow}) in this limit fulfils the differential equation of the continuum gradient flow
\begin{align}
\partial_t \tilde{A}_\mu(x,t)&=\sum\limits_\nu \tilde{D}_\nu\tilde{F}_{\nu\mu},
&
\tilde{A}_\mu(x,0)&=A_\mu(x)
\end{align}
with
\begin{align}
\tilde{D}_\mu&=\partial_\mu+\big[\tilde{A}_\mu(x,t),\; \cdot \;\big],
&
\tilde{F}_{\mu\nu}&=\partial_\mu\tilde{A}_\nu(x,t)-
\partial_\nu\tilde{A}_\mu(x,t)+
\big[\tilde{A}_\mu(x,t),\tilde{A}_\nu(x,t)\big]
\;.
\end{align}
These results agree with the ones given in Refs.~\cite{Luscher:2010iy} and \cite{Luscher:2011bx}.

%
\section{Feynman rules of fermion actions with stout smearing or Wilson flow\label{sec:feynman_rules}}
%
The Feynman rules are obtained by inserting the  expansion of the link variables (\ref{eq:generic_U_expansion})  into the Fermion action.
This results in expressions for {antiquark}-quark-$n$-gluon vertices at each order $g_0^n$. 
These vertices are proportional to $a^{n-1}$ such that only the $\bar{q}qg$ vertex survives in the continuum limit $a\to 0$, as is expected from continuum perturbation theory.
For the purpose of one-loop calculations, we will need the  $\bar{q}qg$-, $\bar{q}qgg$- and $\bar{q}qggg$-vertex, i.e.\ anything up to order $g_0^3$.

For deriving the Feynman rules of the fermion action including stout smearing or Wilson flow, there are two possible  strategies.
One is to simply insert the perturbative expansion of the smeared or flowed link variable into the action and expand to the needed order in $g_0$. 
This ``brute force'' approach leads to very large intermediate expressions and involves many steps where indices have to be renamed etc. 
This is not feasible beyond one step of stout smearing.
The second much more effective method is based on the $SU(N_c)$ expansion of the link variable.
One simply derives the Feynman rules for generic link variables of the form
\begin{align}
U_\mu(x)=1+ig_0T^aA^a_\mu(x)-\frac{g_0^2}{2}T^aT^b A^a_\mu(x)A^b_\mu(x)-i\frac{g_0^3}{6}T^aT^bT^cA^a_\mu(x)A^b_\mu(x)A^c_\mu(x)
\label{eq:generic_U_expansion}
\end{align} 
which results in fairly simple expressions.
One then couples the stout or flow ``form factors'' to these generic Feynman rules to obtain the complete results.
This also practically allows us to easily split lengthy expressions into smaller parts for {\emph{Mathematica}} to work on.  {
A similar procedure has been described in the context of Feynman rules for the HISQ action in Ref.~\cite{Hart:2009nr}.}
The following sections show the derivations of the generic Feynman rules and how to couple them to stout smearing or the Wilson flow.\\

We start with the Wilson quark action 
\begin{align}
S_W=\sum\limits_{x,y}\sum\limits_{\mu=\pm 1}^{\pm 4}\bar{\psi}(x)\frac{1}{2}\bigg[  
U_\mu(x)\delta_{x+\hat{\mu},y}\gamma_\mu-r\Big(U_\mu(x)\delta_{x+\hat{\mu},y}-\delta_{x,y}\Big)
\bigg]\psi(y)
\;.
\end{align}
Links with negative indices are given by $U_{-\mu}(x)=U^\dagger_\mu(x-\hat{\mu})$ and $\gamma_{-\mu}=-\gamma_\mu$.
Up to order $g_0^3$ we get the well{-}known three vertices of a quark {antiquark} pair coupling to one, two, and three gluons respectively.
\begin{align}
V^a_{1W\mu}(p,q)&=-g_0T^a\Big(i\gamma_\mu c(p_\mu+q_\mu)+r s(p_\mu+q_\mu)\Big)
\\
V^{ab}_{2W\mu\nu}(p,q)&=\frac{g_0^2}{2}T^aT^b\delta_{\mu\nu}\Big(i\gamma_\mu s(p_\mu+q_\mu)-r c(p_\mu+q_\mu)\Big)
\\
V^{abc}_{3W\mu\nu\rho}(p,q)&=\frac{g_0^3}{6}T^aT^bT^c\delta_{\mu\nu}\delta_{\mu\rho}\Big(i\gamma_\mu c(p_\mu+q_\mu)+r s(p_\mu+q_\mu)\Big)
\end{align}
These appear in the expansion of the action contracted with the (original) gluon fields
\begin{align}
S_W&\sim \sum\limits_\mu\bar{\psi}(q)T^aV_{1\,\mu}(p,q)A^a_\mu(q-p)\psi(p)
\nonumber\\&
+ \sum\limits_{\mu\nu}\bar{\psi}(q)T^aT^bV_{2\,\mu\nu}(p,q)A^a_\mu(k_1)A^b_\nu(k_2)\delta(q-p-k_1-k_2)\psi(p)
\nonumber\\&
+ \sum\limits_{\mu\nu\rho}\bar{\psi}(q)T^aT^bT^cV_{3\,\mu\nu\rho}(p,q)A^a_\mu(k_1)A^b_\nu(k_2)A^c_\rho(k_3)\delta(q-p-k_1-k_2-k_3)\psi(p)
\end{align}
In order to derive the Feynman rules with smearing or gradient flow, we replace the gluon fields $A^a_\mu$ by $\tilde{A}^{(n)a}_\mu$ or $\tilde{A}^a_\mu(t)$ respectively. Thereby new terms are added at higher orders of $g_0$. For example, replacing the $A^a_\mu$ which couples to $V^a_{1\,\mu}$ by a smeared or flowed field, provides the new $\bar{q}qg$ vertex with a form factor and adds terms to the $\bar{q}qgg$ and $\bar{q}qggg$ vertices proportional to $A^{[ab]}_\mu$ and $A^{abc}_\mu$ respectively.
Specifically, the new stout smeared vertices look like
\begin{align}
V^{(n)a}_{1\mu}(p,q)&=T^a\sum\limits_\nu \tilde{g}^n_{\mu\nu}(\varrho,q-p)V_{1\nu}(p,q)
\\
V^{(n)ab}_{2\mu\nu}(p,q,k_1,k_2)&=
T^aT^b\sum\limits_{\rho\sigma}
\tilde{g}^n_{\mu\rho}(\varrho,k_1) \tilde{g}^n_{\nu\sigma}(\varrho,k_2) V_{2\rho\sigma}(p,q,k_1,k_2)
\nonumber\\&
-\frac{g_0}{2i}[T^a,T^b]\sum\limits_\rho \tilde{g}^{(n)}_{\rho\mu\nu}(\varrho,k_1,k_2)V_{1\rho}(p,q)
\\
V^{(n)abc}_{3\mu\nu\rho}(p,q,k_1,k_2,k_3)&=T^aT^bT^c\sum\limits_{\alpha\beta\gamma}
\tilde{g}^n_{\mu\alpha}(\varrho,k_1) \tilde{g}^n_{\nu\beta}(\varrho,k_2)  \tilde{g}^n_{\rho\gamma}(\varrho,k_3)V_{3\alpha\beta\gamma}(p,q,k_1,k_2,k_3)
\nonumber\\
-\frac{g_0}{2i}\{T^a,[T^b,T^c]\}&\sum\limits_{\alpha\beta} \tilde{g}^n_{\mu\alpha}(\varrho,k_1)\tilde{g}^{(n)}_{\beta\nu\rho}(\varrho,k_2,k_3)
V_{2\alpha\beta}(p,q,k_1,k_2+k_3)
\nonumber\\
-\frac{g_0^2}{6}
\bigg[T^aT^bT^c+T^c&T^bT^a-\frac{1}{N_c}\Tr\big(T^aT^bT^c+T^cT^bT^a\big)\bigg] \sum\limits_\alpha \tilde{g}^{(n)}_{\alpha\mu\nu\rho}(\varrho,k_1,k_2,k_3)V_{1\alpha}(p,q)
\end{align}
with the $\tilde{g}$ given in Eqs.~(\ref{eq:g_tilde_LO}),(\ref{eq:g_tilde_NLO}), and (\ref{eq:g_tilde_NNLO}).
The vertices with Wilson flow are analogously
\begin{align}
V^{a}_{1\mu}(p,q,t)&=T^a\sum\limits_\nu B_{\mu\nu}(q-p,t) V_{1\nu}(p,q)
\\
V^{ab}_{2\mu\nu}(p,q,k_1,k_2,t)&=
T^aT^b\sum\limits_{\rho\sigma}
B_{\mu\rho}(k_1,t)B_{\nu\sigma}(k_2,t) V_{2\rho\sigma}(p,q,k_1,k_2)
\nonumber\\&
-\frac{g_0}{2i}[T^a,T^b]\sum\limits_\rho B_{\rho\mu\nu}(k_1,k_2,t)V_{1\rho}(p,q)
\\
V^{abc}_{3\mu\nu\rho}(p,q,k_1,k_2,k_3,t)&=T^aT^bT^c\sum\limits_{\alpha\beta\gamma}
B_{\mu\alpha}(k_1,t) B_{\nu\beta}(k_2,t)  B_{\rho\gamma}(k_3,t)V_{3\alpha\beta\gamma}(p,q,k_1,k_2,k_3)
\nonumber\\
-\frac{g_0}{2i}\{T^a,[T^b,T^c]\}&\sum\limits_{\alpha\beta} B_{\mu\alpha}(k_1,t)B_{\beta\nu\rho}(k_2,k_3,t)
V_{2\alpha\beta}(p,q,k_1,k_2+k_3)
\nonumber\\
-\frac{g_0^2}{6}
\bigg[T^aT^bT^c+T^c&T^bT^a-\frac{1}{N_c}\Tr\big(T^aT^bT^c+T^cT^bT^a\big)\bigg] \sum\limits_\alpha B_{\alpha\mu\nu\rho}(k_1,k_2,k_3,t)V_{1\alpha}(p,q)
\end{align}
with the $B$ given in Eqs.~(\ref{eq:B_LO}), (\ref{eq:B_NLO}), and (\ref{eq:B_NNLO}).\\

For the calculation of the self-energy below we need the $\bar{q}qg$- and  $\bar{q}qgg$-vertices and we use the Feynman rules of the  Wilson clover action
\begin{align}
V_{1\mu}(p,q)&=V_{1W\mu}(p,q)+V_{1C\mu}(p,q)
\\
V_{2\mu\nu}(p,q,k_1,k_2)&=V_{2W\mu\nu}(p,q)+V_{2C\mu\nu}(p,q,k_1,k_2)
\end{align}
with the clover vertices
\begin{align}
V_{1C\mu}(p,q)&=i\frac{g_0}{2}\csw\sum\limits_{\nu}\sigma_{\mu\nu}c(p_\mu-q_\mu)\bar{s}(p_\nu-q_\nu)
\\
V_{2C\mu\nu}(p,q,k_1,k_2)&=
i g_0^2 \csw\Bigg(\frac{1}{4}\delta_{\mu\nu}
\sum\limits_\rho \sigma_{\mu\rho}s(q_\mu-p_\mu)\big(\bar{s}(k_{2\rho})-\bar{s}(k_{1\rho})\big)
\nonumber\\&
+\sigma_{\mu\nu}\bigg(c(k_{1\nu})c(k_{2\mu})c(q_\mu-p_\mu)c(p_\nu-p_\nu)-\frac{1}{2}c(k_{1\mu})c(k_{2\nu})\bigg)
\Bigg)
\end{align}
where $\sigma_{\mu\nu}=\frac{i}{2}[\gamma_\mu,\gamma_\nu]$ and the  tree-level value $\csw=r$.
%
%
\section{Fermion self-energy with stout smearing or Wilson~flow\label{sec:self_energy}}
%
\begin{figure}[!tb]
\centering
\includegraphics[scale=1.0]{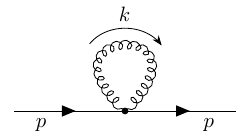}
\includegraphics[scale=1.0]{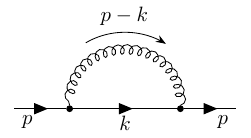}
\caption{Tadpole (left) and sunset (right) diagrams of the quark self-energy.}
\label{fig:self_energy}
\end{figure}

To illustrate the use of the smeared Feynman rules derived in the previous
section, we give in this section the result of the self-energy of a clover
fermion coupled to a gluon background through stout-smeared or
gradient-flowed vertices.
The smoothing thus affects both the covariant derivative (in the unimproved
part of the fermion action) and the field strength tensor (in the part
proportional to $\csw$).

At the one-loop order the fermion self-energy $\Sigma$ receives contributions
from the ``tadpole'' and the ``sunset'' diagrams shown in Fig.~\ref{fig:self_energy}.
We define $\Sigma_0$ through the linearly divergent part and $\Sigma_1$ as
the finite piece in an expansion in the lattice spacing \cite{GonzalezArroyo:1982ts,Hamber:1983qa,Capitani:2002mp}
\begin{align}
\Sigma=\frac{g_0^2C_F}{16\pi^2}\bigg(\frac{\Sigma_0}{a}+\Sigma_1 i\slashed{p}+\mathcal{O}(a)\bigg)
\;.
\label{eq:Sigma}
\end{align}
where $p_\mu$ is the momentum of the external fermion (see Fig.~1), and
the additive mass shift follows as
\begin{align}
am_{\mathrm{crit}}=\frac{g_0^2C_F}{16\pi^2}\Sigma_0
\;.
\end{align}
The contributions to $\Sigma$ from the two diagrams read
\begin{align}
\Sigma^{\mathrm{(tadpole)}}&=\intdk \sum\limits_{\mu,\nu,a} G_{\mu\nu}(k)V_{2\,\mu\nu}^{aa}(p,p,k,-k)\\
\Sigma^{\mathrm{(sunset)}}&=\intdk\sum\limits_{\mu,\nu,a} V_{1\,\mu}^a(p,k)G_{\mu\nu}(p-k)S(k)V_{1\,\nu}^a(k,p)
\end{align}
respectively, and from the sum we obtain $\Sigma_0$ and $\Sigma_1$ by
\begin{align}
\frac{g_0^2 C_F}{16\pi^2}\frac{\Sigma_0}{a}&=\Sigma\big|_{p=0}\\
\frac{g_0^2 C_F}{16\pi^2}\Sigma_1&=\frac{1}{4}\Tr\bigg[\frac{\partial}{\partial p_\mu}\Sigma\cdot \gamma_\mu\bigg]_{p=0}
\end{align}
where $\mu$ is kept fixed, i.e.\ not summed over.
While $\Sigma_0$ is gauge independent, $\Sigma_1$ is part of the fermion field renormalization and does depend on the gauge. We use Feynman gauge and all results for $\Sigma_1$ are given in this gauge.
We refer the reader to Ref.~\cite{Ammer:2023otl} for the explicit
form of the gluon and fermion propagators in this gauge.
Furthermore, for the gluon propagator we employ both the Wilson ``plaquette''
gauge action and the tree-level ``Symanzik improved'' gluon action, since
these are the most popular choices in contemporary {nonperturbative} studies.
These results will be labeled as ``plaq'' and ``sym'', respectively,
in the figures and tables below.
A somewhat technical point is that in Feynman gauge $\Sigma_1$ depends
logarithmically on an external momentum $p\to0$
\begin{align}
\Sigma_1=\log(p^2)+\Sigma_{10}
\label{eq:Sigma_1}
\end{align} 
and this is why we shall plot or quote the constant part $\Sigma_{10}$.

Besides the choice of the gauge action also the numerical values of the
Wilson parameter $r$ and the clover coefficient $c_\mathrm{SW}$ matter and, of course,
the parameter $({\varrho},n_\mathrm{stout})$ or the gradient flow time $t/a^2$. To enhance verbosity, in this section $n_\mathrm{stout}$ is
used to denote the number of stout steps.
It goes without saying that we cannot represent all information in a compact form.
This is why we restrict ourselves to a quick overview on how $\Sigma$ varies as a function
of ${\varrho}$ at fixed $c_\mathrm{SW}=r$, and an overview on how it depends on ($c_\mathrm{SW}=r$) $\csw$ and $r$ 
for a few choices of $({\varrho},n_\mathrm{stout})$ or the flow time $t/a^2$.
Some additional information is shifted to the appendix, in the hope that the selection
made is broad enough to allow the user to make an informed decision when it comes to choosing
a lattice fermion action for future {nonperturbative} studies.

In Ref.~\cite{Capitani:2006ni}  results for $\Sigma_0$ with up to three steps of stout smearing at the specific
value $\varrho=0.1$ (labeled there as $\alpha=0.6$, due to the perturbative matching between stout
and APE smearing) were given, which agree with our results (see our Appendix for details).
In addition, results for $\Sigma_0$ and $\Sigma_1$ with one step of SLINC smearing for
plaquette and L\"uscher-Weisz glue were given in Ref.~\cite{Horsley:2008ap}.
In the SLINC action only the links entering the Wilson part of the fermion action are subject
to one step of stout smearing (with variable $\varrho$), while the clover part stays unsmeared.
Reproducing these results%
\footnote{There is a typo in Eq.~(53) of Ref.~\cite{Horsley:2008ap}, the number 2235.407087 should read 2335.407087.}
served as a cross-check for our code.\\

\begin{figure}[!h]
\centering
\includegraphics[scale=0.55]{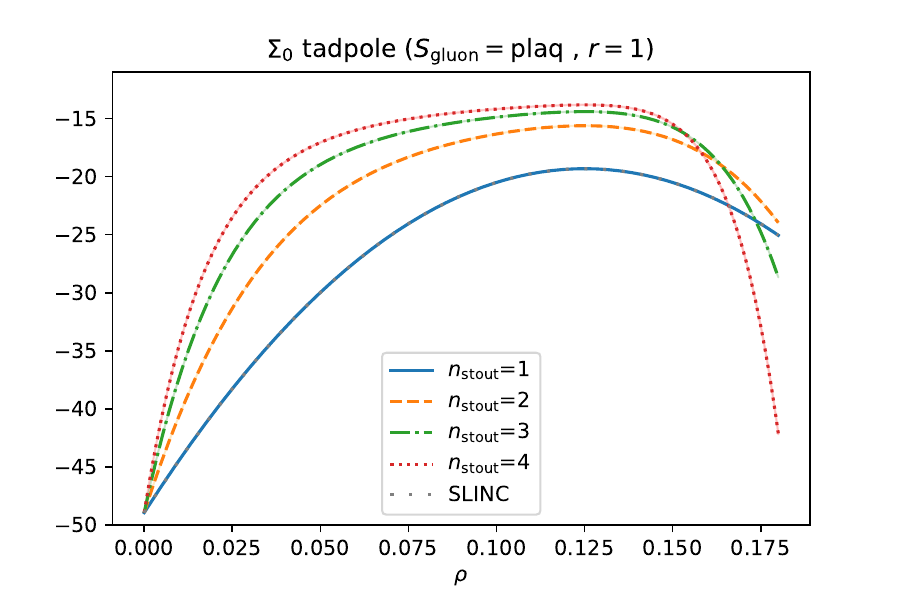}
\includegraphics[scale=0.55]{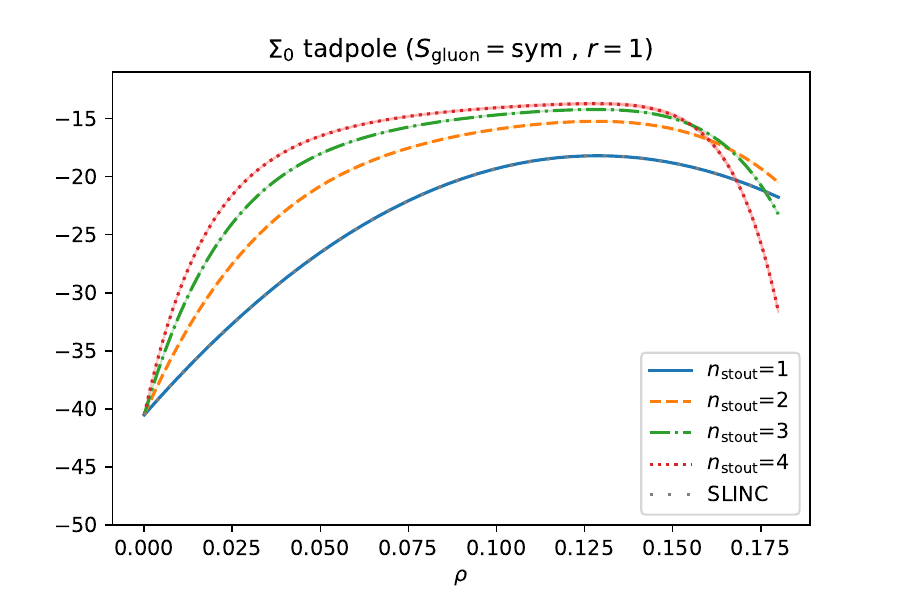}
\includegraphics[scale=0.55]{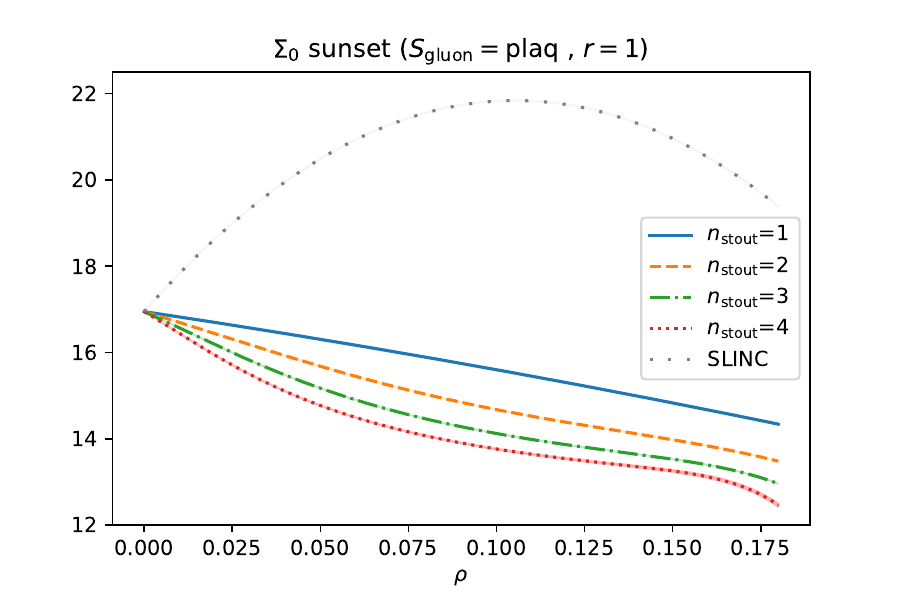}
\includegraphics[scale=0.55]{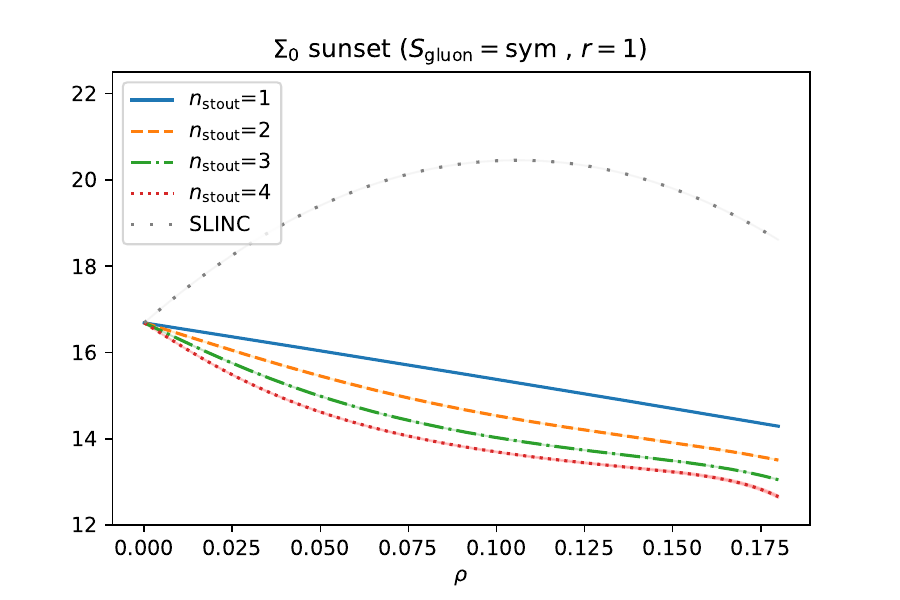}
\includegraphics[scale=0.55]{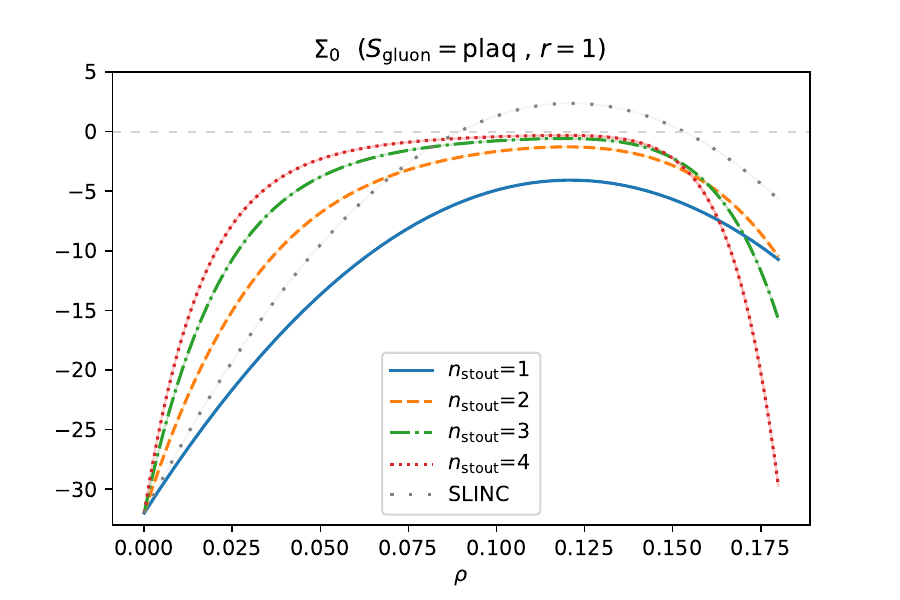}
\includegraphics[scale=0.55]{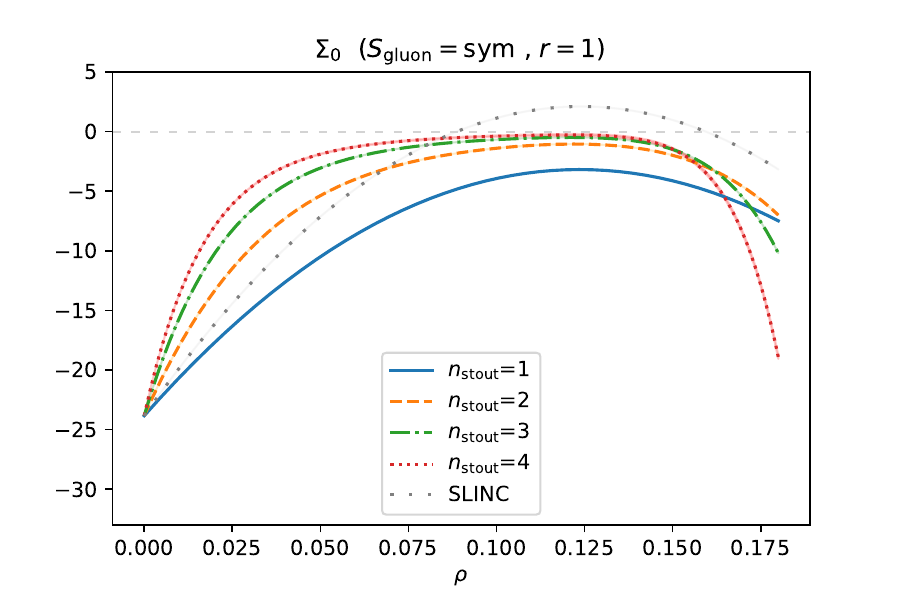}
\caption{
Contributions to the divergent piece $\Sigma_0$ of the clover
($\csw=r=1$) fermion self-energy from the tadpole (top) and sunset
(middle) diagrams and their sum (bottom). Results are shown as a function
of the smearing parameter $\varrho$, with plaquette gluon background (left) and
tree-level Symanzik-improved gluon background (right).
In the top row the results for one step of overall stout
smearing ($n_\mathrm{stout}=1$) coincide with those for SLINC fermions \cite{Horsley:2008ap}.
\label{fig:S0_stout}}
\end{figure}
\begin{figure}[!h]
\centering
\includegraphics[scale=0.53]{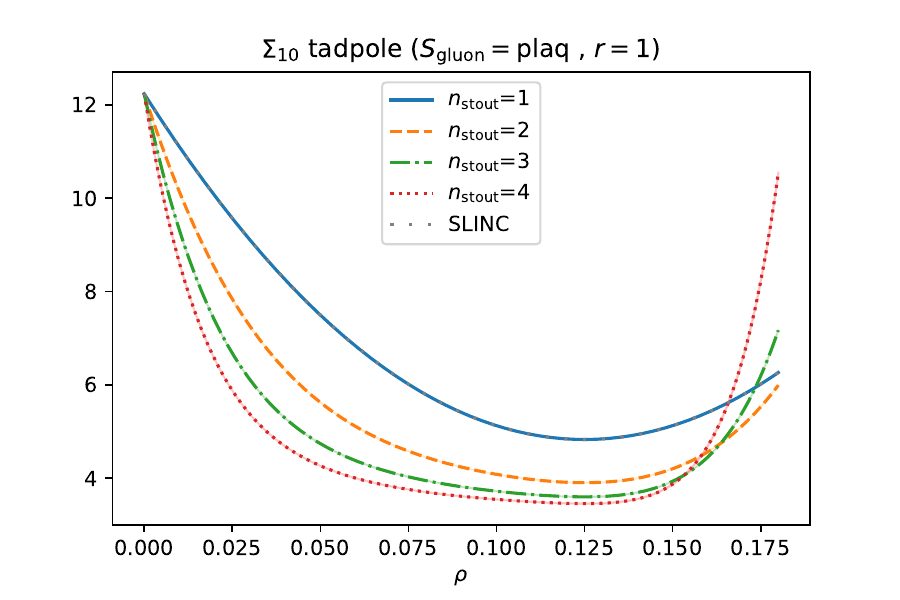}
\includegraphics[scale=0.53]{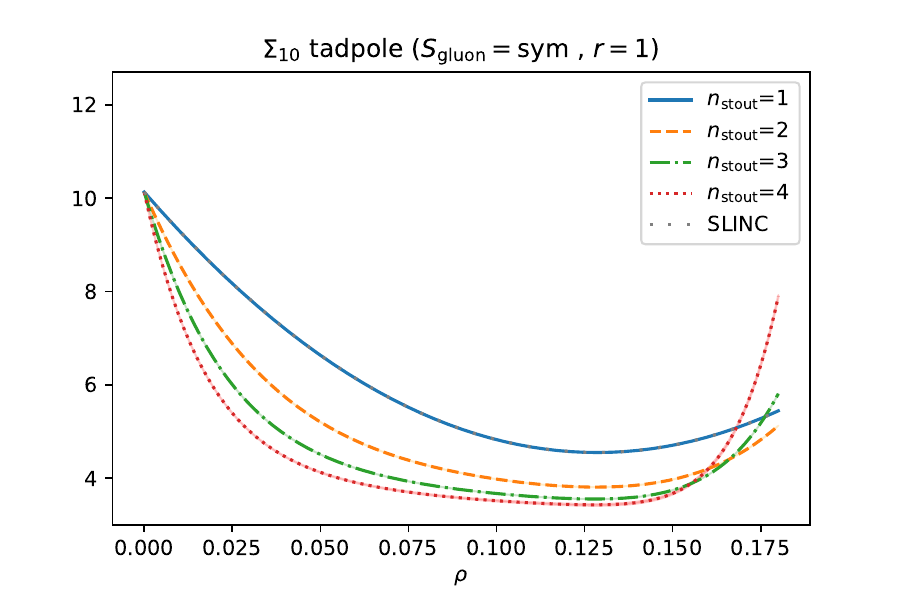}
\includegraphics[scale=0.53]{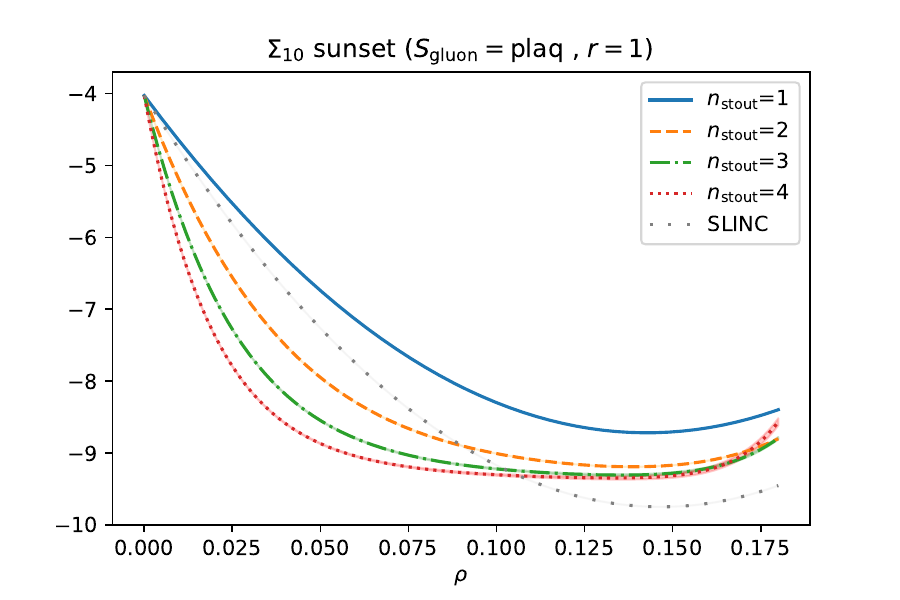}
\includegraphics[scale=0.53]{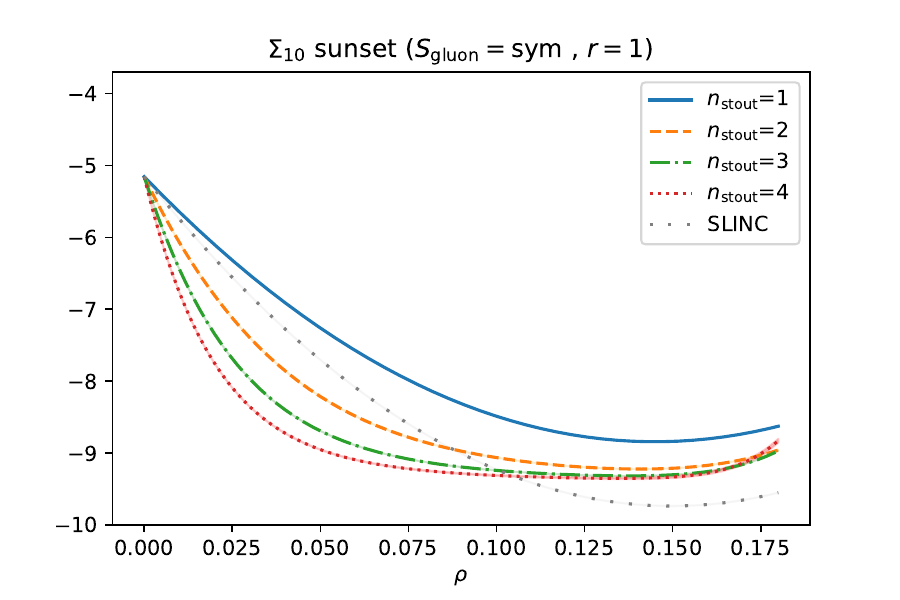}
\includegraphics[scale=0.53]{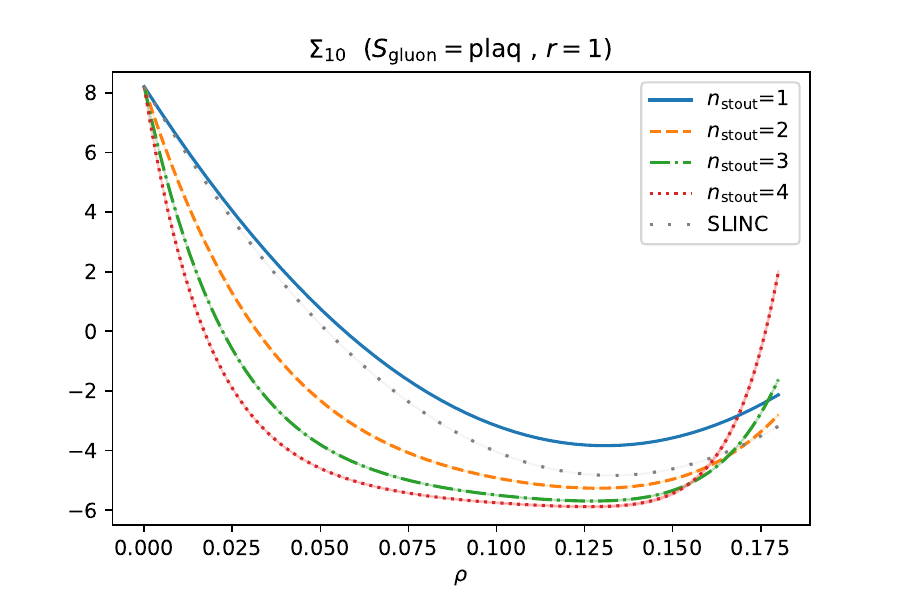}
\includegraphics[scale=0.53]{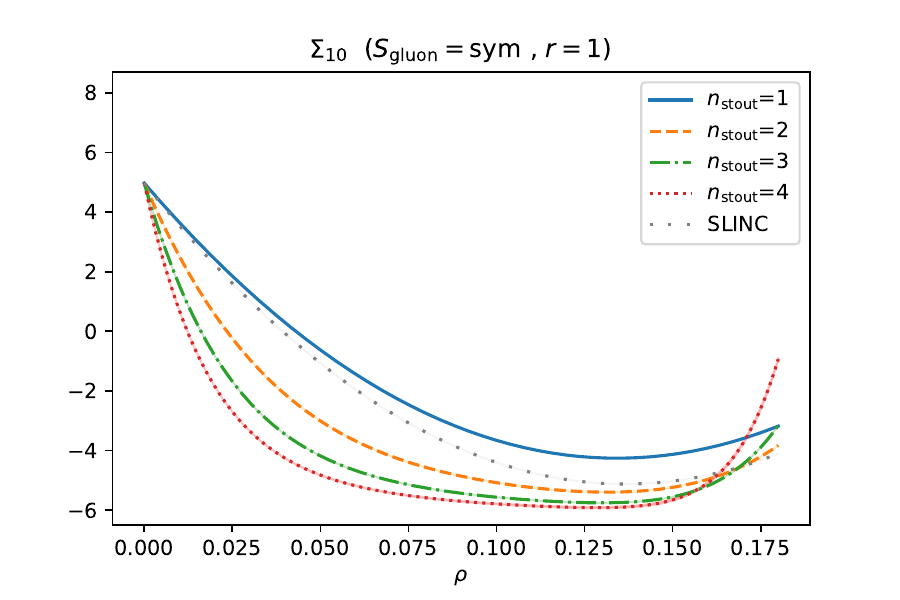}
\caption{ Same as Fig.~\ref{fig:S0_stout}, but for the finite piece $\Sigma_{10}$; see
Eq.~(\ref{eq:Sigma_1}).
\label{fig:S1_stout}}
\end{figure}

Calculating $\Sigma$ and extracting $\Sigma_0$ and $\Sigma_{10}$ from it becomes more challenging
as the number of stout steps (or the overall flow time) increases.
We managed to get reasonably accurate results for up to $n_\mathrm{stout}=4$ smearing steps.\\

\begin{table}[!b]
\centering
\begin{tabular}{|c|c|c|c|c|}
\hline
 & $n=1$ & $n=2$ & $n=3$ & $n=4$ \\
 \hline
 $\varrho_\mathrm{max}$ of $\Sigma_0$ (plaq) &
 0.12096871(3) &
0.11973422(9) &
0.1189366(3) &
0.1187(4)\\
$\varrho_\mathrm{max}$ of $\Sigma_0$ (sym) &
0.12357712(3) &
0.1220287(3) &
0.120999(5) &
0.1204(9)\\
\hline
$\varrho_\mathrm{min}$ of $\Sigma_{10}$ (plaq) &
0.13078397(5) &
0.1285409(2)&
0.1271189(9)&
0.1263(6)\\
$\varrho_\mathrm{min}$ of $\Sigma_{10}$ (sym) &
0.13417728(6)&
0.1315944(2)&
0.129866(6)&
0.1287(6)\\
\hline
\end{tabular}
\caption{Position of the extreme points in the bottom panels of Figs.~\ref{fig:S0_stout}  and \ref{fig:S1_stout}.
\label{tab:extrema}}
\end{table}
Fig{ure}~\ref{fig:S0_stout} displays how the piece $\Sigma_0$ of the self-energy of a clover fermion ($\csw=r$)
depends on $\varrho$ for up to four stout steps.
Regardless of the gauge background adopted (left versus right column), choosing a reasonable $\varrho$ is
found to reduce the additive mass shift (bottom row) quite drastically, while (in Feynman gauge) the
``tadpole'' and ``sunset'' contributions, inspected individually, do not show any noteworthy feature
(top and middle panels).
In the physical result (bottom row) the initial slope is proportional to $n_\mathrm{stout}$ (see below).
Compared to the full curve the SLINC curve (wide-spaced dots) seems to be an improvement for small
values of $\varrho$, but for a value $\varrho\sim0.08$ the critical mass of SLINC fermions changes sign.
On the other hand, the ``overall smearing'' advocated in Ref.~\cite{Capitani:2006ni} seems to be rather benevolent;
the more smearing steps (with $0<\varrho<0.12$) are taken, the more $\Sigma_0$ tends to zero.
Moreover, regardless of $n_\mathrm{stout}$ and the chosen gluon background, the maximum of $\Sigma_0$
is always near $\varrho\sim0.12$, see Tab{le}~\ref{tab:extrema} for details.
This is consistent with the observation made in Ref.~\cite{Capitani:2006ni} that $\varrho$ in the stout recipe should be
kept below 0.125 in order to have a (first order) smearing form factor smaller than one throughout the
Brillouin zone.

Figure~\ref{fig:S1_stout} displays how $\Sigma_{10}$ depends on $\varrho$ for up to four stout steps (again
for $\csw=r$).
Iterated stout smearings with a parameter in the vicinity of $\varrho\simeq 0.1$ seem
to bring this quantity to a value near $-6$.
Again the extremal points are found to be near $\varrho\sim0.12$, see Table~\ref{tab:extrema} for details.\\

\begin{figure}[!h]
\centering
\includegraphics[scale=0.55]{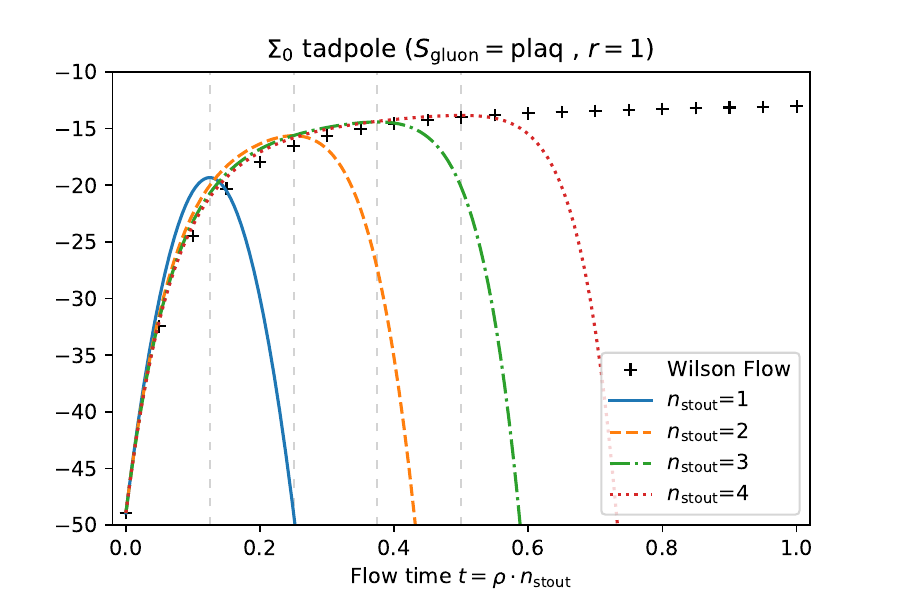}
\includegraphics[scale=0.55]{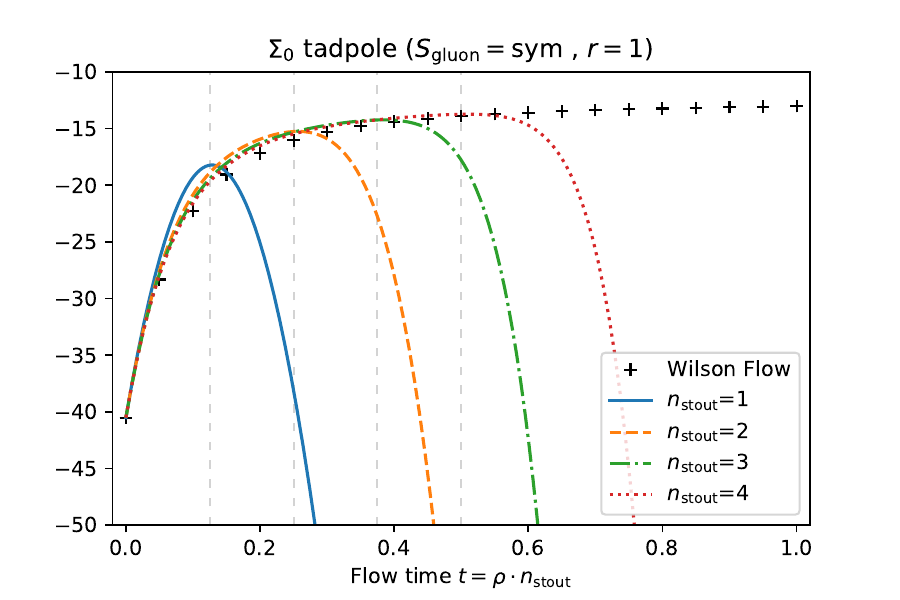}
\includegraphics[scale=0.55]{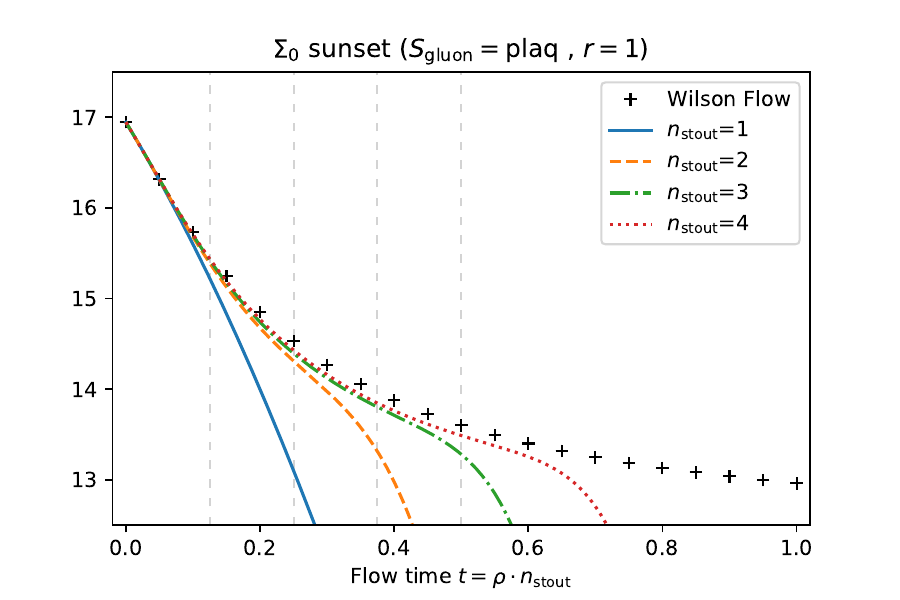}
\includegraphics[scale=0.55]{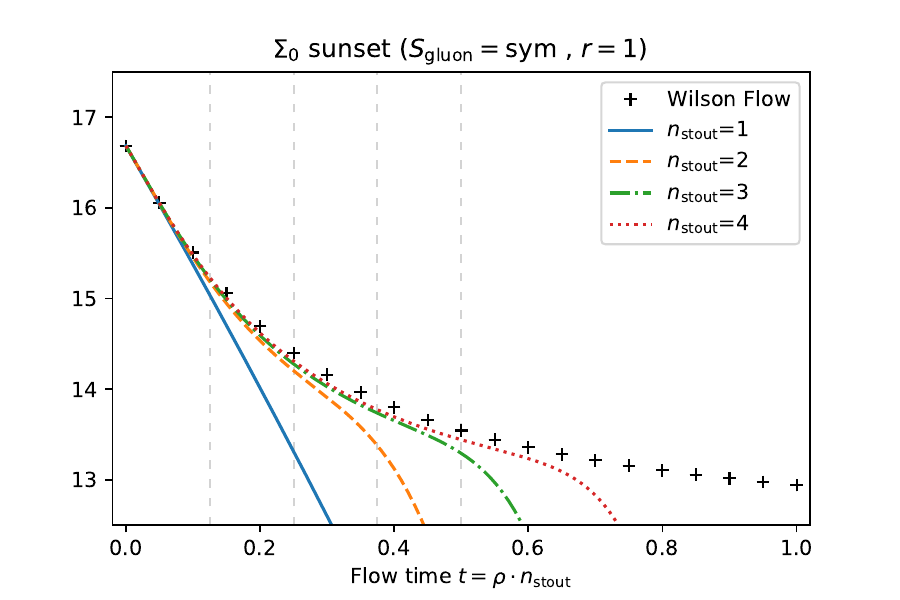}
\includegraphics[scale=0.55]{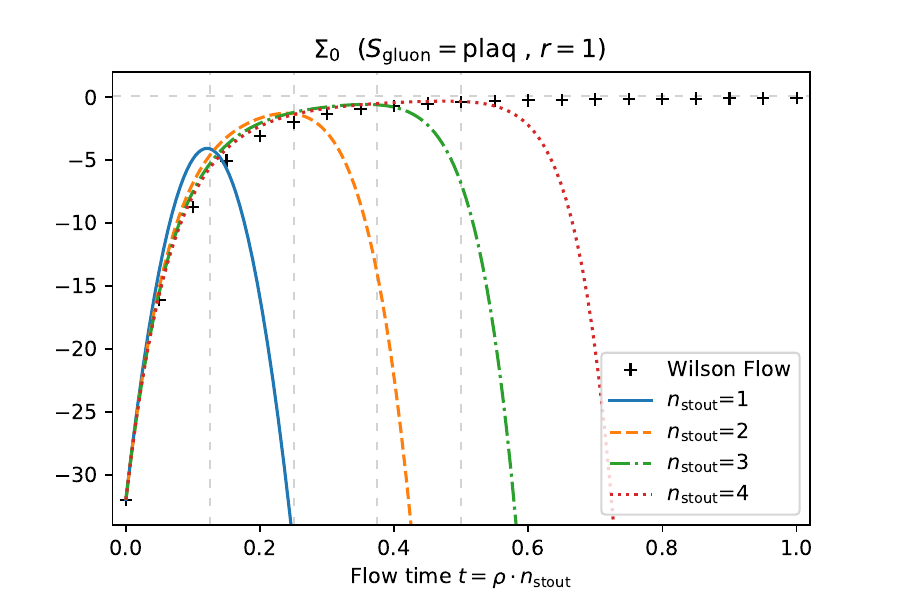}
\includegraphics[scale=0.55]{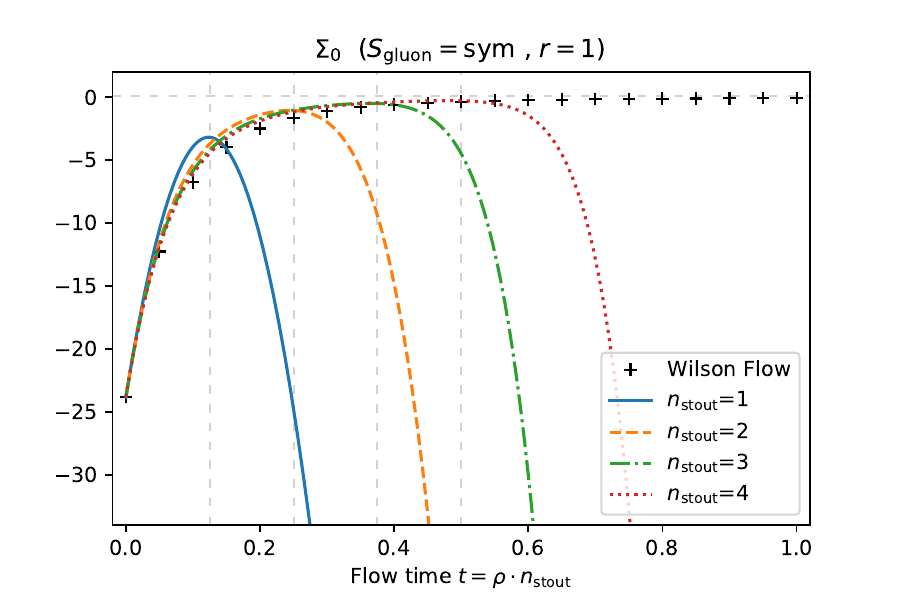}
\caption{Same information as in Fig.~\ref{fig:S0_stout}, but plotted versus the effective flow time $t = \varrho {\cdot} n_\mathrm{stout}$. The vertical dashed lines indicate integer multiples of t=0.125 (in lattice units). The curve of the Wilson
flow (to which these results converge under $n_\mathrm{stout}\to\infty$) is
indicated with black crosses.
\label{fig:S0_flow_and_stout_01}}
\end{figure}
\begin{figure}[!h]
\centering
\includegraphics[scale=0.55]{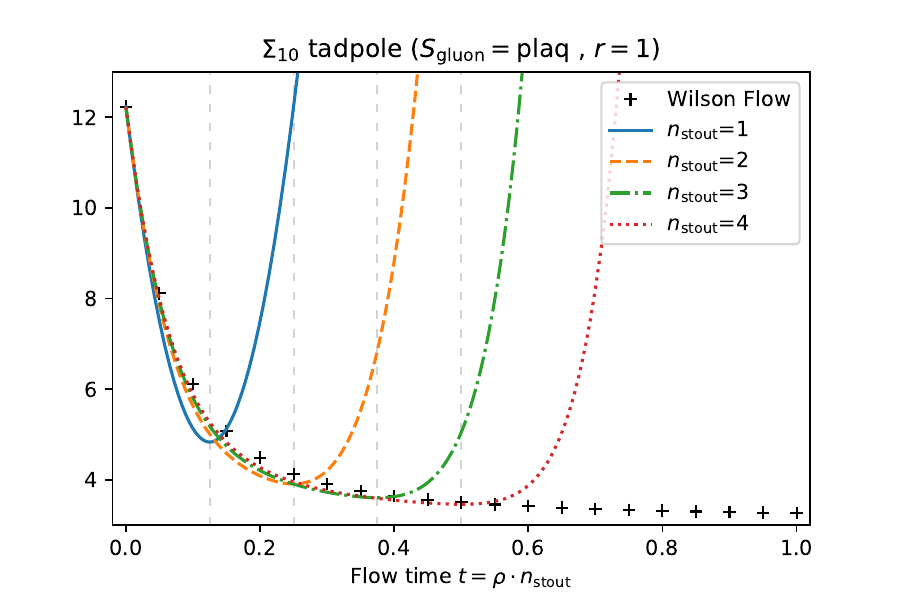}
\includegraphics[scale=0.55]{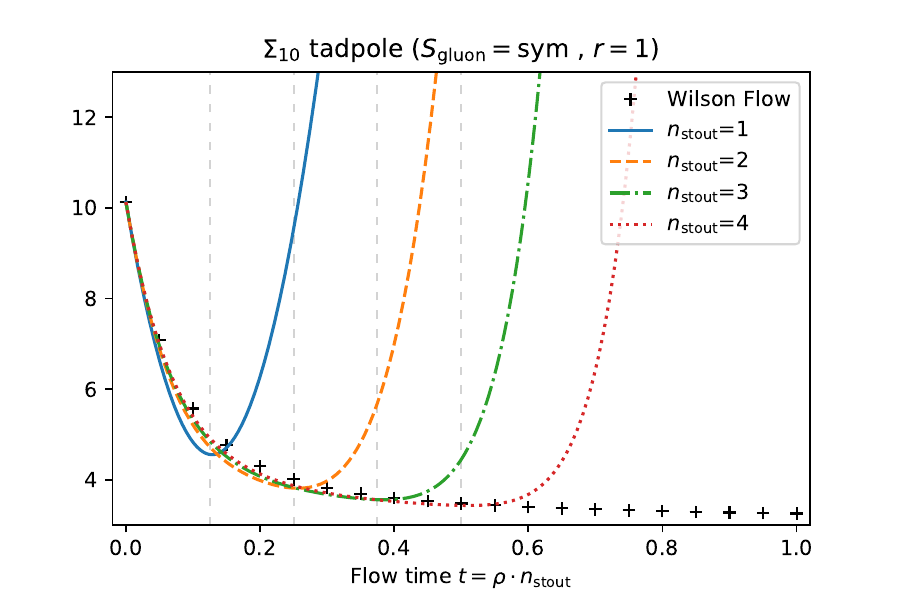}
\includegraphics[scale=0.55]{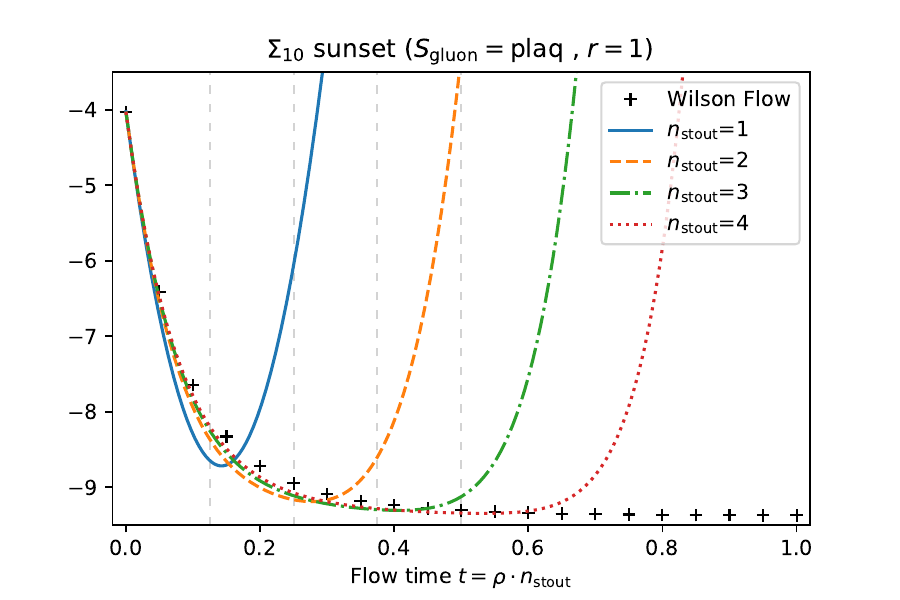}
\includegraphics[scale=0.55]{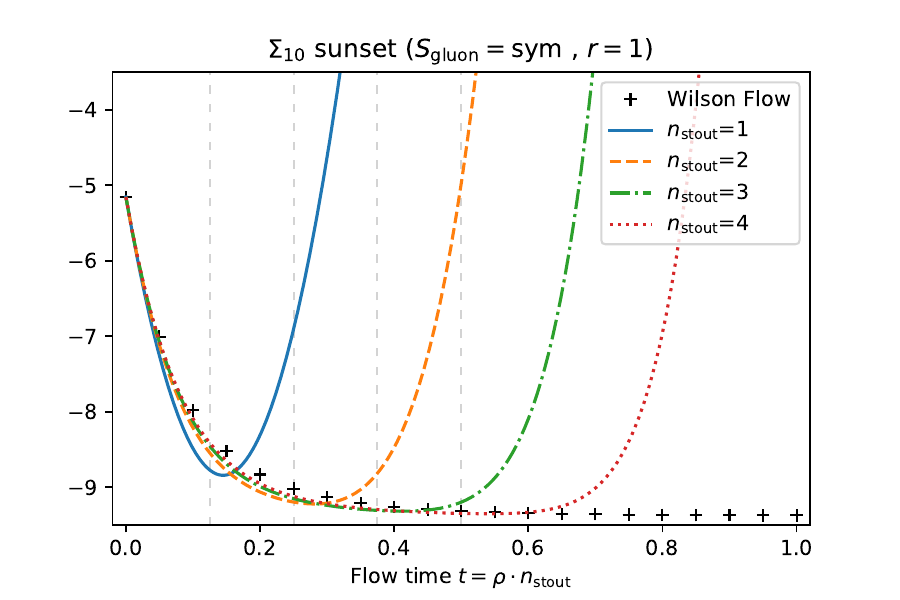}
\includegraphics[scale=0.55]{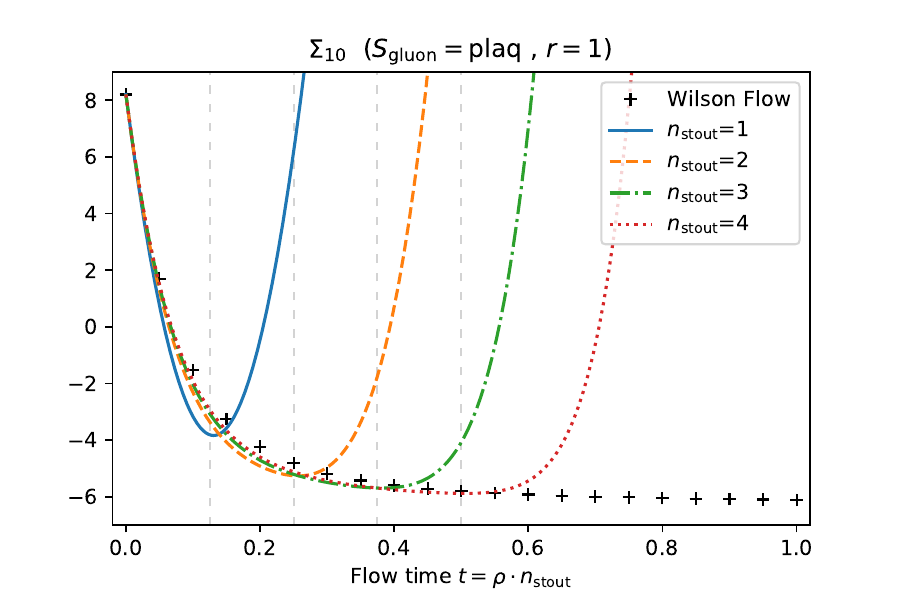}
\includegraphics[scale=0.55]{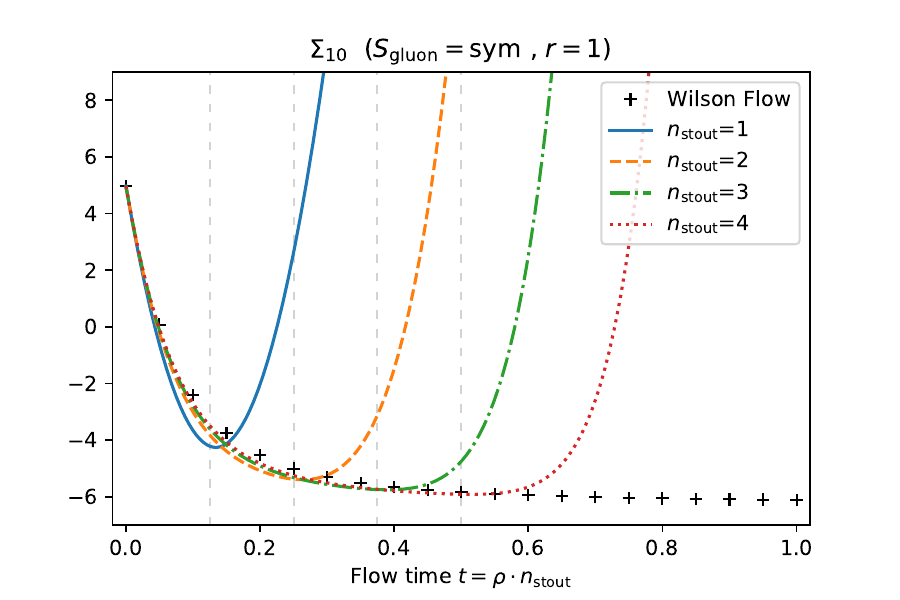}
\caption{Same as Fig.~\ref{fig:S0_flow_and_stout_01}, but for the finite piece $\Sigma_{10}$; see
Eq.~(\ref{eq:Sigma_1}).
\label{fig:S1_flow_and_stout_01}}
\end{figure}
It is interesting to plot the results for iterated stout smearing versus $n_\mathrm{stout}{\cdot}\varrho$ rather
than versus $\varrho$; this is done in Figs.~\ref{fig:S0_flow_and_stout_01} and \ref{fig:S1_flow_and_stout_01}.
With a bit of intuition one can see that for any fixed value of $n_\mathrm{stout}{\cdot\varrho}$ these results
converge to a universal curve, which we indicate through black crosses.
Given our results in Sec.~3, this should not come as a surprise.
The black curves in Figs.~\ref{fig:S0_flow_and_stout_01} and \ref{fig:S1_flow_and_stout_01} indicate the results for the Wilson variety of the gradient flow.
Evidently, what we see is a graphical illustration of the correspondence
\footnote{Corrections to (\ref{eq:stout_gradflow_equivalence}) start at order $t^2/(a^4n_\mathrm{stout})$, see Eqs.~(\ref{eq:flow_time_correction_1}),  (\ref{eq:flow_time_correction_2}), and (\ref{eq:flow_time_correction_3}).
Hence, the left-hand side of (\ref{eq:stout_gradflow_equivalence}) is to be dressed
with a factor $(1+\mathcal{O}(t/(a^2 n_\mathrm{stout})))$.}
\begin{equation}
t/a^2 = n_\mathrm{stout}\cdot\varrho
\label{eq:stout_gradflow_equivalence}
\end{equation}
that links the dimensionless flow time to the cumulative parameter of the iterated stout recipe.
Note that this identification becomes exact under $n_\mathrm{stout}\to\infty$ and $\varrho\to 0$ with the
product $n_\mathrm{stout}\cdot\varrho$ held constant.
This matching has been discussed in
Refs.~\cite{Alexandrou:2017hqw,Nagatsuka:2023jos}, too.

The thin vertical lines in Figs.~\ref{fig:S0_flow_and_stout_01} and \ref{fig:S1_flow_and_stout_01}  indicate integer multiples of 0.125.
They are meant to separate each curve into a part where it gives a reasonable approximation to
the gradient flow (to the left) and a part where it does not (to the right).
For instance the mark at $3\cdot0.125=0.375$ separates the dash-dotted line ($n_\mathrm{stout}=3$)
into an ascending part (where it approximates the crosses quite well), and a descending part
(where it does not).
In the terminology of numerical mathematics one would say that multiple stout smearings implement
a simple integration scheme (with an integration error in the flow time $t/a^2$ which scales like
$1/n_\mathrm{stout}$), and the caveat of Ref.~\cite{Capitani:2006ni}  is meant to limit the integration error.

\begin{figure}[!h]
\centering
\includegraphics[scale=0.55]{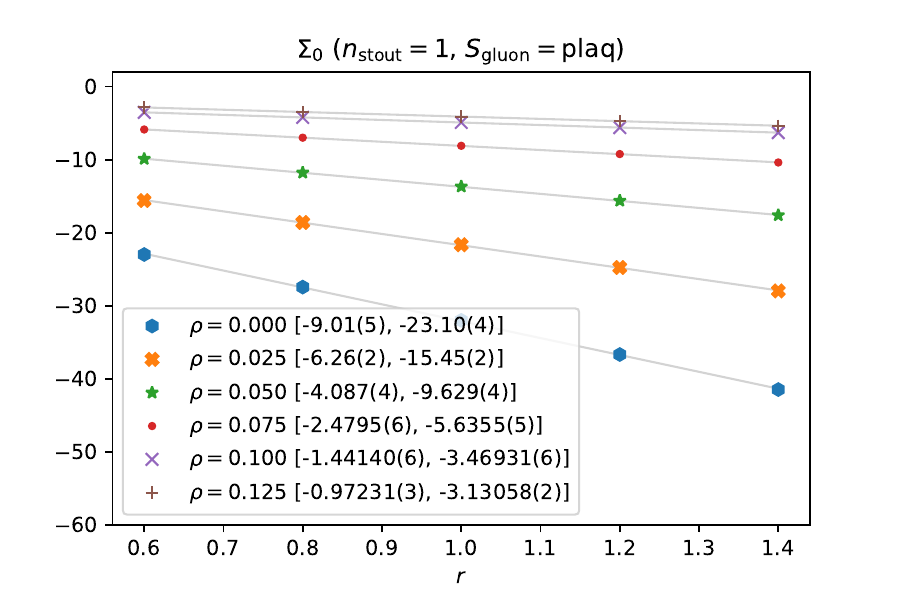}
\includegraphics[scale=0.55]{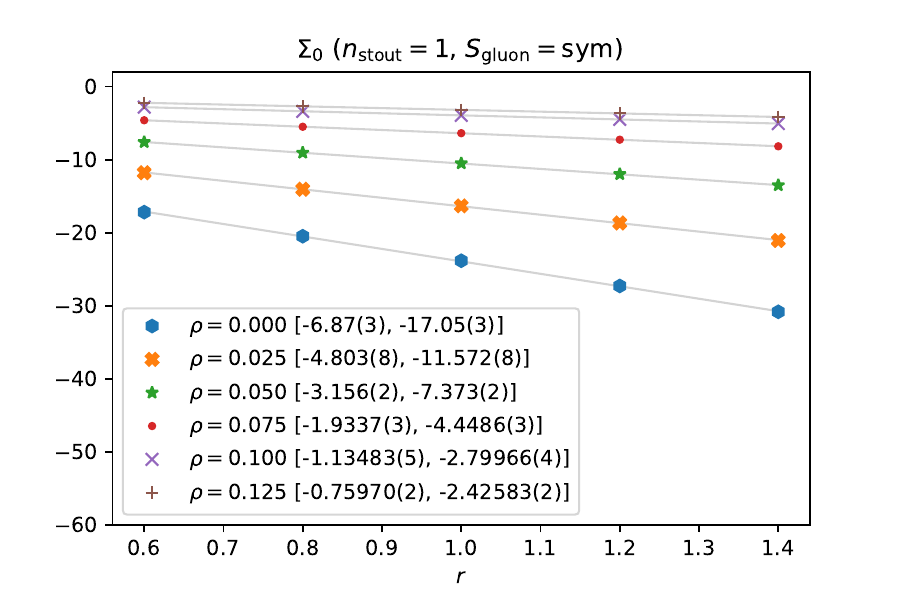}
\includegraphics[scale=0.55]{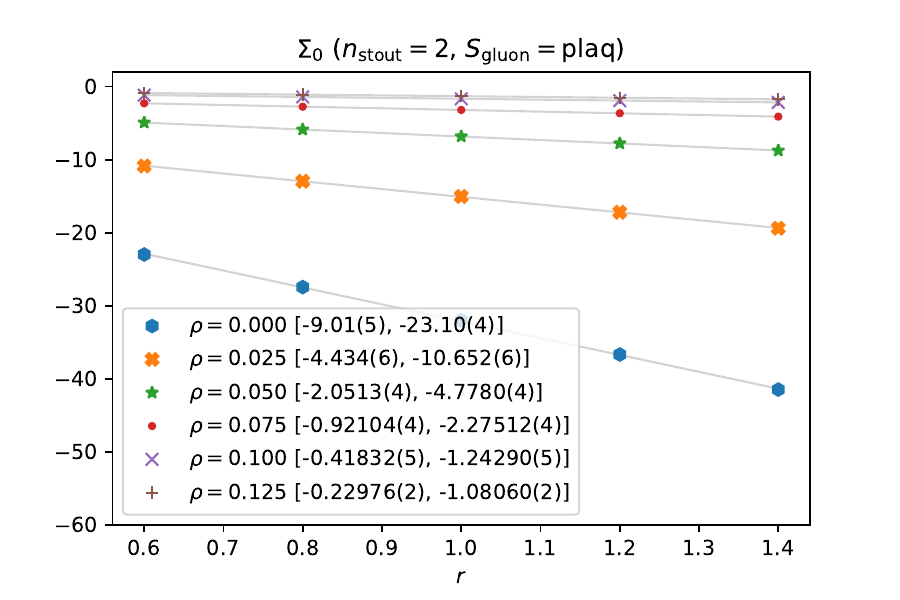}
\includegraphics[scale=0.55]{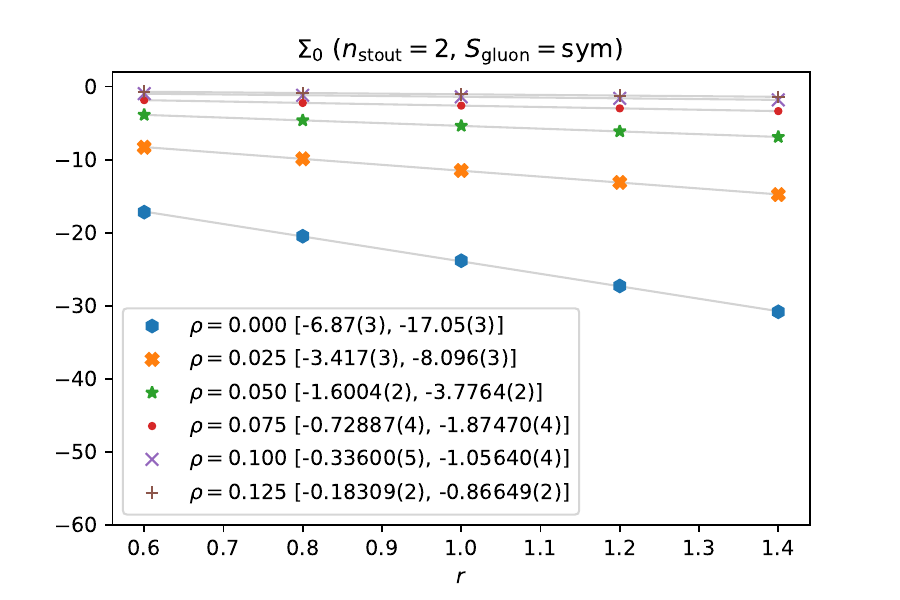}
\includegraphics[scale=0.55]{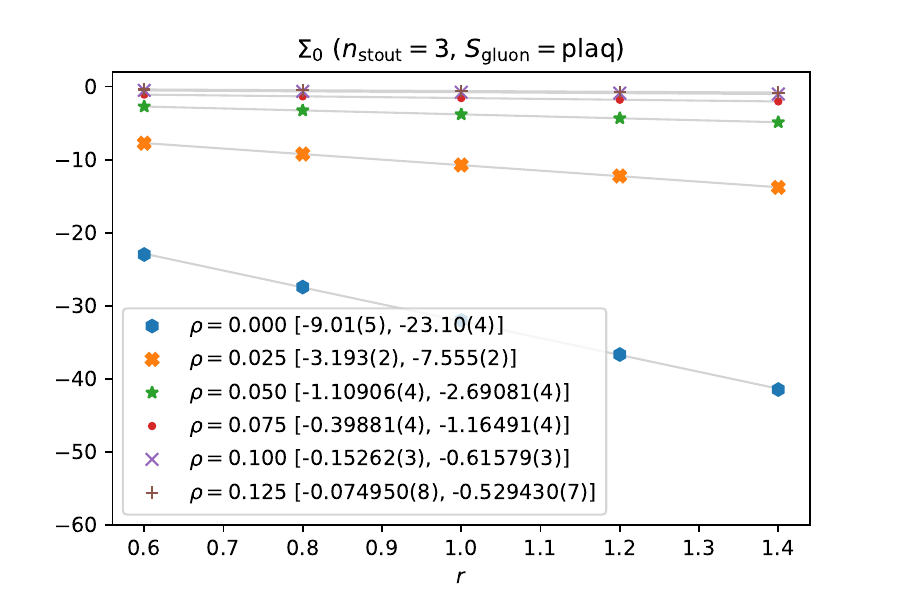}
\includegraphics[scale=0.55]{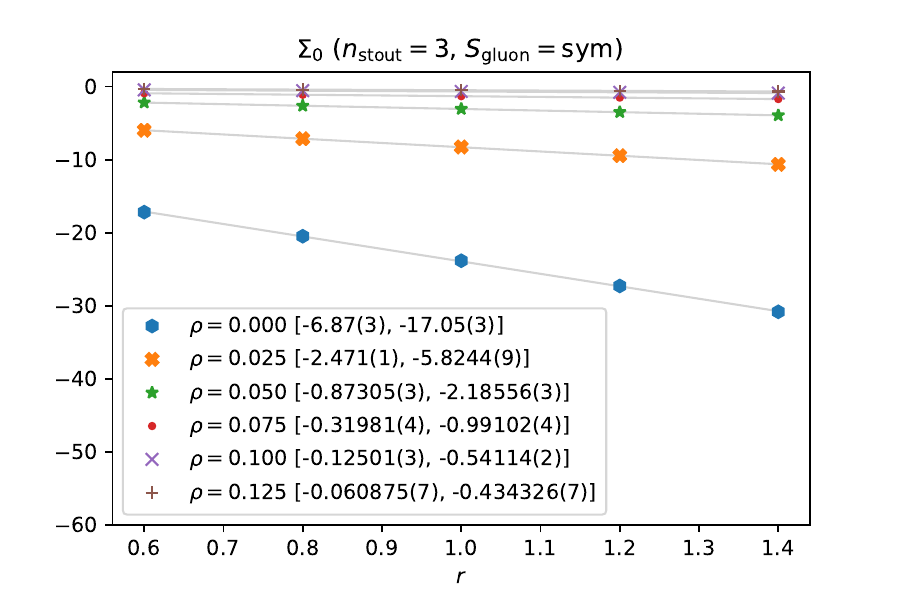}
\includegraphics[scale=0.55]{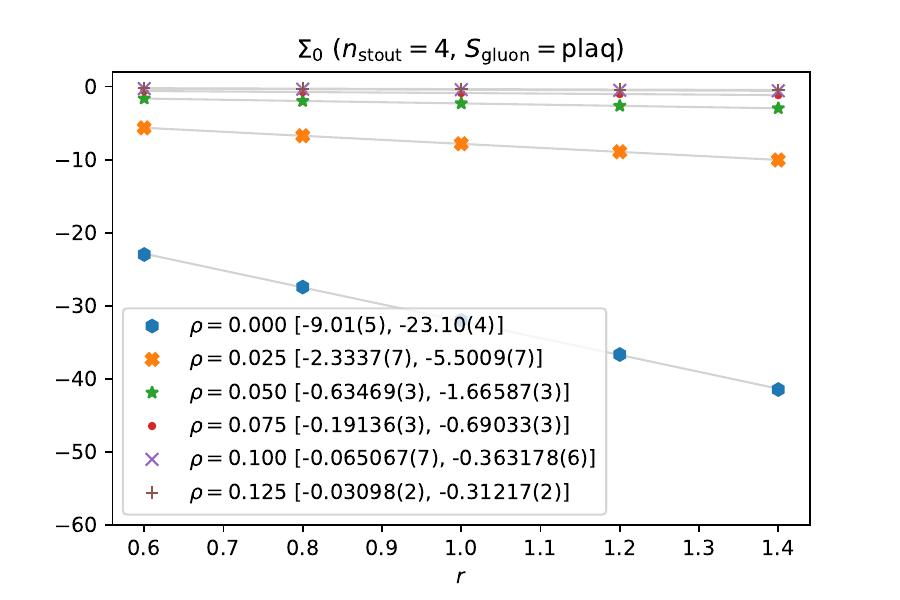}
\includegraphics[scale=0.55]{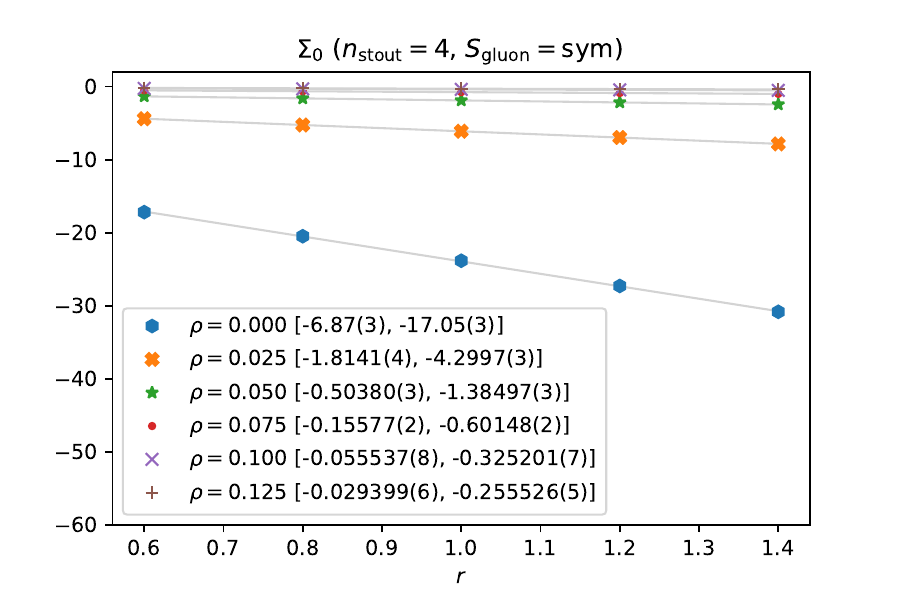}
\caption{The divergent piece $\Sigma_0$ of the clover fermion ($\csw=r$)
self-energy as a function of $r$, with 1 to 4 stout steps (top to bottom).
The background uses plaquette glue (left) or tree-level Symanzik improved
glue (right). The legends give the coefficients $[c_0,c_1]$ of linear least-squares fits of the form $c_0+c_1\cdot r$. The entry for $\varrho=0$ is the unsmeared case.
\label{fig:S0_of_r_stout}}
\end{figure}
\begin{figure}[!h]
\centering
\includegraphics[scale=0.55]{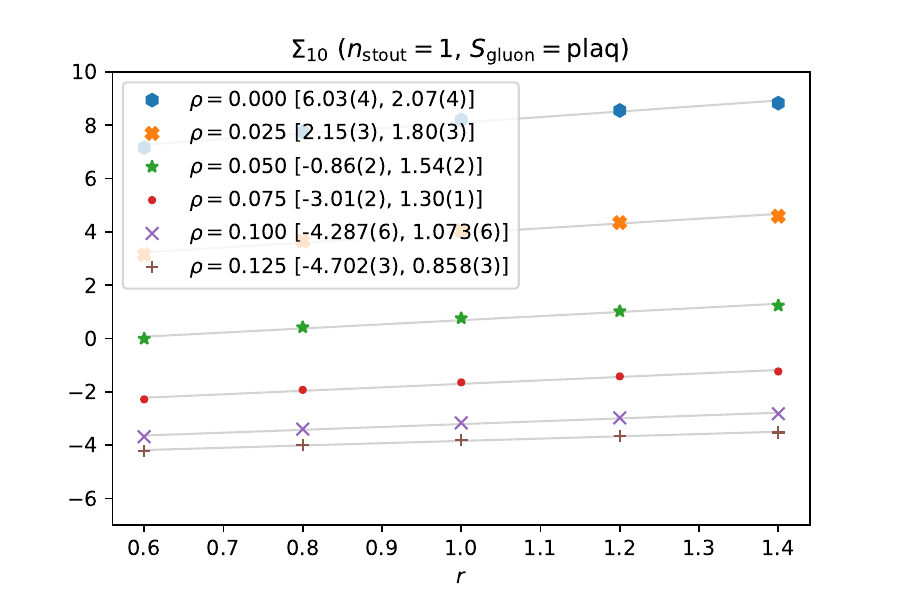}
\includegraphics[scale=0.55]{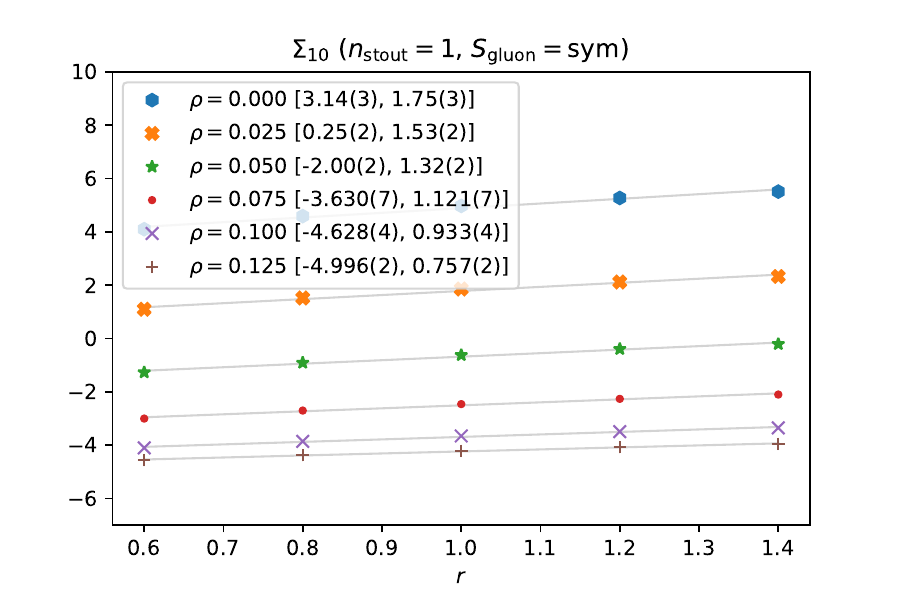}
\includegraphics[scale=0.55]{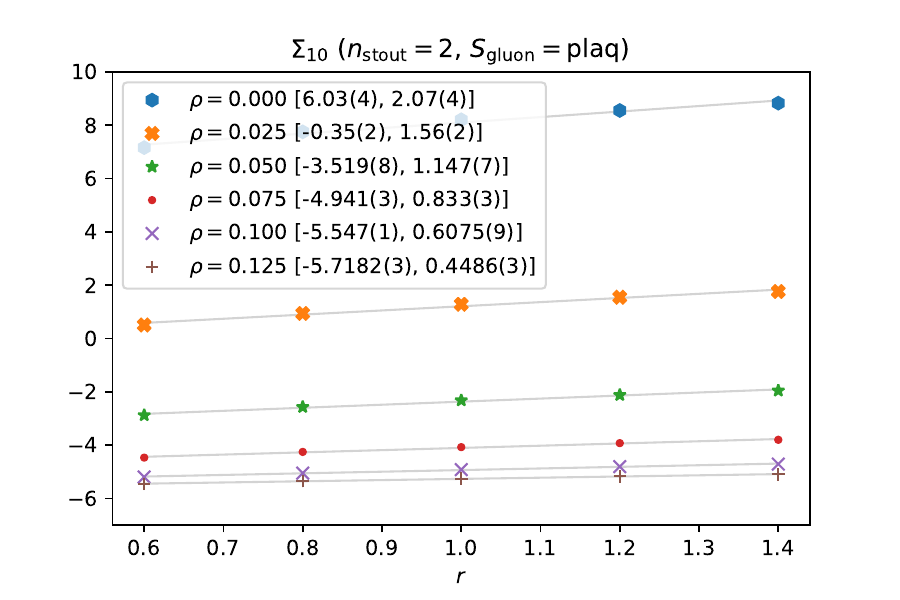}
\includegraphics[scale=0.55]{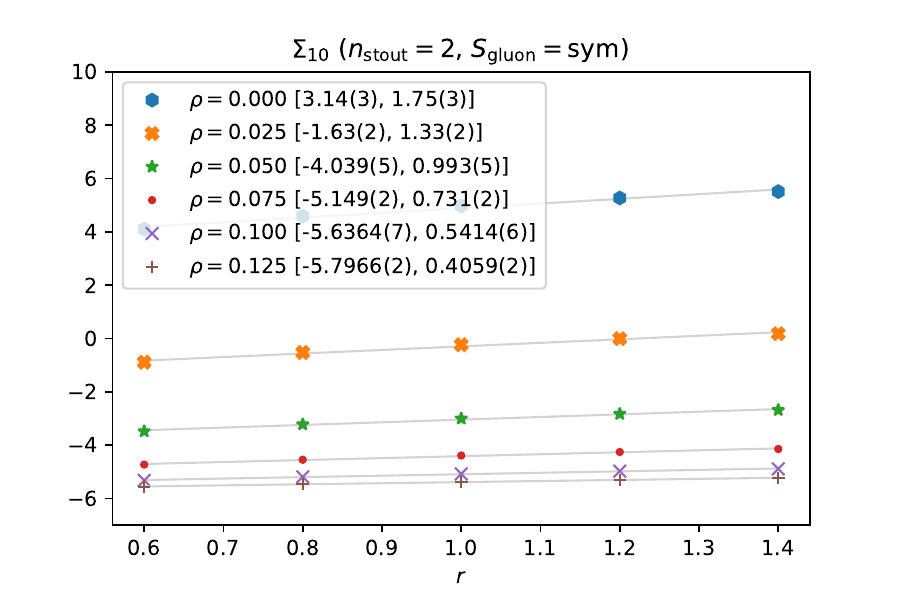}
\includegraphics[scale=0.55]{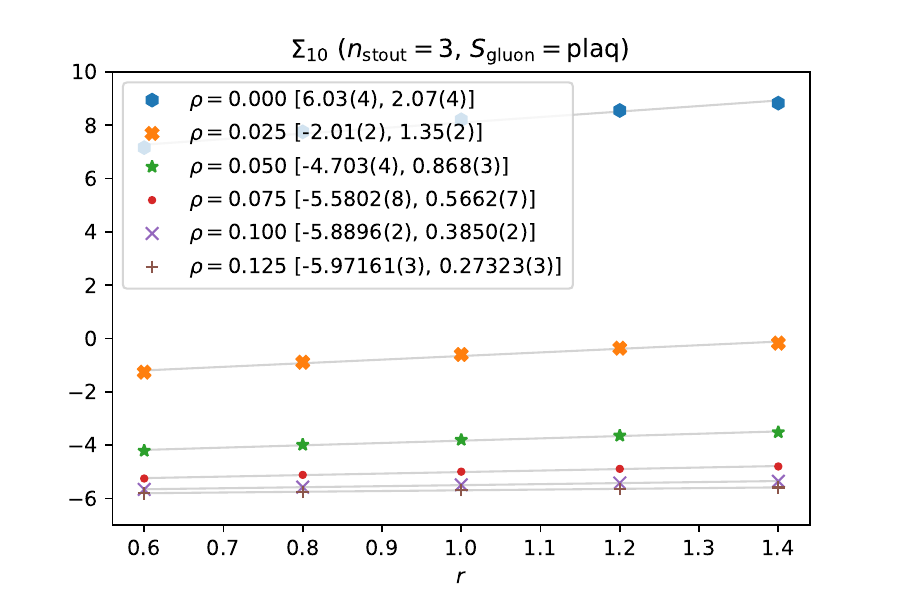}
\includegraphics[scale=0.55]{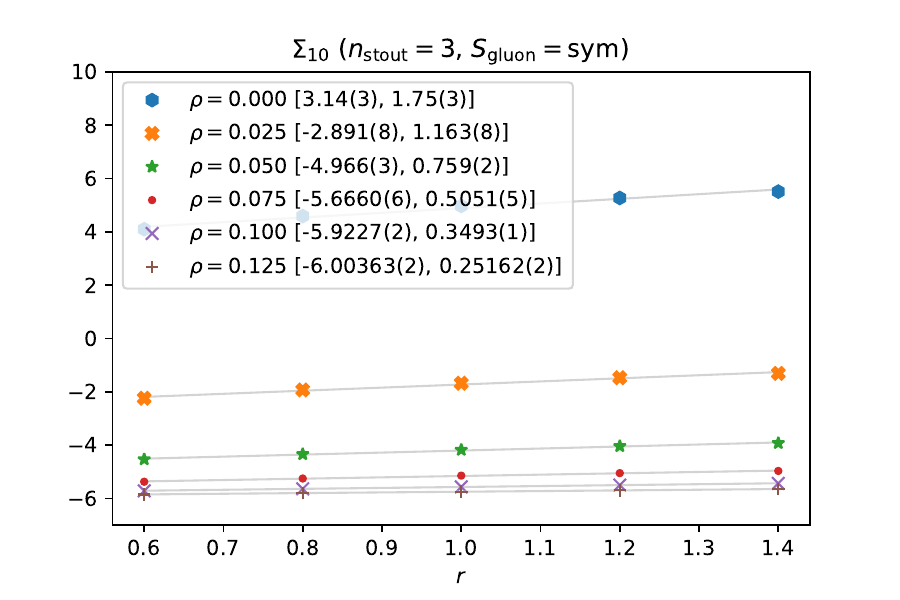}
\includegraphics[scale=0.55]{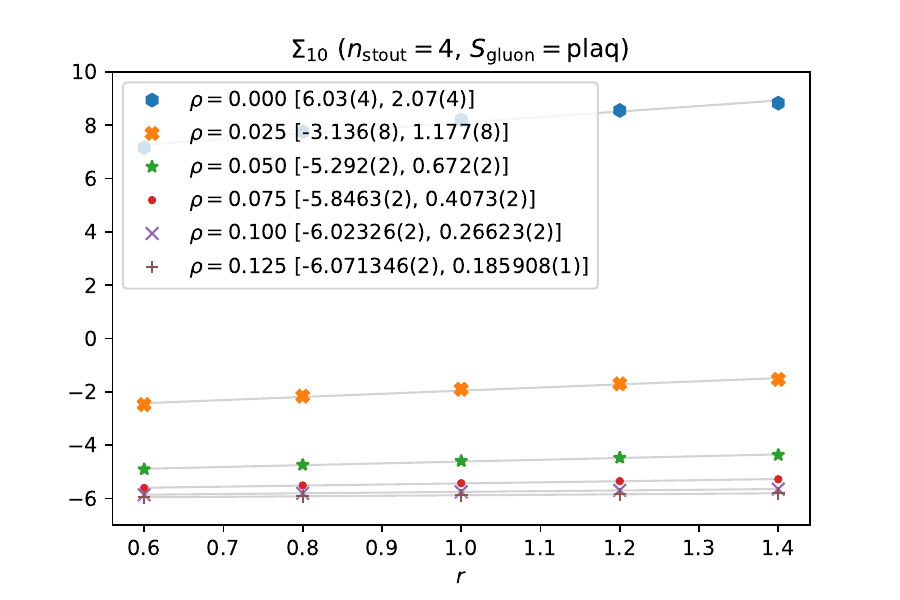}
\includegraphics[scale=0.55]{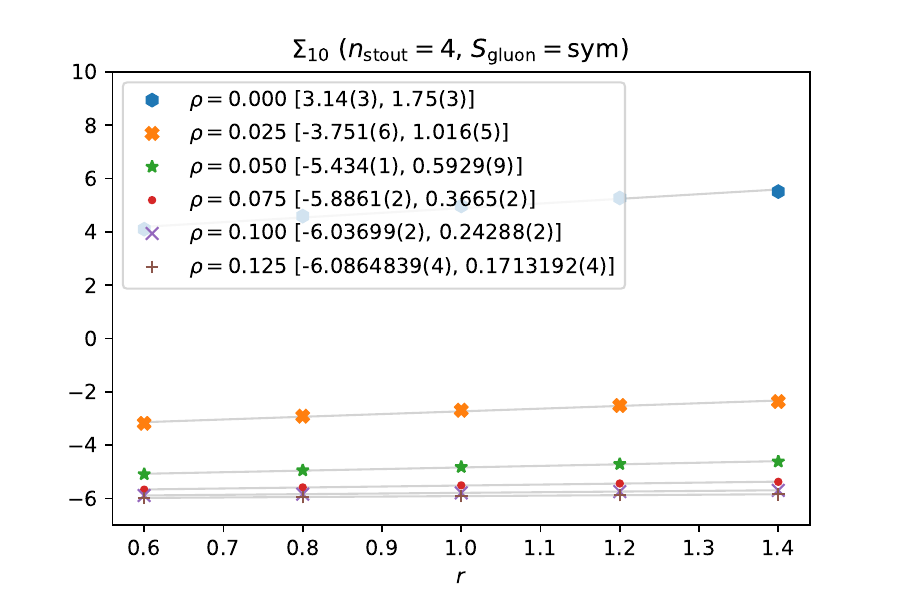}
\caption{Same as Fig.~\ref{fig:S0_of_r_stout}, but for the finite piece $\Sigma_{10}$; see Eq.~(\ref{eq:Sigma_1}).
\label{fig:S1_of_r_stout}}
\end{figure}
Let us consider the behavior of the self-energy as a function of the Wilson parameter $r$ (with
$\csw=r$) as shown in Figs.~\ref{fig:S0_of_r_stout} and \ref{fig:S1_of_r_stout}. Even though the self-energy at one-loop order is not a
linear function in $r$ analytically (since $r$ appears in the denominator of the fermion propagator),
$\Sigma_0$ and $\Sigma_{10}$ seem almost linear over the plotted range.
The former quantity decreases and the latter one increases with increasing $r$.
For $\varrho$ approaching 0.125 and/or more smearing steps the behavior becomes gradually flatter.
In practical terms this means that the Wilson parameter $r$ is not a useful knob to engineer the
self-energy of tree-level clover fermions.\\

\begin{figure}[!h]
\centering
\includegraphics[scale=0.55]{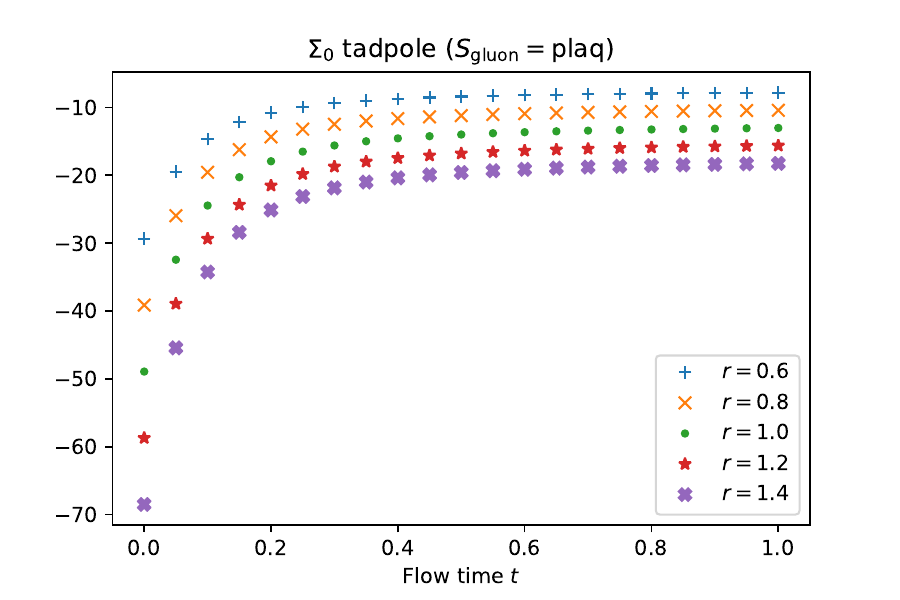}
\includegraphics[scale=0.55]{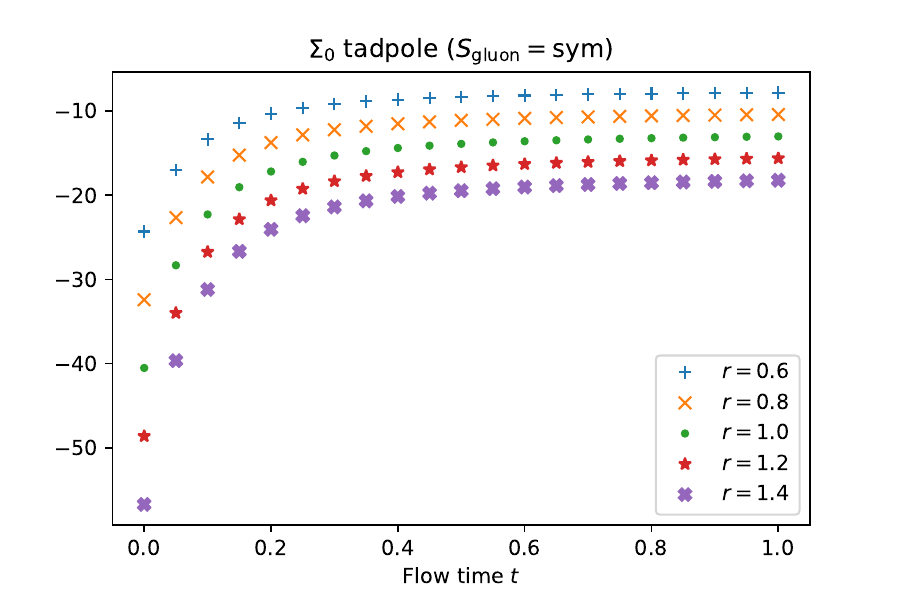}
\includegraphics[scale=0.55]{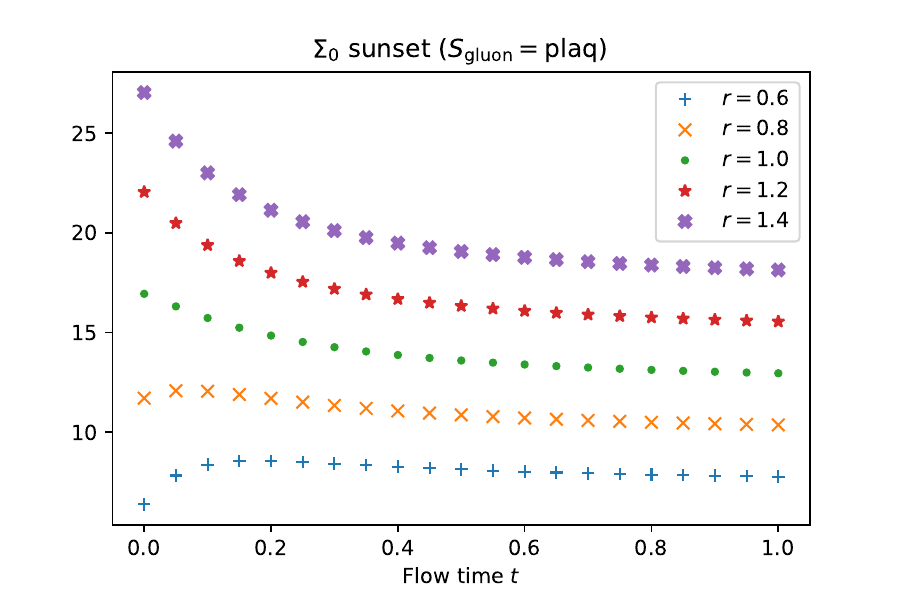}
\includegraphics[scale=0.55]{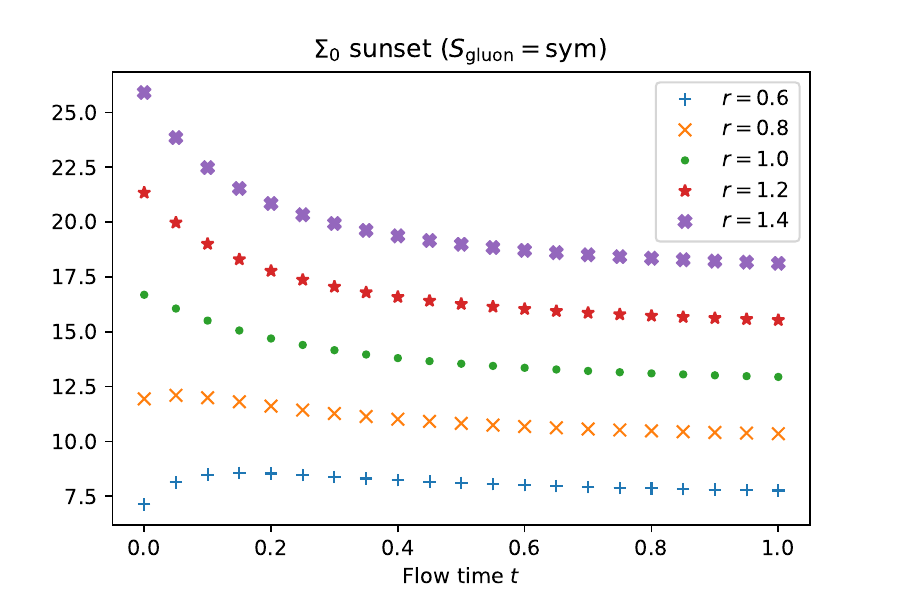}
\includegraphics[scale=0.55]{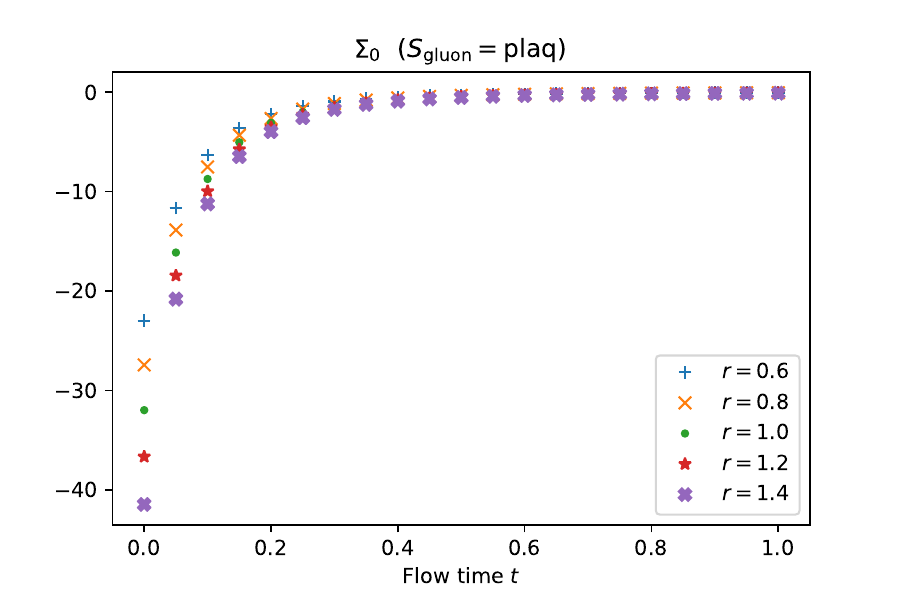}
\includegraphics[scale=0.55]{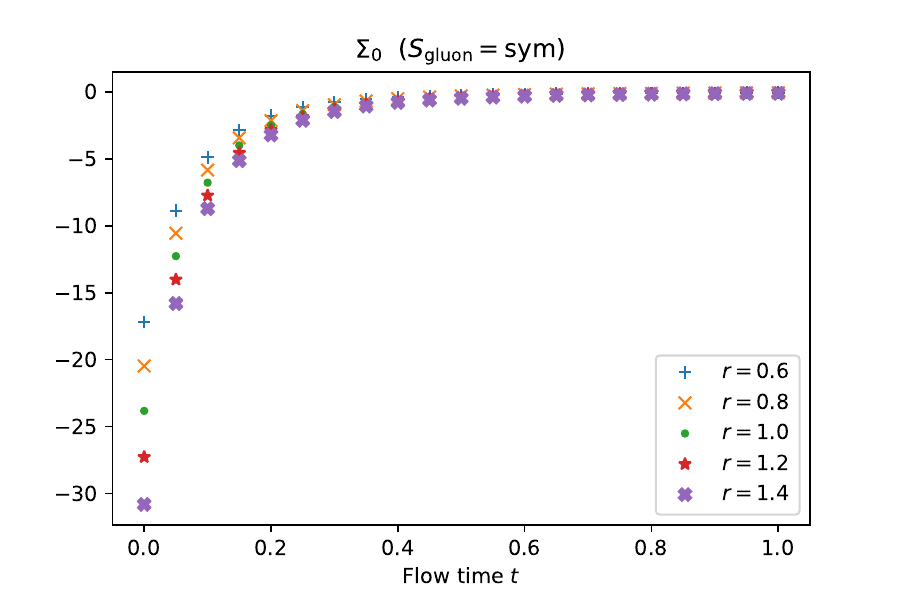}
\caption{Contributions to the divergent piece $\Sigma_0$ of the clover
($\csw=r=1$) fermion self-energy from the tadpole (top) and sunset
(middle) diagrams and their sum (bottom). Results for several values of $r$
are given as a function of the gradient-flow time $t$ in lattice units.
\label{fig:S0_flow_r}}
\end{figure}
\begin{figure}[!h]
\centering
\includegraphics[scale=0.55]{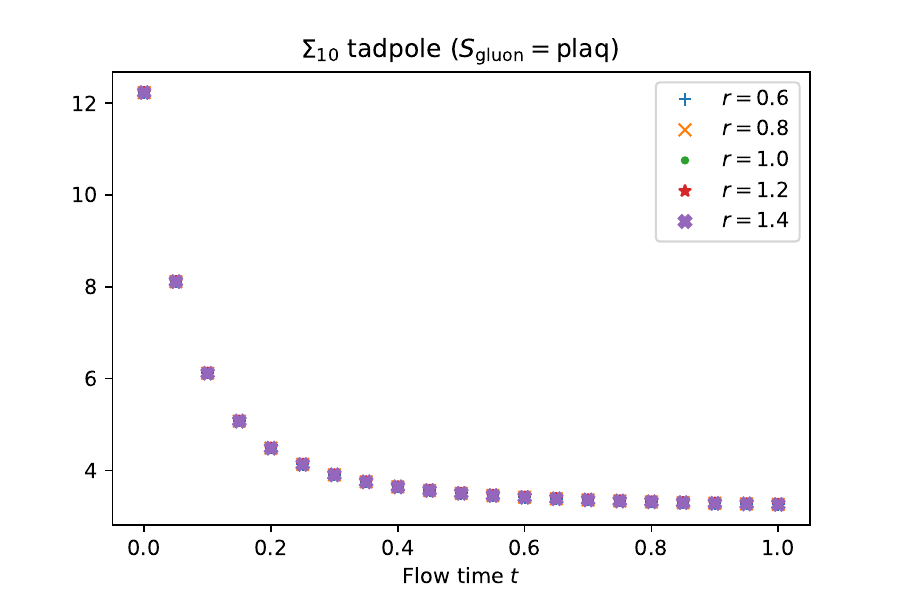}
\includegraphics[scale=0.55]{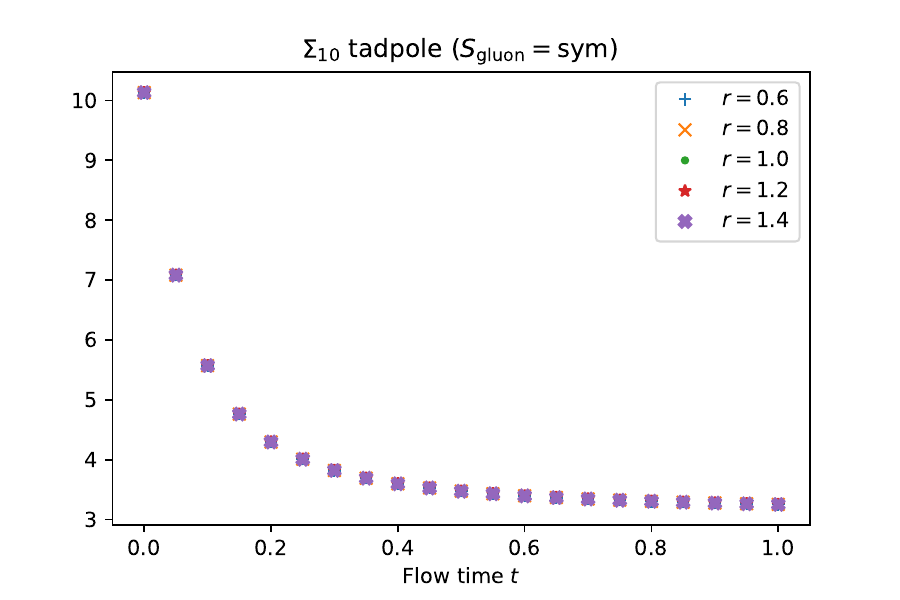}
\includegraphics[scale=0.55]{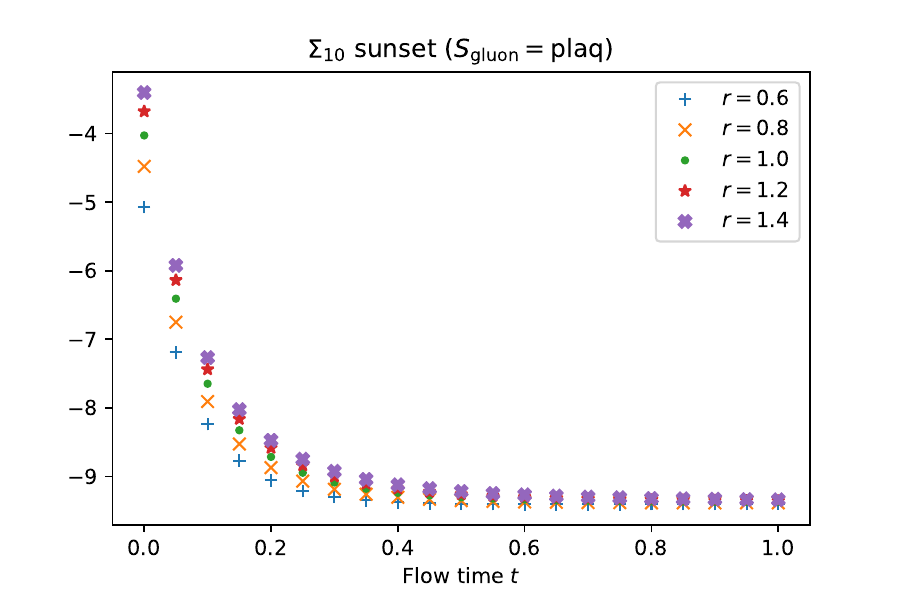}
\includegraphics[scale=0.55]{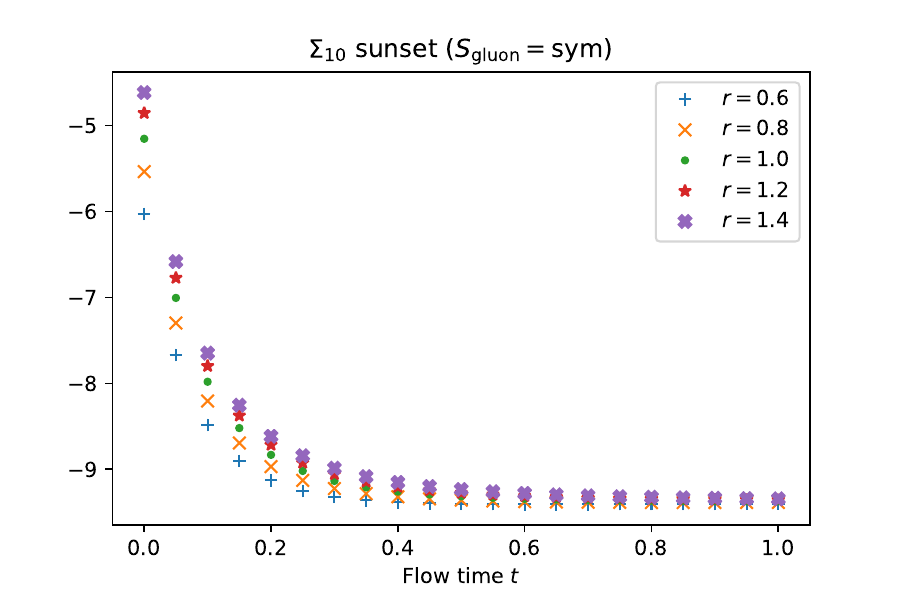}
\includegraphics[scale=0.55]{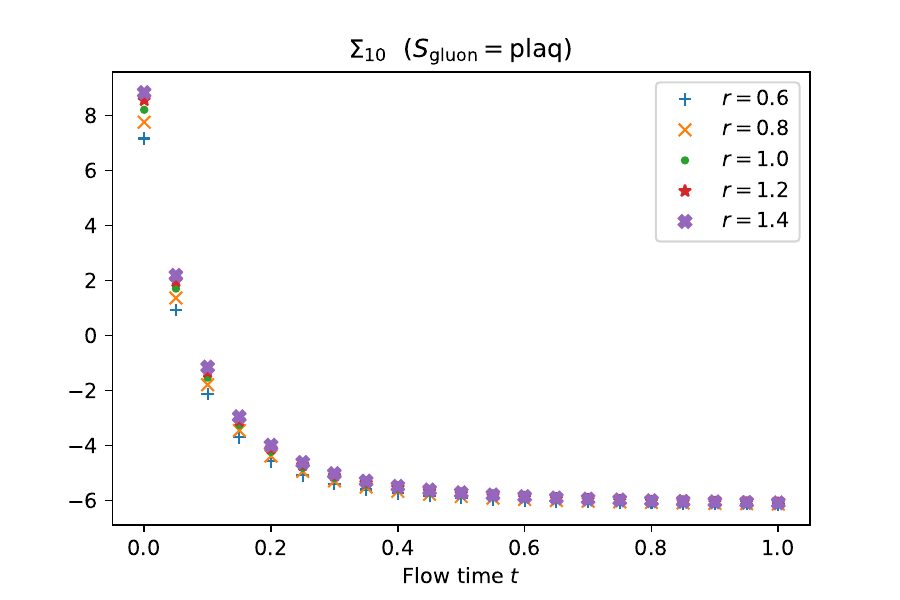}
\includegraphics[scale=0.55]{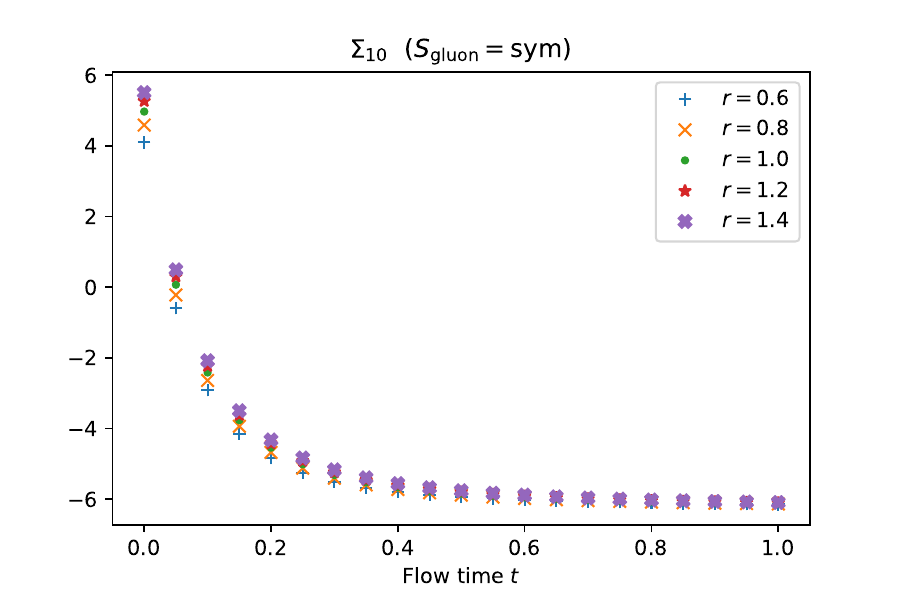}
\caption{Same as Fig.~\ref{fig:S0_flow_r}, but for the finite piece $\Sigma_{10}$; see
Eq.~(\ref{eq:Sigma_1}).
\label{fig:S1_flow_r}}
\end{figure}
In Figs.~\ref{fig:S0_flow_r} and \ref{fig:S1_flow_r}  we finally shift to results with the gradient flow recipe.
For $r=1$ (green dots) the ``tadpole'' and the ``sunset'' contribution seem to tend to unsuspicious
finite values, as $t/a^2$ is taken large, and the physical sum tends to zero.
Varying $r$ (and synchronously $c_\mathrm{SW}=r$) seems to affect only the splitting between the
``tadpole'' and the ``sunset'' contributions, but barely the final (physical) result.
And the former observation is only pronounced for $\Sigma_0$; for $\Sigma_{10}$ it is rather mild, here  the ``tadpole'' contribution does not depend on $r$ at all.
\begin{figure}[!h]
\centering
\includegraphics[scale=0.5]{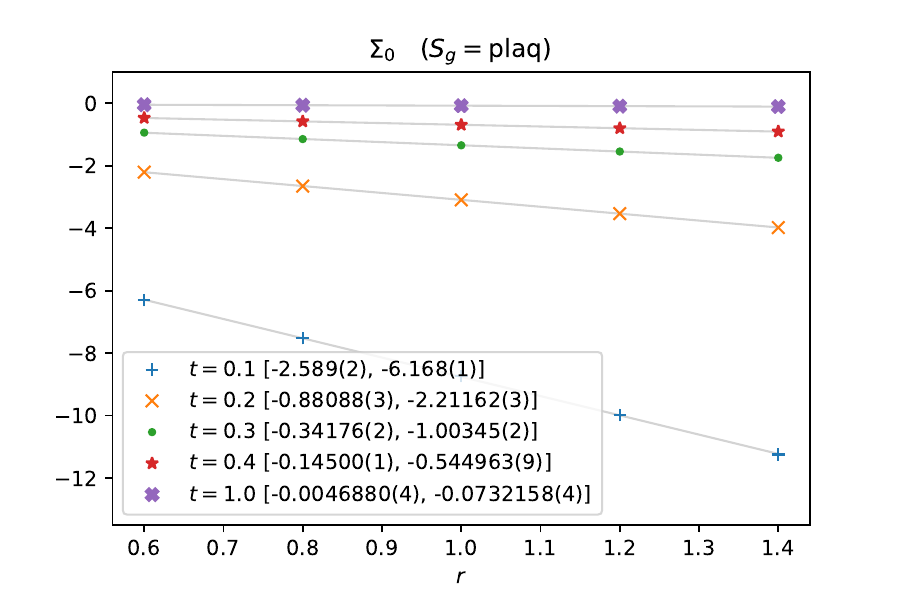}
\includegraphics[scale=0.5]{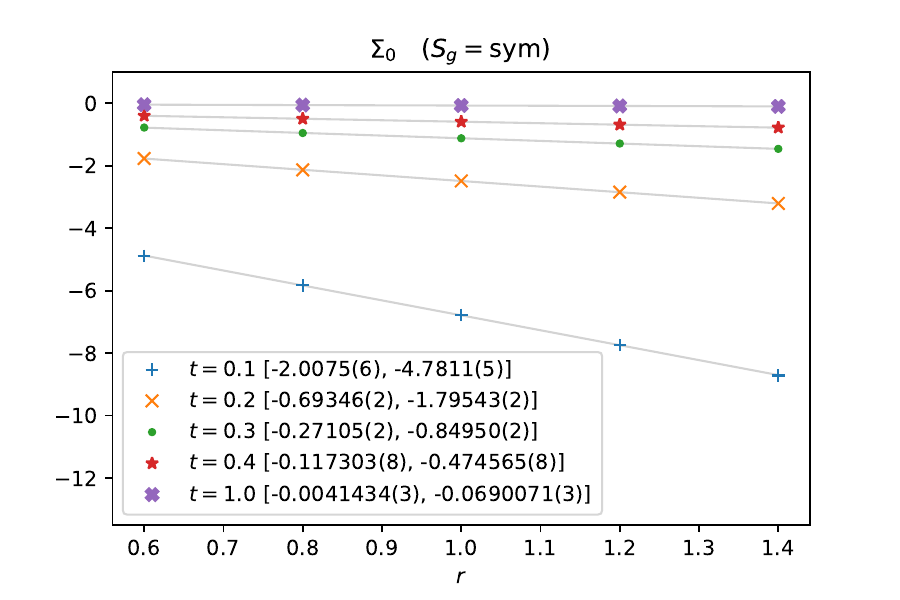}
\includegraphics[scale=0.5]{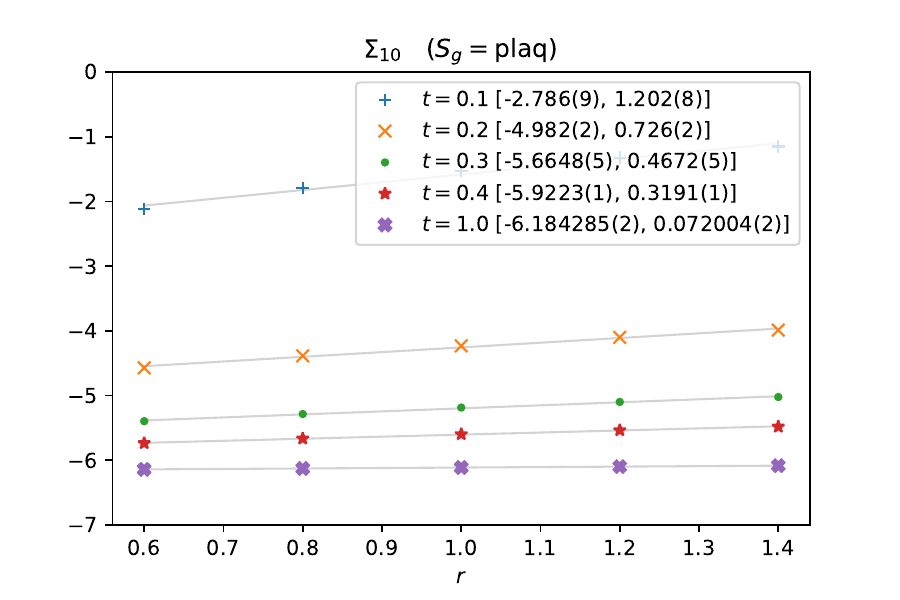}
\includegraphics[scale=0.5]{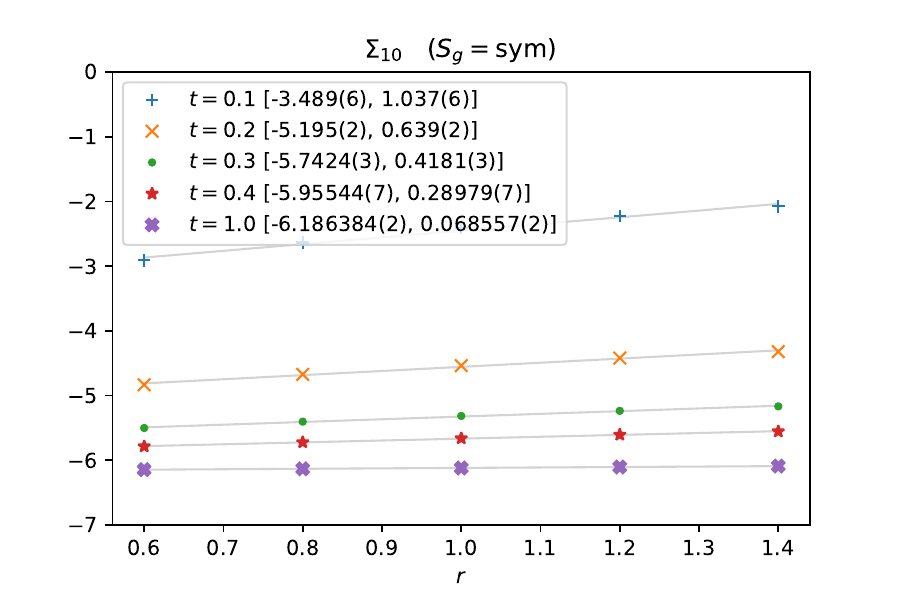}
\caption{The dependence on the Wilson paramter $r$ of $\Sigma_0$ (top) and $\Sigma_{10}$ (bottom) at different flow times $t$ of the Wilson flow. The coefficients $[c_0,c_1]$ of linear least-square fits of the form $c_0+c_1\cdot r$ are given in brackets.\label{fig:flow_r}}
\end{figure}
For completeness we show in Fig.~\ref{fig:flow_r} how $\Sigma_0$ and $\Sigma_{10}$ depend on $r$ for a few
Wilson flow times $t/a^2$.
Any dependence that is still present at tiny flow times ($t/a^2\simeq0.1$) is found
to quickly disappear for moderate flow times ($t/a^2\simeq1$).\\

\newpage

In our view, the takeaway message from this section is that there is a perturbative backing
of the well-known benefit that comes from multiple stout smearing. One step reduces the absolute value of
$\Sigma_0$ by about 85\%, two steps by 95\%, three by 98\% and four steps by 99\%.

\section{Conclusions}

In this paper we have worked out a systematic approach for deriving Feynman rules for
lattice fermion actions with smeared gauge links (to be used interchangeably in the covariant
derivative, in the species-separating covariant Laplacian and/or a possible clover term).

The basis for such calculations is provided by a perturbative expansion of the stout
smearing recipe of Ref.~\cite{Morningstar:2003gk} and the Wilson variety of the gradient
flow introduced in Refs.~\cite{Luscher:2010iy,Luscher:2011bx,Luscher:2013cpa};
these expansions have been presented in Sec.~2 and Sec.~3, respectively.

With these results in hand, one may derive the Feynman rules for a fermion action
with an arbitrary parameter combination $(\varrho,n_\mathrm{stout})$ in case of stout
smearing or an arbitrary flow time $t/a^2\in\mathbb{R}$ in case of the gradient flow.
In Sec.~4 we argue that the ``brute force'' approach, i.e.\ inserting the perturbative
expansion of the smeared/flowed variable into the underlying action, is not
recommendable, since it leads to excessively long intermediate expressions.
We present an approach based on the $SU(N_c)$ expansion of the original (``unsmeared'')
links $U_\mu(x)$ to the required order in $g_0$ (typically $g_0^3$ for one-loop
calculations), which keeps intermediate expressions much shorter.
This holds true for both stout smearing and the gradient flow, and confirms, as
a by-product, the perturbative matching (\ref{eq:stout_gradflow_equivalence}) between these recipes.

To give an immediate application of our recipe, we derive in Sec.~4 the Feynman rules
of a Wilson fermion with clover improvement for arbitrary Wilson parameter $r$ and
Sheikholeslami-Wohlert parameter $c_\mathrm{SW}$.
With these rules in hand, we calculate in Sec.~5 the self-energy of the clover fermion
for $c_\mathrm{SW}=r$, for variable $\varrho$ and $1 \leq n_\mathrm{stout} \leq 4$ in case
of stout smearing and the range $0 \leq t/a^2\leq 1$ in case of gradient 
flowing.

The cross-check with Ref.~\cite{Capitani:2006ni}, where numerical results obtained in
the ``brute force'' approach were presented, proves particularly enlightening.
Of course, reproducing these results felt like a relief, but the difference is in the details.
In Ref.~\cite{Capitani:2006ni} results for $\Sigma_0$ were given for a single stout parameter
($\varrho=0.1$ in our terminology), restricted to $n_\mathrm{stout}\leq 3$ due to the numerical burden.
In our approach we managed to go up to $n_\mathrm{stout}=4$ and to
evaluate the generic result for arbitrary $\varrho$ (using rather modest
computational resources).
This illustrates that the more elegant approach implies not just aesthetic pleasures but actual savings compared to the ``brute force'' approach.
Still, the bottom line is an affirmation of the finding of Ref.~\cite{Capitani:2006ni}
that combining tree-level clover improvement with stout smearing or gradient flowing with a
cumulative flow time of the order $t/a^2\simeq1$ yields an undoubled fermion action with
surprisingly good chiral properties (see the discussion at the end of our Sec.~5).

A logical next step would be to work out the one-loop values of
$c_\mathrm{SW}$ for Wilson and/or Brillouin fermions with a limited number
of stout steps (hopefully again up to four iterations) and/or a few values
of the flow time $t/a^2$, for instance $t/a^2\in\{0.1,0.2,0.4,0.7,1.0\}$.
This amounts to combining the work we did in Ref.~\cite{Ammer:2023otl} with the
framework for smeared/flowed Feynman rules presented in this paper, and we hope to
present such results in the not so distant future.
In a further perspective, one might also work out the improvement factors $Z_{S,P,V,A}$
to one-loop (and perhaps even two-loop) order in stouted/flowed perturbation theory.

The alert reader might interject that a fully nonperturbative improvement strategy
is more desirable than a perturbative one, and the link smearing may imply only mild
technical complications in the nonperturbative approach.
However, as mentioned in the introduction and detailed in Ref.~\cite{Sommer:1997jg}, the
nonperturbative approach builds, in practical terms, on existing perturbative results
to one-loop (or even better: two-loop) order, as such results provide essential constraints
in the final step of summarizing the results through a rational function in $g_0^2$.
To us this indicates that perturbative work continues to be useful in lattice gauge theory.

\appendix
\newpage
\section{Clover fermion self-energy details}
In this appendix we give more details for the self-energy of a Wilson
fermion (with and without a clover term) in the presence of stout smearing
or gradient flow.

We start with the tree-level improved case ($\csw=r=1$). The pieces
$\Sigma_0$ and $\Sigma_{10}$, see Eqs. (\ref{eq:Sigma}) and (\ref{eq:Sigma_1}), for up to four
stout steps are listed in Tab.~\ref{tab:S0_S1_values} for a selection of $\varrho$ values. We refer
to Ref.~\cite{Capitani:2006ni} for an argument why choosing $\varrho$ larger
than 0.125 should be avoided. We believe that a spline interpolation
will suffice for practical purposes. In the event the reader requests
higher precision, Tabs.~\ref{tab:S0_poly_coeffs} and \ref{tab:S1_poly_coeffs} hold the necessary information to
reconstruct $\Sigma_0$ and $\Sigma_{10}$ for arbitrary $\varrho$.
The first respective lines (i.e.\ those where the entry does not depend on
$n$) hold the unsmeared values for comparison.

\begin{figure}[!hb]
\centering
\includegraphics[scale=0.5]{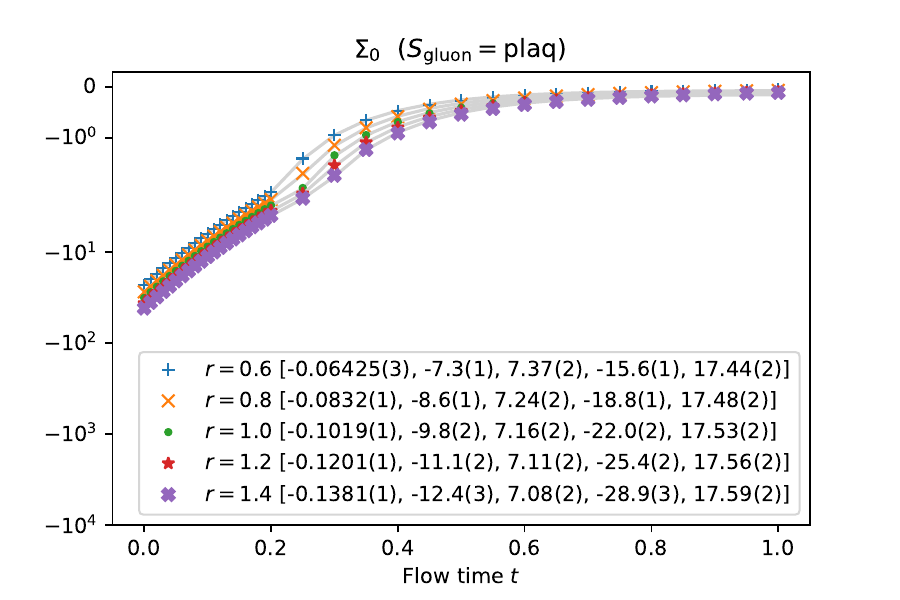}
\includegraphics[scale=0.5]{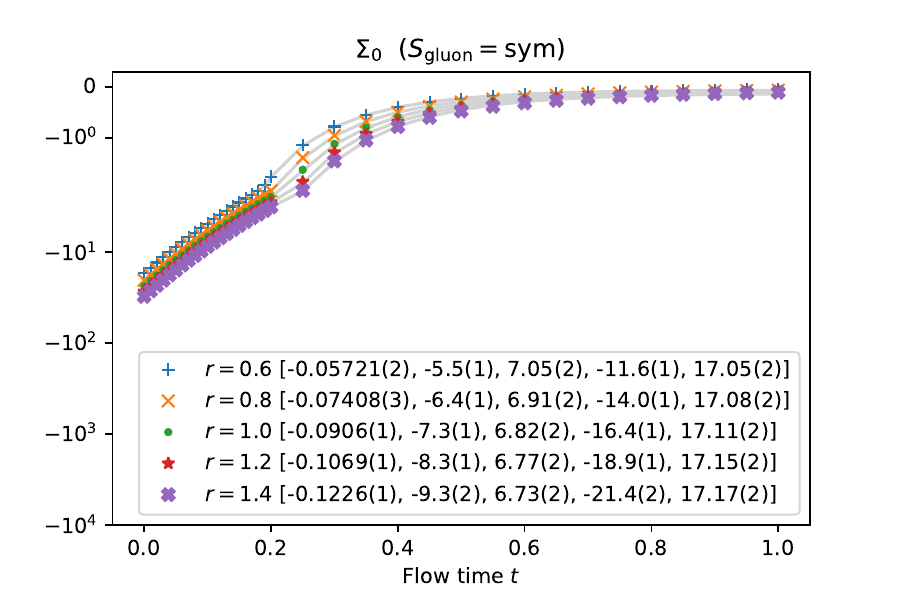}
\includegraphics[scale=0.5]{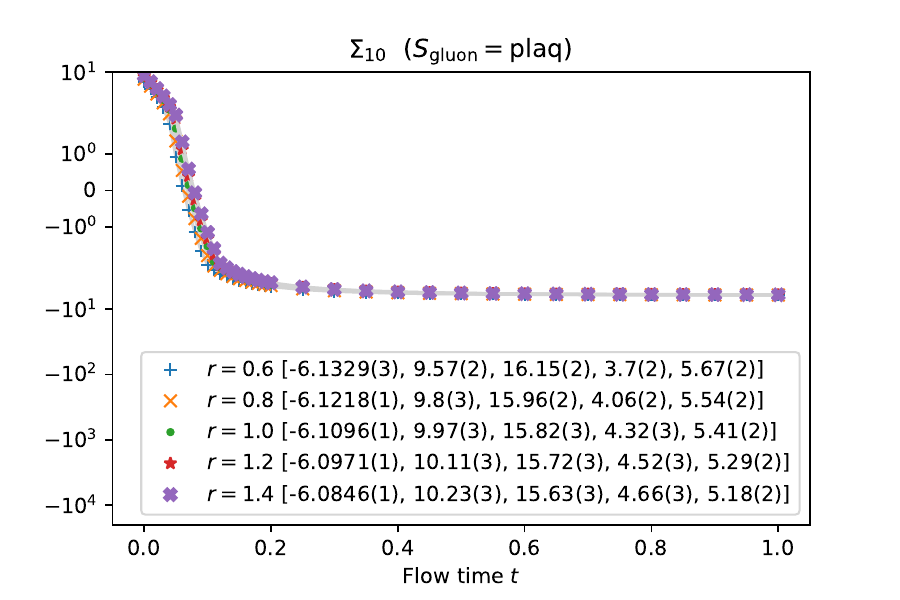}
\includegraphics[scale=0.5]{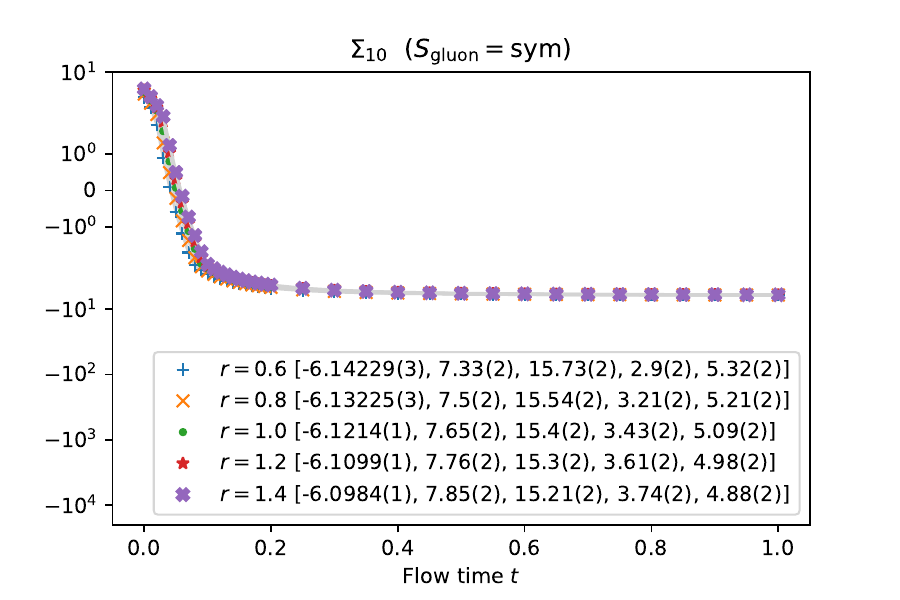}
\caption{Double exponential fits of the form $c_0+c_1e^{-c_2 t}+c_3e^{-c_4 t}$ of $\Sigma_0$ (top) and $\Sigma_{10}$ (bottom) with Wilson flow for different values of $r$ (with $\csw=r$) plotted on a logarithmic $y$ axis. The fit coefficients $[c_0,c_1,c_2,c_3,c_4]$ are given in brackets. \label{fig:S0_S1_flow_fits}}
\end{figure}

The results with Wilson flow are more cumbersome to communicate.
Fig.~\ref{fig:S0_S1_flow_fits} shows $\Sigma_0$ and $\Sigma_{10}$ for plaquette and L\"uscher-Weisz
glue as a function of the flow time in lattice units. It turns out that a
two-exponential fit represents the data with reasonable accuracy over the
range shown. The plots contain this information in the legends. Note that
these results reflect the tree-level choice $\csw=r=1$.\\

\begin{table}[!b]
\centering
\begin{tabular}{|c|l|c|c|c|c|}
\hline
 & $\varrho$ &$n=1$ & $n=2$& $n=3$ & $n=4$ \\
\hline
\multirow{6}{*}{$\Sigma_{0}$ (plaq)}	
& 0.05 & -13.67830(4) & -6.81841(4) & -3.79814(4) & -2.30(1)\\
 & 0.09 & -5.89958(4) & -2.12071(5) & -1.00136(6) & -0.56(2)\\
 & 0.11 & -4.29949(4) & -1.37888(5) & -0.62450(6) & -0.34(2)\\
 & 0.12 & -4.07175(4) & -1.27800(6) & -0.57513(6) & -0.32(3)\\
 & 0.125 & -4.10097(4) & -1.31229(6) & -0.60582(7) & -0.34(3)\\
 & 0.13 & -4.22557(4) & -1.41647(6) & -0.69334(7) & -0.42(3)\\
\hline
\multirow{6}{*}{$\Sigma_{0}$ (sym)}	
& $0.05$ 	 &-10.50283(3)&-5.36991(3)  &-3.05827(4)  &-1.89(4)  \\
& $0.09$ 	 &-4.70711(3)  &-1.76141(5  &-0.85802(8)  &-0.49(6)  \\
& $0.11$ 	 &-3.43187(3)  &-1.15565(5)  &-0.5431(2)  &-0.31(7)  \\
& $0.12$ 	 &-3.19992(3)  &-1.04722(6)  &-0.4871(2)  &-0.28(8)  \\
& $0.125$&-3.18535(3)  &-1.05160(6) &-0.4966(2) &-0.29(8)\\
& $0.13$  &-3.23840(3)  &-1.10355(6) &-0.5435(3) &-0.33(9)\\
\hline
\hline
\multirow{6}{*}{$\Sigma_{10}$ (plaq)}	
& $0.05$ 	 &0.75710(4)  &-2.32322(4)  &-3.80405(4)  &-4.599(7)  \\
& $0.09$ 	 &-2.66716(4)  &-4.65582(5)  & -5.34360(5) & -5.66(2) \\
& $0.11$ 	 &-3.53426(4) &-5.11323(5)  &-5.60877(5)  & -5.83(2) \\
& $0.12$ 	 &-3.75656(4)  &-5.23165(5)  &-5.67941(6)  &-5.88(2)  \\
& $0.125$&-3.81490(4) &-5.26045(5) &-5.69546(6) &-5.89(2)\\
& $0.13$  &-3.83802(4)  &-5.26563(5) &'-5.69366(7) &-5.88(3)\\
\hline
\multirow{6}{*}{$\Sigma_{10}$ (sym)}	
& $0.05$ 	 &-0.625159(8)  &-3.00765(1)  &-4.18067(2)  &-4.824(9)  \\
& $0.09$ 	 &-3.25805(1)  &-4.86118(2)  &-5.43657(5)  & -5.71(2) \\
& $0.11$ 	 &-3.95912(2)  &-5.24318(2)  &-5.66454(9)  &-5.86(2)  \\
& $0.12$ 	 &-4.15581(2)  &-5.35110(2)  &-5.7305(2)  & -5.90(3) \\
& $0.125$&-4.21569(2)  &-5.38307(2) &-5.7497(2) &-5.92(3)\\
& $0.13$  &-4.24993(2)  &-5.39832(2) &-5.7566(2) &-5.92(3)\\
\hline
\end{tabular}
\caption{Values of $\Sigma_0$ and $\Sigma_{10}$ for some values of the smearing parameter $\varrho$ and smearing steps $n\in[1,2,3,4]$ with plaquette gauge action (plaq) and L\"uscher-Weisz gauge action (sym). 
All information refers to the choice $\csw=r=1$. \label{tab:S0_S1_values}}
\end{table}
\newpage
Sometimes the reader may choose another value of $\csw$, e.g.\ no improvement at all  ($\csw=0$). Since (to one{-}loop order) $\Sigma_0$
and $\Sigma_{10}$ are quadratic in $\csw$, we split them according to
\begin{align}
\Sigma_0&=\Sigma_0^{(0)}+\csw\Sigma_0^{(1)}+\csw^2\Sigma_0^{(2)}
\label{eq:S0_split}\\
\Sigma_{10}&=\Sigma_{10}^{(0)}+\csw\Sigma_{10}^{(1)}+\csw^2\Sigma_{10}^{(2)}
\;.
\label{eq:S1_split}
\end{align}
and give $\Sigma_0^{(0)}$, $\Sigma_0^{(1)}$ and  $\Sigma_0^{(2)}$
in Tab.~\ref{tab:S00_S01_S02_values} and 
$\Sigma_{10}^{(0)}$, $\Sigma_{10}^{(1)} $ and $\Sigma_{10}^{(2)} $ in Tab.~\ref{tab:S10_S11_S12_values} for the same selection of $\varrho$ values as
before (the parameter in the Wilson term is always $r=1$).
Again, we expect that a simple spline interpolation will suffice
to obtain useful numbers for a different choice of $\varrho$. In the event this
expectation is wrong, Tabs.~\ref{tab:S00_coeffs} through \ref{tab:S12_coeffs} hold the necessary information to
reconstruct $\Sigma_0$ and $\Sigma_{10}$ for arbitrary $\varrho$.

\begin{figure}[p]
\centering
\includegraphics[scale=0.5]{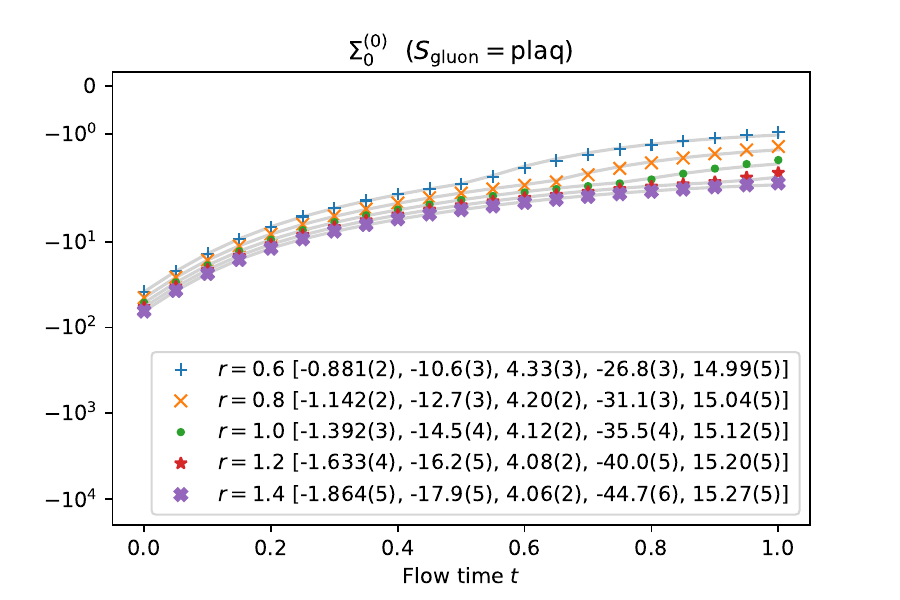}
\includegraphics[scale=0.5]{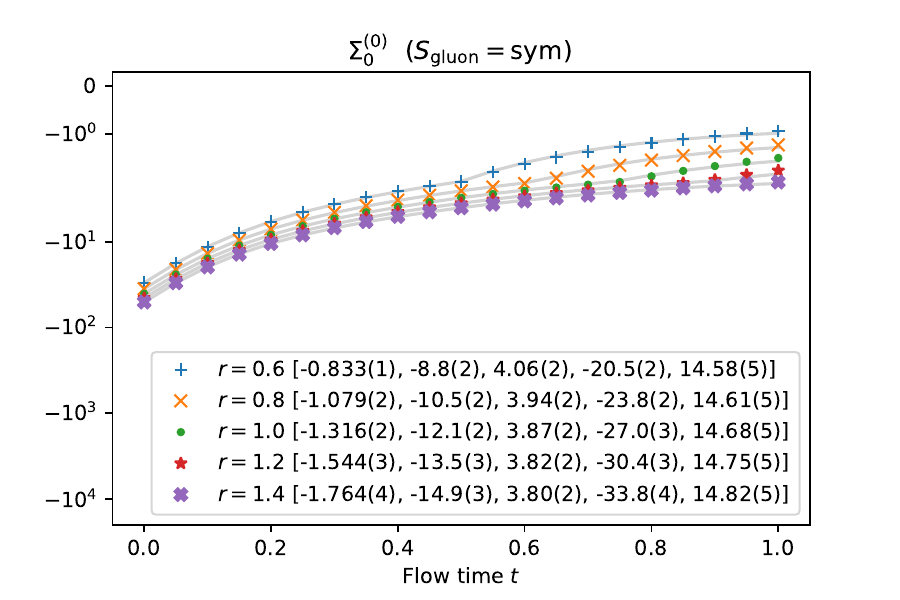}
\includegraphics[scale=0.5]{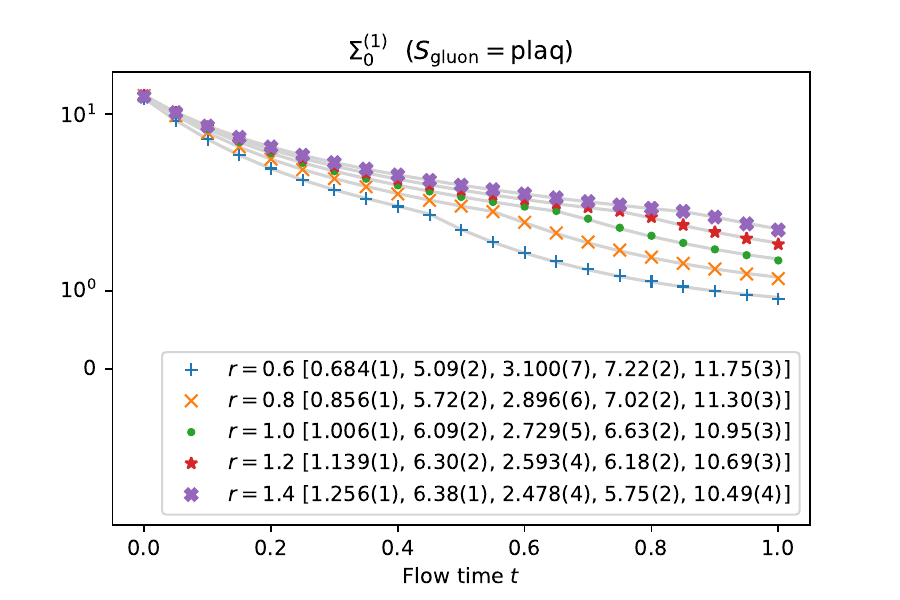}
\includegraphics[scale=0.5]{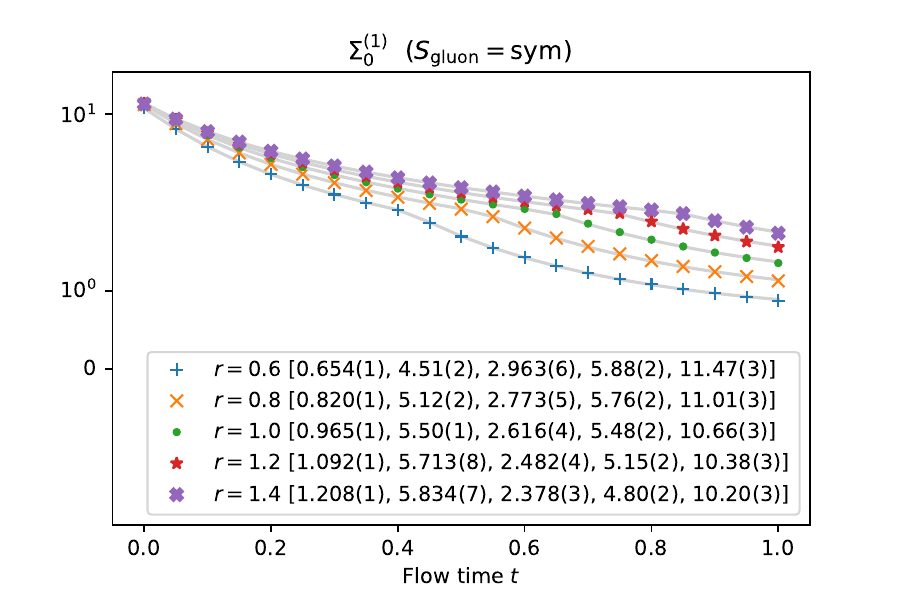}
\includegraphics[scale=0.5]{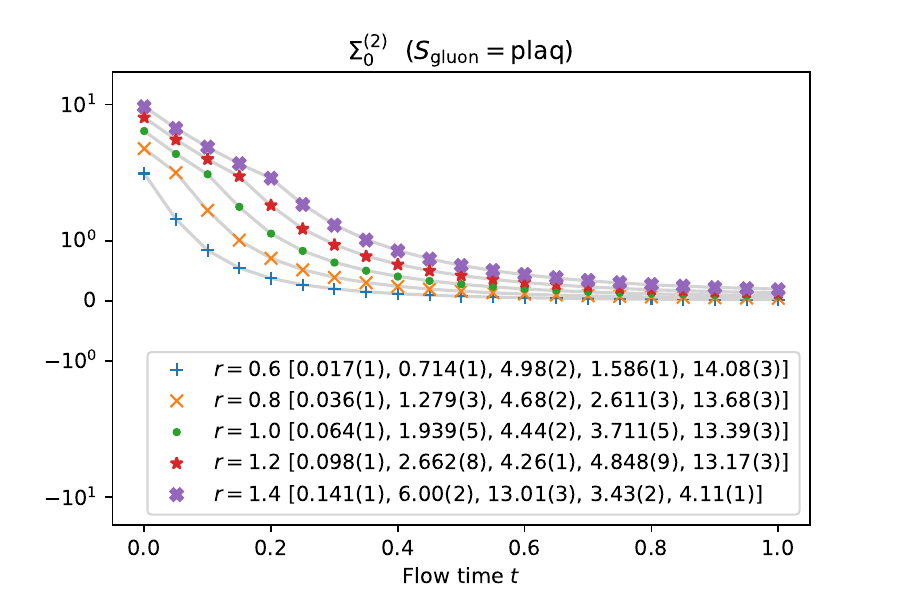}
\includegraphics[scale=0.5]{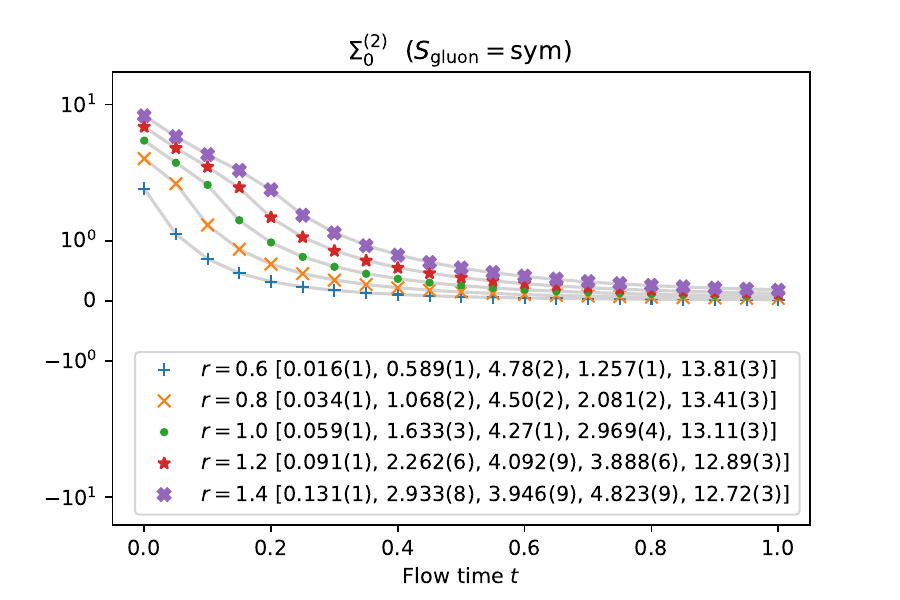}
\caption{Double exponential fits of the form $c_0+c_1e^{-c_2 t}+c_3e^{-c_4 t}$ of $\Sigma^{(0)}_0$, $\Sigma^{(1)}_0$, and $\Sigma^{(2)}_{0}$ (top to bottom) with Wilson flow for different values of $r$ plotted on a logarithmic $y$ axis. The fit coeffcients $[c_0,c_1,c_2,c_3,c_4]$ are given in brackets. \label{fig:S0_flow_splits_fits}}
\end{figure}

\begin{figure}[p]
\centering
\includegraphics[scale=0.5]{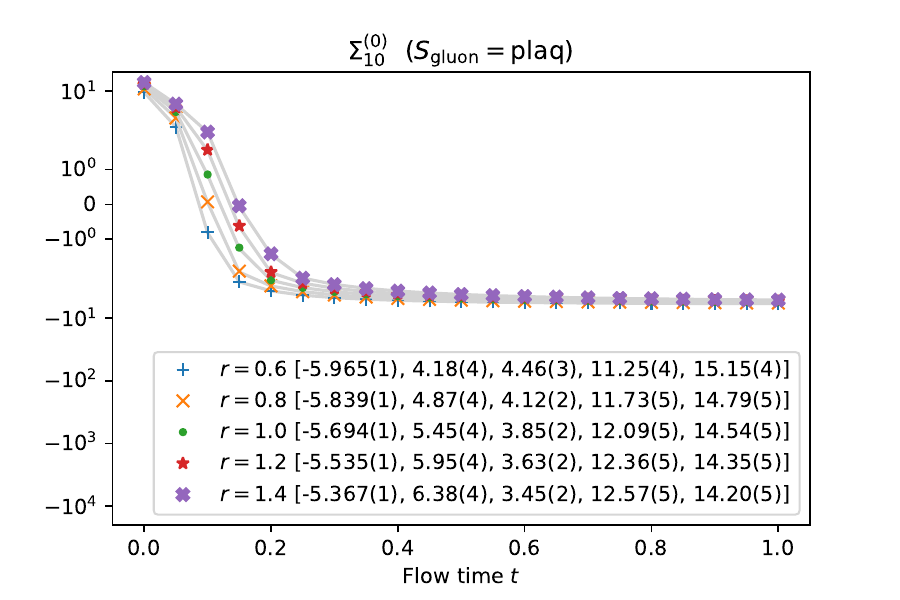}
\includegraphics[scale=0.5]{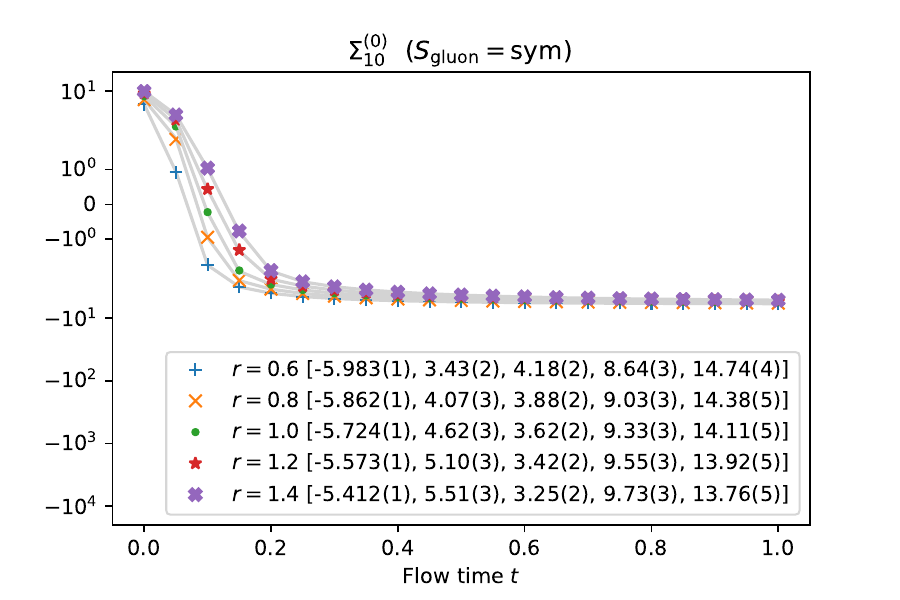}
\includegraphics[scale=0.5]{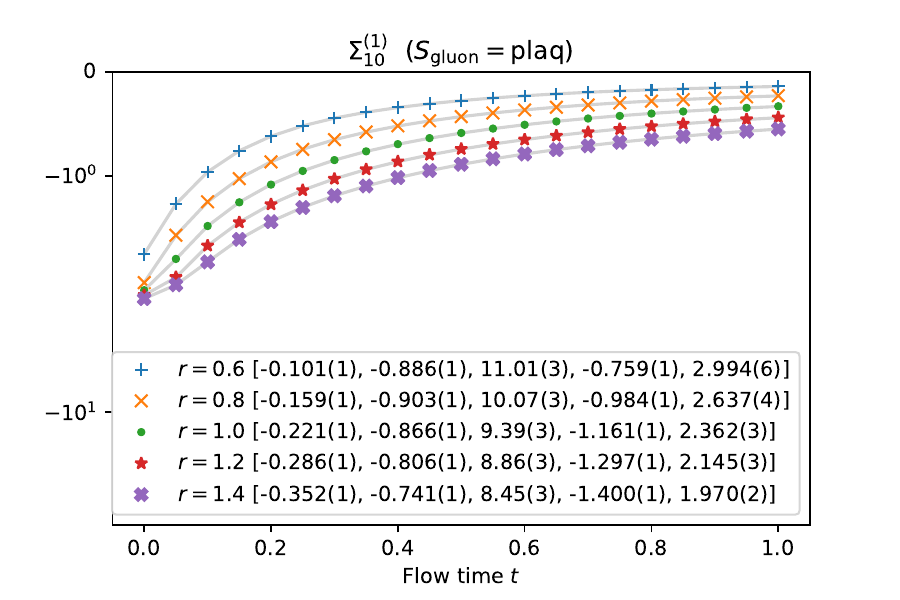}
\includegraphics[scale=0.5]{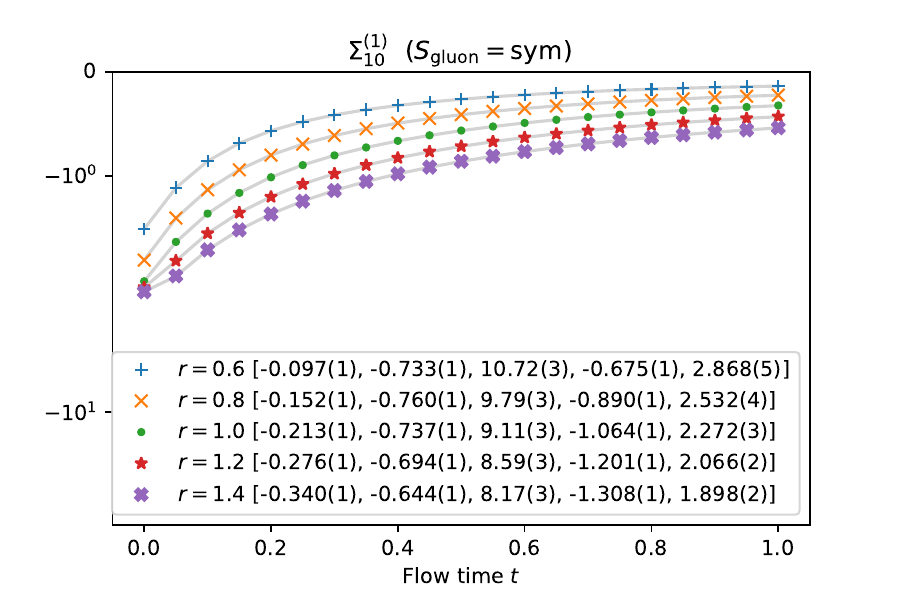}
\includegraphics[scale=0.5]{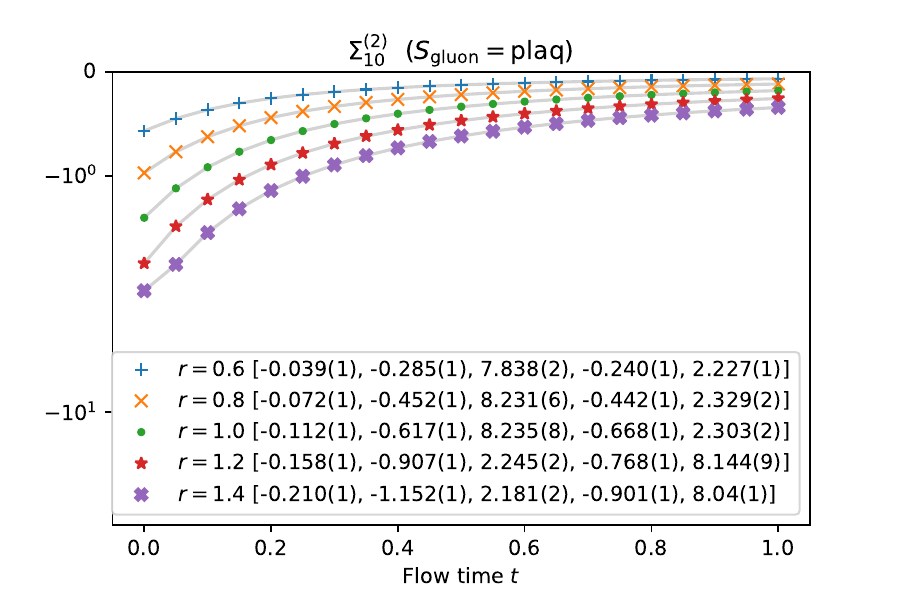}
\includegraphics[scale=0.5]{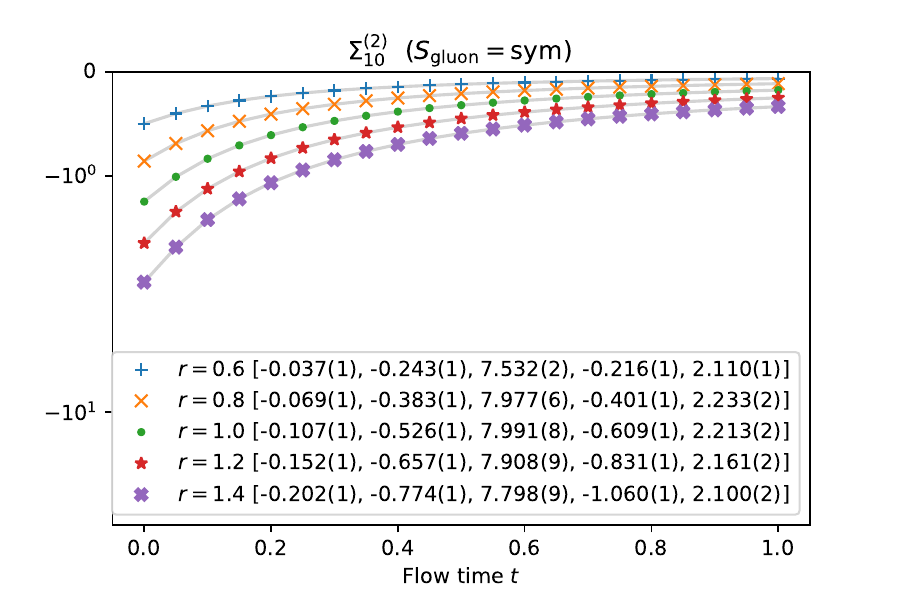}
\caption{Same as Fig.~\ref{fig:S0_flow_splits_fits} but for $\Sigma^{(0)}_{10}$, $\Sigma^{(1)}_{10}$, and $\Sigma^{(2)}_{10}$ (top to bottom).
\label{fig:S1_flow_splits_fits}}
\end{figure}

The results with Wilson flow are represented in Figs.~\ref{fig:S0_flow_splits_fits} and~\ref{fig:S1_flow_splits_fits} in the same style
as before. Again a
two-exponential fit is found to reproduce the data well over the range
shown. \\

Physics wise, the most impressive lesson to be learned from this
appendix comes from comparing the top panels of Fig.~\ref{fig:S0_S1_flow_fits} to  Fig.~\ref{fig:S0_flow_splits_fits}. With tree-level improvement (Fig.~\ref{fig:S0_S1_flow_fits}) the additive mass
shift (which is proportional to $\Sigma_0$) approaches zero quite quickly
with increasing flow time. For unimproved clover fermions this is not the
case (top row in Fig.~\ref{fig:S0_flow_splits_fits}); in fact the nice feature seen in Fig.~\ref{fig:S0_S1_flow_fits} stems
from a cancellation between $\Sigma_0^{(0)}$, $\Sigma_0^{(1)}$ and $\Sigma_0^{(2)}$ of Fig.~\ref{fig:S0_flow_splits_fits}.
\begin{table}[!h]
\centering
$\Sigma_{0}$\\
\vspace{5pt}
\begin{tabular}{|c|c|c|c|c|c|}
\hline
&& $n=1$ & $n=2$ & $n=3$ & $n=4$ \\
\hline
\multirow{2}{*}{$\varrho^0$}
					& plaq & -31.98644(3)  & -31.98644(3) & -31.98644(3) & -31.98644(3) \\
                   & sym  & -23.83234(3)  & -23.83234(3) & -23.83234(3) & -23.83234(3) \\
\hline
\multirow{2}{*}{$\varrho^1$}
					& plaq & 461.5488(1)  &  923.0975(2) & 1384.645(4)  &  1846.19(6)\\
                    & sym  &  334.1999(1)  & 668.3997(1) & 1002.5996(2) & 1336.8(6) \\
\hline
\multirow{2}{*}{$\varrho^2$}
					& plaq & -1907.719658(3) & -11446.31795(2)  & -28615.795(2)  &  -53416.150(3)\\
                    & sym  & -1352.19157(4) & -8113.149(2)  & -20282.874(1) &  -37861.370(2) \\
\hline
\multirow{2}{*}{$\varrho^3$}
					& plaq &   & 69587.70312(7) & 347938.5(8) &  974228(2)\\
                    & sym  &   & 48476.67898(2) & 242383.40(2) &  678673.4(2) \\
\hline
\multirow{2}{*}{$\varrho^4$}
					& plaq &   & -171121.8788(3)  & -2566828(2)  & -11978531(7)\\
                    & sym  &   & -117482.6334(5)  & -1762239.50(1)  &  -8223785(5) \\
\hline
\multirow{2}{*}{$\varrho^5$}
					& plaq &   &   & 10728080(5)  &  100128744(41)\\
                    & sym  &   &  & 7273681(5) &   67887689(29)\\
\hline
\multirow{2}{*}{$\varrho^6$}
					& plaq &   &   & -19627720(22)  &  -549575266(290)\\
                    & sym  &   &  & -13163619.12(3) &  -368581513(347) \\
\hline
\multirow{2}{*}{$\varrho^7$}
					& plaq &   &   &   &  1795933838(5106)\\
                    & sym  &   &  &  &  1193026157(2578) \\
\hline
\multirow{2}{*}{$\varrho^8$}
					& plaq &   &   &   &  -2658101497(19216)\\
                    & sym  &   &  &  &   -1750940862(13091)\\
\hline
\end{tabular}
\caption{Coefficients of the polynomials in the smearing parameter $\varrho$ for $\Sigma_0$ for smearing steps $n\in[1,2,3,4]$ with plaquette gauge action (plaq) and L\"uscher-Weisz gauge action (sym).\label{tab:S0_poly_coeffs}}
\end{table}
\begin{table}[h]
\centering
$\Sigma_{10}$\\
\vspace{5pt}
\begin{tabular}{|c|c|c|c|c|c|}
\hline
& & $n=1$ & $n=2$ & $n=3$ & $n=4$ \\
\hline
\multirow{2}{*}{$\varrho^0$}
					& plaq & 8.20627(3)  & 8.20627(3)   & 8.20627(3)   & 8.20627(3)   \\
                  	& sym & 4.973633(5)  & 4.973633(5) & 4.973633(5)& 4.973633(5)\\
\hline
\multirow{2}{*}{$\varrho^1$}
					& plaq  & -184.19270(6)  &  -368.3854(2)& -552.58(5)  &  -736.77(7))\\
                    & sym &  -137.61668(5) & -275.2334(1) & -412.85002(9) & -550.5(2) \\
\hline
\multirow{2}{*}{$\varrho^2$}
					& plaq  & 704.18682(5) & 4225.1209(3)  & 10562.8(2)  &  19717.2(3)\\
                   & sym  & 512.81660(2)& 3076.8996(2) & 7692.2490(3) &  14358.9(2) \\
\hline
\multirow{2}{*}{$\varrho^3$}
					& plaq  &   					& -24227.54502(9) & -121137.7(5) &  -339186(1)\\
                    & sym &   					& -17289.6263(2) & -86448.13(4) &  -242054.9(4) \\
\hline
\multirow{2}{*}{$\varrho^4$}
					& plaq  &   					& 56866.2480(2)  & 852993.70(3) & 3980637(6)\\
                    & sym &   					& 39893.8762(1)  & 598408.143(2) &  2792571.48(4) \\
\hline
\multirow{2}{*}{$\varrho^5$}
					& plaq  &  				 	&   							& -3431931(1)  &  -32031358(9)\\
                    & sym &   					&  							& -2372221(1) &   -22140728(2)\\
\hline
\multirow{2}{*}{$\varrho^6$}
					& plaq  &   					&   							& 6083759(7)	  &  170345240(173)\\
                    & sym &  					&  							& 4150857.326(4)&  116224020(336) \\
\hline
\multirow{2}{*}{$\varrho^7$}
					& plaq  &   					&   							&   							& -542124208(1774)\\
                   & sym  &   					&  							&  							&  -365651055(67) \\
\hline
\multirow{2}{*}{$\varrho^8$}
					& plaq  &   					&   							&   							& 784643279(12269)\\
                    & sym &   					&  							&  							&  523847362(2021)\\
\hline
\end{tabular}
\caption{Coefficients of the polynomials in the smearing parameter $\varrho$ for $\Sigma_{10}$ for smearing steps $n\in[1,2,3,4]$ with plaquette gauge action (plaq) and L\"uscher-Weisz gauge action (sym).\label{tab:S1_poly_coeffs}}
\end{table}
\begin{table}[p]
\centering
\begin{tabular}{|c|l|c|c|c|c|}
\hline
 & $\varrho$ &$n=1$ & $n=2$& $n=3$ & $n=4$ \\
\hline
\multirow{6}{*}{$\Sigma_{0}^{(0)}$ (plaq)}	
 & 0.05 & -26.65808(5) & -16.28605(5) & -11.14760(7) & -8.268(9)\\
 & 0.09 & -15.24425(5) & -8.17389(7) & -5.4550(2) & -4.08(2)\\
 & 0.11 & -12.33982(5) & -6.44793(8) & -4.3213(2) & -3.26(2)\\
 & 0.12 & -11.58822(6) & -5.96517(9) & -3.9821(2) & -3.00(2)\\
 & 0.125 & -11.38758(6) & -5.83167(9) & -3.8853(3) & -2.92(2)\\
 & 0.13 & -11.30371(6) & -5.7827(1) & -3.8560(3) & -2.90(2)\\
\hline
\multirow{6}{*}{$\Sigma_{0}^{(0)}$ (sym)}	
 & 0.05 & -21.88153(4) & -13.84082(5) & -9.73971(9) & -7.38(4)\\
 & 0.09 & -13.10032(4) & -7.34394(7) & -5.0335(2) & -3.83(6)\\
 & 0.11 & -10.73243(5) & -5.88197(8) & -4.0428(3) & -3.09(7)\\
 & 0.12 & -10.05417(5) & -5.43642(9) & -3.7247(4) & -2.85(8)\\
 & 0.125 & -9.84145(5) & -5.29131(9) & -3.6179(4) & -2.76(8)\\
 & 0.13 & -9.71302(5) & -5.2058(1) & -3.5577(5) & -2.72(8)\\
\hline
\multirow{6}{*}{$\Sigma_{0}^{(1)}$ (plaq)}	
 & 0.05 & 9.7541(2) & 7.4558(4) & 5.9965(6) & 5.001(4)\\
 & 0.09 & 7.4250(4) & 5.0954(7) & 3.882(1) & 3.138(6)\\
 & 0.11 & 6.5452(4) & 4.3673(8) & 3.288(2) & 2.641(8)\\
 & 0.12 & 6.1764(5) & 4.0752(9) & 3.054(2) & 2.45(1)\\
 & 0.125 & 6.0098(5) & 3.9442(9) & 2.950(2) & 2.36(2)\\
 & 0.13 & 5.8551(5) & 3.8227(9) & 2.853(2) & 2.28(2)\\
\hline
\multirow{6}{*}{$\Sigma_{0}^{(1)}$ (sym)}	
 & 0.05 & 8.6852(2) & 6.7573(3) & 5.5094(4) & 4.644(2)\\
 & 0.09 & 6.7474(3) & 4.7368(5) & 3.6600(7) & 2.987(3)\\
 & 0.11 & 6.0027(3) & 4.0976(6) & 3.1260(9) & 2.532(4)\\
 & 0.12 & 5.6865(4) & 3.8381(7) & 2.913(1) & 2.353(5)\\
 & 0.125 & 5.5423(4) & 3.7209(7) & 2.818(1) & 2.273(5)\\
 & 0.13 & 5.4075(4) & 3.6115(7) & 2.729(1) & 2.198(6)\\
\hline
\multirow{6}{*}{$\Sigma_{0}^{(2)}$ (plaq)}	
 & 0.05 & 3.2256(3) & 2.0118(3) & 1.3529(3) & 0.9637(9)\\
 & 0.09 & 1.9196(3) & 0.9578(3) & 0.5719(3) & 0.380(2)\\
 & 0.11 & 1.4952(3) & 0.7018(3) & 0.4088(3) & 0.267(3)\\
 & 0.12 & 1.3400(3) & 0.6120(3) & 0.3527(3) & 0.229(4)\\
 & 0.125 & 1.2768(3) & 0.5751(3) & 0.3296(3) & 0.213(5)\\
 & 0.13 & 1.2230(3) & 0.5436(3) & 0.3097(3) & 0.199(5)\\
\hline
\multirow{6}{*}{$\Sigma_{0}^{(2)}$ (sym)}	
 & 0.05 & 2.6935(2) & 1.7136(2) & 1.1721(2) & 0.8470(3)\\
 & 0.09 & 1.6458(2) & 0.8457(2) & 0.5155(2) & 0.348(2)\\
 & 0.11 & 1.2978(2) & 0.6287(2) & 0.3736(2) & 0.248(3)\\
 & 0.12 & 1.1678(2) & 0.5511(2) & 0.3240(2) & 0.214(5)\\
 & 0.125 & 1.1138(2) & 0.5188(2) & 0.3034(2) & 0.200(6)\\
 & 0.13 & 1.0671(2) & 0.4907(2) & 0.2853(2) & 0.187(7)\\
\hline
\end{tabular}
\caption{Same as Tab.~\ref{tab:S0_S1_values} but for $\Sigma_0^{(0)}$, $\Sigma_0^{(1)}$ and $\Sigma_0^{(2)}$  from Equation (\ref{eq:S0_split}).\label{tab:S00_S01_S02_values} }
\end{table}
\begin{table}[p]
\centering
\begin{tabular}{|c|l|c|c|c|c|}
\hline
 & $\varrho$ &$n=1$ & $n=2$& $n=3$ & $n=4$ \\
\hline
\multirow{6}{*}{$\Sigma_{10}^{(0)}$ (plaq)}	
 & 0.05 & 3.60121(2) & -0.01039(3) & -1.86437(3) & -2.934(4)\\
 & 0.09 & -0.33688(3) & -2.95019(5) & -4.00186(5) & -4.551(6)\\
 & 0.11 & -1.41830(3) & -3.61776(5) & -4.45331(7) & -4.886(7)\\
 & 0.12 & -1.73711(3) & -3.82460(6) & -4.59974(9) & -4.999(9)\\
 & 0.125 & -1.84104(3) & -3.89407(6) & -4.65020(9) & -5.039(9)\\
 & 0.13 & -1.90798(3) & -3.93775(7) & -4.6807(1) & -5.06(1)\\
\hline
\multirow{6}{*}{$\Sigma_{10}^{(0)}$ (sym)}	
 & 0.05 & 1.956497(6) & -0.882122(8) & -2.38041(2) & -3.266(8)\\
 & 0.09 & -1.114339(7) & -3.26513(2) & -4.16661(3) & -4.65(2)\\
 & 0.11 & -2.000100(8) & -3.83374(2) & -4.56328(5) & -4.95(2)\\
 & 0.12 & -2.280566(8) & -4.02084(2) & -4.69847(6) & -5.06(2)\\
 & 0.125 & -2.380196(9) & -4.08938(2) & -4.74919(7) & -5.10(2)\\
 & 0.13 & -2.452757(9) & -4.13934(2) & -4.78580(8) & -5.13(2)\\
\hline
\multirow{6}{*}{$\Sigma_{10}^{(0)}$ (plaq)}	
 & 0.05 & 3.60121(2) & -0.01039(3) & -1.86437(3) & -2.934(4)\\
 & 0.09 & -0.33688(3) & -2.95019(5) & -4.00186(5) & -4.551(6)\\
 & 0.11 & -1.41830(3) & -3.61776(5) & -4.45331(7) & -4.886(7)\\
 & 0.12 & -1.73711(3) & -3.82460(6) & -4.59974(9) & -4.999(9)\\
 & 0.125 & -1.84104(3) & -3.89407(6) & -4.65020(9) & -5.039(9)\\
 & 0.13 & -1.90798(3) & -3.93775(7) & -4.6807(1) & -5.06(1)\\
\hline
\multirow{6}{*}{$\Sigma_{10}^{(0)}$ (sym)}	
 & 0.05 & 1.956497(6) & -0.882122(8) & -2.38041(2) & -3.266(8)\\
 & 0.09 & -1.114339(7) & -3.26513(2) & -4.16661(3) & -4.65(2)\\
 & 0.11 & -2.000100(8) & -3.83374(2) & -4.56328(5) & -4.95(2)\\
 & 0.12 & -2.280566(8) & -4.02084(2) & -4.69847(6) & -5.06(2)\\
 & 0.125 & -2.380196(9) & -4.08938(2) & -4.74919(7) & -5.10(2)\\
 & 0.13 & -2.452757(9) & -4.13934(2) & -4.78580(8) & -5.13(2)\\
\hline
\multirow{6}{*}{$\Sigma_{10}^{(2)}$ (plaq)}	
 & 0.05 & -1.09264(2) & -0.88380(2) & -0.73530(2) & -0.6262(7)\\
 & 0.09 & -0.89206(2) & -0.64168(2) & -0.49704(2) & -0.404(2)\\
 & 0.11 & -0.80614(2) & -0.55755(2) & -0.42364(2) & -0.341(2)\\
 & 0.12 & -0.76678(2) & -0.52223(2) & -0.39401(2) & -0.316(3)\\
 & 0.125 & -0.74800(2) & -0.50598(2) & -0.38060(3) & -0.305(3)\\
 & 0.13 & -0.72981(2) & -0.49059(2) & -0.36803(3) & -0.295(3)\\
\hline
\multirow{6}{*}{$\Sigma_{10}^{(2)}$ (sym)}	
 & 0.05 & -0.986186(4) & -0.807676(5) & -0.678888(6) & -0.5831(4)\\
 & 0.09 & -0.815953(5) & -0.597649(7) & -0.468699(8) & -0.3850(8)\\
 & 0.11 & -0.742363(6) & -0.523351(7) & -0.40253(1) & -0.327(2)\\
 & 0.12 & -0.708450(6) & -0.491894(8) & -0.37558(1) & -0.304(2)\\
 & 0.125 & -0.692213(6) & -0.477374(8) & -0.36334(1) & -0.294(2)\\
 & 0.13 & -0.676458(6) & -0.463584(8) & -0.35183(2) & -0.284(2)\\
\hline
\end{tabular}
\caption{Same as Tab.~\ref{tab:S0_S1_values} but for $\Sigma_{10}^{(0)}$, $\Sigma_{10}^{(1)}$ and $\Sigma_{10}^{(2)}$  from Equation (\ref{eq:S1_split}).\label{tab:S10_S11_S12_values} }
\end{table}
\begin{table}[p]
\centering
$\Sigma^{(0)}_{0}$\\
\vspace{5pt}
\begin{tabular}{|c|c|c|c|c|c|c|}
\hline
& & $n=1$ & $n=2$ & $n=3$ & $n=4$ \\
\hline
\multirow{2}{*}{$\varrho^0$}
 & plaq & -51.43471(4) & -51.43471(4) & -51.43471(4) & -51.435(8)\\
 & sym & -40.44323(3) & -40.44323(3) & -40.44323(3) & -40.443(7)\\
\hline
\multirow{2}{*}{$\varrho^1$}
 & plaq & 612.3029(1) & 1224.6058(2) & 1836.9087(3) & 2449.21(3)\\
 & sym & 455.5139(1) & 911.0278(2) & 1366.5417(9) & 1822.1(6)\\
\hline
\multirow{2}{*}{$\varrho^2$}
 & plaq & -2335.4071(3) & -14012.442(2) & -35031.106(4) & -65391.6(2)\\
 & sym & -1685.5973(3) & -10113.584(3) & -25283.960(4) & -47196.889(4)\\
\hline
\multirow{2}{*}{$\varrho^3$}
 & plaq &  & 81277.32696(7) & 406386.6344(6) & 1137883.43(5)\\
 & sym &  & 57399.24192(3) & 286996.21(2) & 803589.90(8)\\
\hline
\multirow{2}{*}{$\varrho^4$}
 & plaq &  & -193630.40(3) & -2904456.0(4) & -13554132(2)\\
 & sym &  & -134388.50(2) & -2015827.5(3) & -9407196.9(4)\\
\hline
\multirow{2}{*}{$\varrho^5$}
 & plaq &  &  & 11863189.778(7) & 110723094(4)\\
 & sym &  &  & 8114869(5) & 75738794(14)\\
\hline
\multirow{2}{*}{$\varrho^6$}
 & plaq &  &  & -21329904.75(2) & -597236493(336)\\
 & sym &  &  & -14410242.41(3) & -403487007(319)\\
\hline
\multirow{2}{*}{$\varrho^7$}
 & plaq &  &  &  & 1925422531(5277)\\
 & sym &  &  &  & 1286849516(2631)\\
\hline
\multirow{2}{*}{$\varrho^8$}
 & plaq &  &  &  & -2819282113(19240)\\
 & sym &  &  &  & -1866581334(13091)\\
\hline
\end{tabular}
\caption{Same as Tab. \ref{tab:S0_poly_coeffs} but for $\Sigma^{(0)}_0$ from Eq.~(\ref{eq:S0_split}).
\label{tab:S00_coeffs}}
\end{table}
\begin{table}[p]
\centering
$\Sigma^{(1)}_{0}$\\
\vspace{5pt}
\begin{tabular}{|c|c|c|c|c|c|c|}
\hline
& & $n=1$ & $n=2$ & $n=3$ & $n=4$ \\
\hline
\multirow{2}{*}{$\varrho^0$}
 & plaq & 13.73313(4) & 13.73313(4) & 13.73313(4) & 13.732(2)\\
 & sym & 11.94822(3) & 11.94822(3) & 11.94822(3) & 11.947(2)\\
\hline
\multirow{2}{*}{$\varrho^1$}
 & plaq & -91.442(4) & -182.884(7) & -274.33(1) & -365.76(3)\\
 & sym & -74.603(3) & -149.206(5) & -223.808(8) & -298.403(4)\\
\hline
\multirow{2}{*}{$\varrho^2$}
 & plaq & 237.247540(3) & 1423.48523(3) & 3558.71310(5) & 6642.94(9)\\
 & sym & 186.845029(2) & 1121.070182(3) & 2802.6754(3) & 5231.66(3)\\
\hline
\multirow{2}{*}{$\varrho^3$}
 & plaq &  & -6094.07640(6) & -30470.3820(4) & -85317(2)\\
 & sym &  & -4689.94425(4) & -23449.721432(8) & -65659.2(8)\\
\hline
\multirow{2}{*}{$\varrho^4$}
 & plaq &  & 11191.09479(3) & 167866.4219(4) & 783374(2)\\
 & sym &  & 8465.23735(4) & 126978.5602(6) & 592567(2)\\
\hline
\multirow{2}{*}{$\varrho^5$}
 & plaq &  &  & -543236.523(8) & -5070208(26)\\
 & sym &  &  & -405150.278(4) & -3781420.66(5)\\
\hline
\multirow{2}{*}{$\varrho^6$}
 & plaq &  &  & 789161.8462(9) & 22096532(104)\\
 & sym &  &  & 581383.242(7) & 16278817(190)\\
\hline
\multirow{2}{*}{$\varrho^7$}
 & plaq &  &  &  & -58433357(1223)\\
 & sym &  &  &  & -42575391.8(3)\\
\hline
\multirow{2}{*}{$\varrho^8$}
 & plaq &  &  &  & 71058490(163)\\
 & sym &  &  &  & 51248188(2163)\\
\hline
\end{tabular}
\caption{Same as Tab. \ref{tab:S0_poly_coeffs} but for $\Sigma^{(1)}_0$ from Eq.~(\ref{eq:S0_split}).
\label{tab:S01_coeffs}}
\end{table}
\begin{table}[p]
\centering
$\Sigma^{(2)}_{0}$\\
\vspace{5pt}
\begin{tabular}{|c|c|c|c|c|c|c|}
\hline
& & $n=1$ & $n=2$ & $n=3$ & $n=4$ \\
\hline
\multirow{2}{*}{$\varrho^0$}
 & plaq & 5.7151(3) & 5.7151(3) & 5.7151(3) & 5.7150(5)\\
 & sym & 4.6627(2) & 4.6627(2) & 4.6627(2) & 4.66255(5)\\
\hline
\multirow{2}{*}{$\varrho^1$}
 & plaq & -59.3118850(8) & -118.623770(2) & -177.935654(4) & -237.248(4)\\
 & sym & -46.7112574(3) & -93.4225147(6) & -140.1337727(3) & -186.845(1)\\
\hline
\multirow{2}{*}{$\varrho^2$}
 & plaq & 190.439888(2) & 1142.639327(9) & 2856.59832(3) & 5332.32(7)\\
 & sym & 146.560758(2) & 879.36454(1) & 2198.41137(2) & 4103.70(5)\\
\hline
\multirow{2}{*}{$\varrho^3$}
 & plaq &  & -5595.54740(2) & -27977.73695(9) & -78337.66(7)\\
 & sym &  & -4232.61867(2) & -21163.09330(3) & -59256.7(2)\\
\hline
\multirow{2}{*}{$\varrho^4$}
 & plaq &  & 11317.4275(3) & 169761.412(4) & 792220(5)\\
 & sym &  & 8440.6307(2) & 126609.461(2) & 590846.979(8)\\
\hline
\multirow{2}{*}{$\varrho^5$}
 & plaq &  &  & -591871.37(2) & -5524133(26)\\
 & sym &  &  & -436037.438(2) & -4069704(48)\\
\hline
\multirow{2}{*}{$\varrho^6$}
 & plaq &  &  & 913021.209(7) & 25564594(2)\\
 & sym &  &  & 665240.039(6) & 18626734(383)\\
\hline
\multirow{2}{*}{$\varrho^7$}
 & plaq &  &  &  & -71058490(163)\\
 & sym &  &  &  & -51248188(2163)\\
\hline
\multirow{2}{*}{$\varrho^8$}
 & plaq &  &  &  & 90125153(5113)\\
 & sym &  &  &  & 64387254(3655)\\
\hline
\end{tabular}
\caption{Same as Tab. \ref{tab:S0_poly_coeffs} but for $\Sigma^{(2)}_0$ from Eq.~(\ref{eq:S0_split}).
\label{tab:S02_coeffs}}
\end{table}
\begin{table}[p]
\centering
$\Sigma^{(0)}_{10}$\\
\vspace{5pt}
\begin{tabular}{|c|c|c|c|c|c|c|}
\hline
& & $n=1$ & $n=2$ & $n=3$ & $n=4$ \\
\hline
\multirow{2}{*}{$\varrho^0$}
 & plaq & 11.852396(8) & 11.852396(8) & 11.852396(8) & 11.852(2)\\
 & sym & 8.231254(4) & 8.231254(4) & 8.231254(4) & 8.231(2)\\
\hline
\multirow{2}{*}{$\varrho^1$}
 & plaq & -202.0079(2) & -404.0158(3) & -606.0237(2) & -808.03(4)\\
 & sym & -152.56416(3) & -305.12832(6) & -457.6925(2) & -610.3(2)\\
\hline
\multirow{2}{*}{$\varrho^2$}
 & plaq & 739.6836(2) & 4438.102(1) & 11095.255(3) & 20711.14(3)\\
 & sym & 541.38051(4) & 3248.2831(3) & 8120.7077(3) & 15158.66(4)\\
\hline
\multirow{2}{*}{$\varrho^3$}
 & plaq &  & -24956.6947(4) & -124783.4729(5) & -349393.7(5)\\
 & sym &  & -17858.09191(3) & -89290.4593(2) & -250013.3(2)\\
\hline
\multirow{2}{*}{$\varrho^4$}
 & plaq &  & 57973.619(4) & 869604.29(5) & 4058153.0(5)\\
 & sym &  & 40735.013(3) & 611025.19(5) & 2851450.82(9)\\
\hline
\multirow{2}{*}{$\varrho^5$}
 & plaq &  &  & -3477653.965(2) & -32458091(8)\\
 & sym &  &  & -2406174(1) & -22457620(2)\\
\hline
\multirow{2}{*}{$\varrho^6$}
 & plaq &  &  & 6141663(3) & 171966439(99)\\
 & sym &  &  & 4192996.284(4) & 117403911(55)\\
\hline
\multirow{2}{*}{$\varrho^7$}
 & plaq &  &  &  & -545947087(1804)\\
 & sym &  &  &  & -368383101(142)\\
\hline
\multirow{2}{*}{$\varrho^8$}
 & plaq &  &  &  & 788877857(5503)\\
 & sym &  &  &  & 526824035(1817)\\
\hline
\end{tabular}
\caption{Same as Tab. \ref{tab:S1_poly_coeffs} but for $\Sigma^{(0)}_{10}$ from Eq.~(\ref{eq:S1_split}).
\label{tab:S10_coeffs}}
\end{table}
\begin{table}[p]
\centering
$\Sigma^{(1)}_{10}$\\
\vspace{5pt}
\begin{tabular}{|c|c|c|c|c|c|c|}
\hline
& & $n=1$ & $n=2$ & $n=3$ & $n=4$ \\
\hline
\multirow{2}{*}{$\varrho^0$}
 & plaq & -2.248869(6) & -2.248869(6) & -2.248869(6) & -2.249(2)\\
 & sym & -2.015426(5) & -2.015426(5) & -2.015426(5) & -2.01543(5)\\
\hline
\multirow{2}{*}{$\varrho^1$}
 & plaq & 11.123836(8) & 22.24767(2) & 33.37151(3) & 44.497(3)\\
 & sym & 9.346879(5) & 18.693758(9) & 28.040642(8) & 37.388(3)\\
\hline
\multirow{2}{*}{$\varrho^2$}
 & plaq & -23.518665(6) & -141.11198(3) & -352.77996(7) & -658.52(8)\\
 & sym & -18.958350(2) & -113.75010(2) & -284.37525(2) & -530.83(5)\\
\hline
\multirow{2}{*}{$\varrho^3$}
 & plaq &  & 525.716776(9) & 2628.58386(7) & 7360.035(6)\\
 & sym &  & 411.66550(4) & 2058.3275(2) & 5763.32(2)\\
\hline
\multirow{2}{*}{$\varrho^4$}
 & plaq &  & -878.20974(7) & -13173.1458(7) & -61475(3)\\
 & sym &  & -672.98584(5) & -10094.788(2) & -47107.302(7)\\
\hline
\multirow{2}{*}{$\varrho^5$}
 & plaq &  &  & 39917.11(2) & 372559.6(2)\\
 & sym &  &  & 30072.058(6) & 280672.57(9)\\
\hline
\multirow{2}{*}{$\varrho^6$}
 & plaq &  &  & -55331.94(5) & -1549295(216)\\
 & sym &  &  & -41099.241(6) & -1150778.8(7)\\
\hline
\multirow{2}{*}{$\varrho^7$}
 & plaq &  &  &  & 3959099.6(6)\\
 & sym &  &  &  & 2904907(3)\\
\hline
\multirow{2}{*}{$\varrho^8$}
 & plaq &  &  &  & -4693170(2)\\
 & sym &  &  &  & -3406008(4)\\
\hline
\end{tabular}
\caption{Same as Tab. \ref{tab:S1_poly_coeffs} but for $\Sigma^{(1)}_{10}$ from Eq.~(\ref{eq:S1_split}).
\label{tab:S11_coeffs}}
\end{table}
\begin{table}[p]
\centering
$\Sigma^{(2)}_{10}$\\
\vspace{5pt}
\begin{tabular}{|c|c|c|c|c|c|c|}
\hline
& & $n=1$ & $n=2$ & $n=3$ & $n=4$ \\
\hline
\multirow{2}{*}{$\varrho^0$}
 & plaq & -1.397261(9) & -1.397261(9) & -1.397261(9) & -1.3973(3)\\
 & sym & -1.242203(4) & -1.242203(4) & -1.242203(4) & -1.24223(9)\\
\hline
\multirow{2}{*}{$\varrho^1$}
 & plaq & 6.69138(4) & 13.38277(7) & 20.07415(9) & 26.762(8)\\
 & sym & 5.60061(2) & 11.20122(4) & 16.80183(6) & 22.399(5)\\
\hline
\multirow{2}{*}{$\varrho^2$}
 & plaq & -11.978147(2) & -71.868869(6) & -179.67220(2) & -335.388(8)\\
 & sym & -9.6055598(7) & -57.633356(1) & -144.083397(9) & -268.980(2)\\
\hline
\multirow{2}{*}{$\varrho^3$}
 & plaq &  & 203.43279(2) & 1017.16394(6) & 2848.1(5)\\
 & sym &  & 156.799846(8) & 783.99926(3) & 2195.2(4)\\
\hline
\multirow{2}{*}{$\varrho^4$}
 & plaq &  & -229.16130(5) & -3437.4195(4) & -16041.293(2)\\
 & sym &  & -168.15048(2) & -2522.2572(4) & -11770.532(2)\\
\hline
\multirow{2}{*}{$\varrho^5$}
 & plaq &  &  & 5804.3207(6) & 54173.62(4)\\
 & sym &  &  & 3880.9479(9) & 36222.19(2)\\
\hline
\multirow{2}{*}{$\varrho^6$}
 & plaq &  &  & -2569.25(2) & -71939.0(4)\\
 & sym &  &  & -1039.729(6) & -29112.3(3)\\
\hline
\multirow{2}{*}{$\varrho^7$}
 & plaq &  &  &  & -136222(83)\\
 & sym &  &  &  & -172862(90)\\
\hline
\multirow{2}{*}{$\varrho^8$}
 & plaq &  &  &  & 458914(149)\\
 & sym &  &  &  & 428921(211)\\
\hline
\end{tabular}
\caption{Same as Tab. \ref{tab:S1_poly_coeffs} but for $\Sigma^{(2)}_{10}$ from Eq.~(\ref{eq:S1_split}).
\label{tab:S12_coeffs}}
\end{table}

\clearpage
\printbibliography

\end{document}